\documentclass{aa} 
\usepackage{graphicx}
\usepackage{txfonts}
\usepackage{color}
\usepackage{ulem}
\usepackage{tablefootnote}
\usepackage{float} 
\usepackage{longtable}
\usepackage[labelfont=small,bf]{caption}
\usepackage{amsmath}
\usepackage[english]{babel}
\usepackage{comment}
\usepackage{multirow}
\usepackage{wasysym}
\usepackage{hyperref}
\usepackage{natbib}
\usepackage{lscape}
\bibpunct{(}{)}{;}{a}{}{,} 
\hypersetup{colorlinks=true,citecolor=blue}

\begin{document}

 \title{Reaching the boundary between stellar kinematic groups and very wide binaries}

 \subtitle{IV. The widest Washington Double Star systems with \\$\rho\geq 1000$\,arcsec in {\it Gaia} DR3\thanks{Tables B.1, B.2, B.3, and B.4 are only available in electronic form at the CDS via anonymous ftp to cdsarc.cds.unistra.fr (130.79.128.5) or via https://cdsarc.cds.unistra.fr/cgi-bin/qcat?J/A+A/}}

 \titlerunning{The widest Washington Double Star systems with $\rho\geq 1000$\,arcsec in {\it Gaia} DR3}

 \authorrunning{J.~Gonz\'alez-Payo et al.}
 
 \author{J.~Gonz\'alez-Payo\inst{1}, J.\,A.~Caballero\inst{2},
 M.~Cort\'es-Contreras\inst{2}}

 \institute{Departamento de F\'isica de la Tierra y Astrof\'isica, Facultad de Ciencias F\'isicas, Universidad Complutense de Madrid, E-28040 Madrid, Spain.
  \email{fcojgonz@ucm.es}
 \and
 Centro de Astrobiolog\'ia, CSIC-INTA, Camino Bajo del Castillo s/n, campus ESAC, E-28692 Villanueva de la Ca\~nada, Spain
  }

 \date{Received 16 November 2022 / Accepted 20 December 2022}
 
 \abstract
 {}
 {
With the latest {\it Gaia} DR3 data, we analyse the widest pairs in the Washington Double Star (WDS) catalogue with angular separations, $\rho$, greater than 1000\,arcsec.
 }
 {
We confirmed the pairs' membership to stellar systems based on common proper motions, parallaxes, and (when available) radial velocities, together with the locii of the individual components in colour-magnitude diagrams.
We also looked for additional closer companions to the ultrawide pairs, either reported by WDS or found by us with a new {\it Gaia} astrometric search.
In addition, we determined masses for each star (and white dwarf) and, with the projected physical separation, computed the gravitational potential energy, $|U^*_g|$, of the systems.
 }
{Of the 155\,159 pairs currently catalogued by WDS, there are 504 with $\rho >$ 1000\,arcsec.
Of these,
only 2 ultrawide pairs have  not been identified, 
10 do  not have any available  astrometry, 
339 have not passed a conservative filtering in proper motion or parallax,
59 are members of young stellar kinematic groups, associations or open clusters,
and only 94 remain as bona fide ultrawide pairs in the galactic field.
Accounting for the additional members at shorter separations identified in a complementary astrometric and bibliographic search, we found 79 new stars (39 reported, plus 40 not reported by WDS) in 94 ultrawide stellar systems. This sample is expanded when including new close binary candidates with large {\it Gaia} DR3 {\tt RUWE}, $\sigma_{Vr}$, or a proper motion anomaly.
Furthermore, the large fraction of subsystems and the non-hierarchical configurations of many wide systems with three or more stars is remarkable. In particular, we found 14 quadruple, 2 quintuple, 3 sextuple, and 2 septuple systems.
The minimum computed binding energies, $|U^*_g| \sim 10^{33}$\,J, are in line with theoretical predictions of tidal destruction by the Galactic gravitational potential.
The most fragile and massive systems have huge projected physical separations of well over 1\,pc. Therefore, they are either in the process of disruption or they are part of unidentified juvenile stellar kinematic groups.
 }
 {}

 \keywords{surveys -- virtual observatory tools -- astrometry -- stars: binaries: general, visual} 

 \maketitle

\section{Introduction}
\label{sec:introduction}

Multiple stars have been observed since ancient times, but it has been accepted for millenia that the proximity of two stars was mainly due to chance \citep{fracastoro88}. 
In the 17th century, Galileo was the first to propose the association between stars when trying to measure stellar parallaxes, based on the recommendations of Tycho Brahe \citep{hirshfeld01}. 
The first visual binary, Mizar A and B ($\zeta$ Ursae Majoris), was discovered by Benedetto Castelli, who asked Galileo for his observations of it in 1616, although the discovering was falsely attributed to Giovanni Battista Riccioli in 1650 \citep{allen1899,burnham78}. The first catalogue of binary stars was published in 1781 by Christian Mayer, who speculated about the possibility of them being physical systems, as predicted by Isaac Newton \citep{niemela01}. 
Nevertheless, in the following century, William~F. Herschel questioned that idea, considering that multiple systems could have a gravitational link only when their orbital motion were proven \citep{fracastoro88,niemela01}.
Some years later, he published a work based on his observations \citep{herschel1802}, where he demonstrated that some real star systems were ruled by the universal gravitation laws \citep{niemela01}.
It was the first time that science confirmed that Newton’s laws are also valid outside the Solar System, which sparked a new revolution.

A double or binary system contains two stars that describe closed orbits around their common centre of gravity \citep{batten73}, while multiple systems contain three or more stars with different hierarchical levels \citep{tokovinin97,tokovinin08,eggleton08,duchene13}.
The components of wide multiple systems have large separations between them and, therefore, relatively low gravitational energies.
The classical maximum separation between components in wide systems rarely exceeds 0.1\,pc, driven by the dynamic processes of the stars formation and evolution \citep{tolbert64,kraicheva85,abt88,weinberg88,close90,latham91,wasserman91,garnavich93,allen98,caballero09} and strongly depends on their mass (i.e. spectral type), age, and kinematics \citep{duquennoy91,jensen93,patience02,zapatero05,kraus09}. 
There are newer studies that increase this maximum separation up to 1\,pc \citep{jiang10,caballero10} or even to 1--8\,pc \citep{shaya11,kirkpatrick16,gonzalezpayo21a}. 
At these separations, the pairs are less likely bound for extended lifetimes \citep{retterer82,weinberg87,dhital10}.

In this work, we perform a detailed characterisation of the widest pairs in the Washington Double Star (WDS) catalogue \citep{mason01} by making use of the latest \textit{Gaia} DR3 data \citep{gaiacollaboration22}.
The WDS, which is maintained by the United States Naval Observatory, is the world's principal database of astrometric double and multiple star information.
For each system, we ascertain their actual gravitational binding and search for additional companions.
Since we are investigating pairs with angular separations, $\rho$, greater than 1000\,arcsec, this work can be understood as a {\it Gaia} update of that by \citet{caballero09}, who also used $\rho =$ 1000\,arcsec as the minimum separation between the widest WDS pairs at that time, but had only {\it Hipparcos} \citep{perryman97} parallaxes for a few bright stars and relatively insufficient proper motions for the faintest components.
Furthermore, this work is the fourth item in the series initiated by \citet{caballero09}, which aims to shed light from an observational perspective on the formation and evolution of the most separated and fragile multiple stellar systems in the Milky Way.
Although young systems play an important role in our analysis, here, we focus on field systems that are relatively evolved, old, and at the brink of disruption by the galactic gravitational potential.

This paper is structured as follows: In Sect.~\ref{sec:sample}, we describe the stellar sample. 
Section~\ref{sec:analysis} shows the analysis that we followed to filter, classify, and characterise WDS pairs, as well as to carry out our search for other possible members of the multiple systems. 
We present our results, along with a discussion in Sect.~\ref{sec:results_discussion}.
Finally, we summarise our work in Sect.~\ref{sec:summary}.

\section{Sample}
\label{sec:sample}

\begin{figure}[]
 \centering
 \includegraphics[width=0.99\linewidth, angle=0]{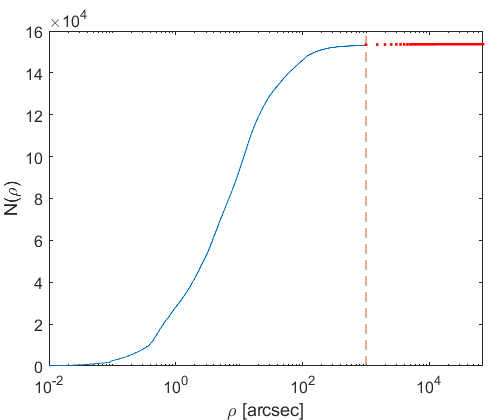}
 \caption{Cumulative number of WDS pairs as a function of $\rho$. 
 Red data points with $\rho>1000$\,arcsec (to the right of the orange vertical dashed line) mark the 504 WDS pair candidates investigated by us. 
This figure can be compared with Fig.~1 in \cite{caballero09}.
}
 \label{fig:acumhist}
\end{figure}

We built our sample from the latest version of the WDS\footnote{\url{http://www.astro.gsu.edu/wds/Webtextfiles/wds_precise.txt}, accessed on 12 November 2022.}.
For each of the 155\,159 resolved pairs, WDS tabulates the WDS identifier (based on J2000 position), 
discoverer code and number, 
number of observations and of components (when there are more than two),
date, position angle ($\theta$, i.e. orientation on the celestial plane of the companion with respect to the primary), and $\rho$ of the first and last observations,
magnitudes, and proper motions of the two components,
along with the equatorial coordinates of the primary of the pair 
and notes about the pair.
In a few cases, WDS also tabulates the Durchmusterung number (Bonn, C\'ordoba, Cape -- \citealt{schonfeld1886}; \citealt{argelander03}) and the spectral type of the primary or companion (or both).
There are numerous pairs that take part of multiple systems with, usually, the same primary star; in general, they share the same WDS identifier, but not always.

In Fig.\,\ref{fig:acumhist}, the cumulative number of WDS pair angular separations increases with a power law between $\rho \sim0.4$\,arcsec and $\rho \sim100$\,arcsec.
This distribution follows \"Opik's law \citep{opik24} for binaries with projected physical separations greater than 25\,au \citep{allen97}.
Outside the $\rho \sim$ 0.4--100\,arcsec range, the distribution flattens at both sides.
This flattening is an observational bias at short angular separations, as micrometer, speckle, lucky imaging, adaptive optics, and even imaging from space are limited by the atmospheric seeing, telescope size, or optical quality (but there seems to be a slight overabundance of close pairs of $\rho \sim$ 0.4--4.0\,arcsec with respect to more separated ones).

\begin{figure}[]
 \centering
 \includegraphics[width=0.99\linewidth, angle=0]{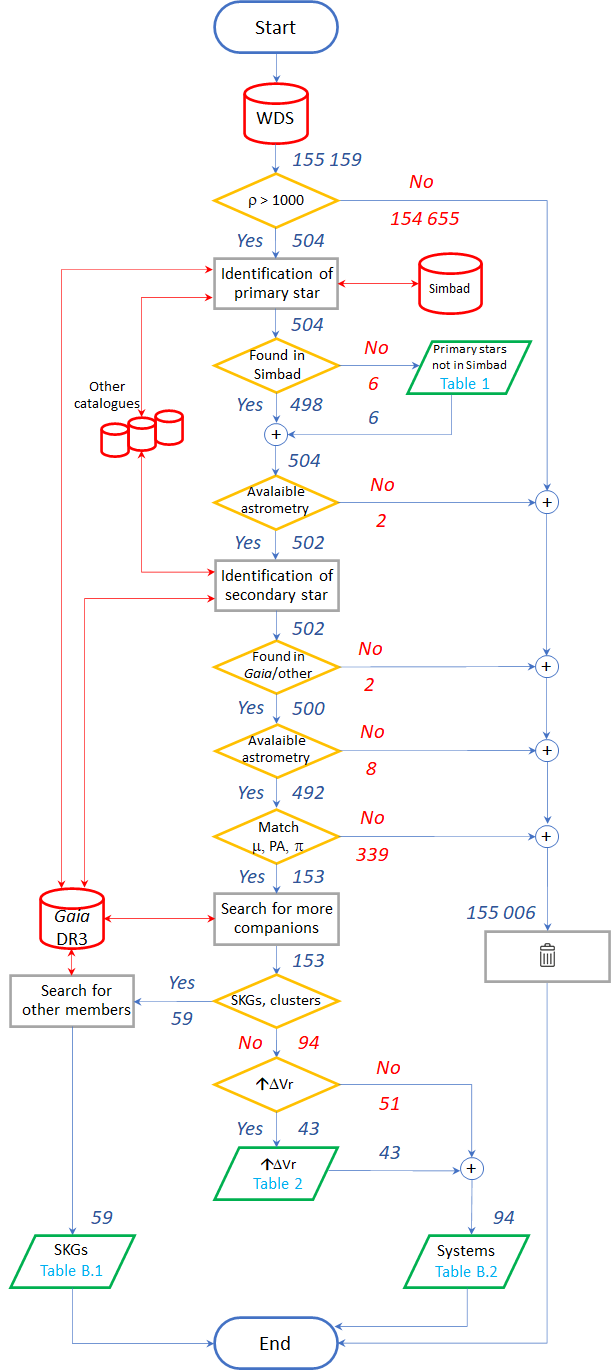}
 \caption{Flowchart describing our analysis. }
 \label{fig:flowchart}
\end{figure}

The flattening of the $\rho$ distribution at wide separations, especially at $\rho \sim$ 200\,arcsec, is mainly due to the actual formation and evolution of multiple stellar systems, although there may also be a contribution from another observational bias: until the advent of \textit{Gaia} \citep{gaiacollaboration16,gaiacollaboration18,gaiacollaboration21}, accurate proper motion and parallax measurements were available only for a tiny fraction of stars, while most wide WDS pairs come from pre-{\it Gaia} common proper motion surveys (e.g. \citealt{allen00}; \citealt{chaname04}; \citealt{lepine07a}; \citealt{dhital10}; \citealt{raghavan10}; \citealt{tokovinin12}; and references therein\footnote{There are also relevant unpublished contributions to the WDS, such as that of the Observatori Astron\`omic del Garraf \citep{caballero13}. See further details at \url{http://www.astro.gsu.edu/wds/wdstext.html\#intro}.}).
In spite of numerous common proper motion surveys, the observational bias remains at $\rho \gtrsim$ 200\,arcsec because most of them looked for companions at angular separations of up to a few arcminutes only, mostly due to past computational limitations. 
However, this difficulty is starting to be alleviated thanks to new {\it Gaia} surveys (e.g. \citealt{kervella22,sarro22}).
Due to the observational bias or the actual difficulty in forming wide binaries \citep{kouwenhoven10,reipurth12,lee17,tokovinin17}, the distribution of ultrawide WDS pairs with $\rho \gtrsim$ 1000\,arcsec becomes extremely flat (Fig.~\ref{fig:acumhist}).

At the time of our analysis, WDS contained 504 pairs separated by more than 1000\,arcsec.
For comparison, \citet{caballero09} investigated about 105\,000 WDS pairs, of which only 35 had $\rho >$ 1000\,arcsec.
Together with the {\it Gaia} DR3 data, a sample about 15 times larger represents a qualitative and quantitative leap with respect to the first item of this paper series.

\section{Analysis}
\label{sec:analysis}

Our analysis procedure is illustrated by the flowchart in Fig.~\ref{fig:flowchart}, with the following elements: ovals represent the initial and final status of the process, diamonds are Boolean questions for which only possible answers are `Yes' or `No', 
rectangles are specific actions, trapezia are partial or final obtained results, and cylinders are databases (or catalogues) from where the data are collected or consulted. 
In Fig.~\ref{fig:flowchart}, the incoming numbers in every block represent the number of processed WDS pairs in that block, and the outgoing numbers represent the number of pairs that match the condition or pass through to the next block of the flowchart.

\subsection{Primary stars in \textit{Gaia} DR3}
\label{sec:primary_stars}

\begin{table*}
 \centering
 \caption[]{Primary stars without a Simbad entry.}
 \begin{tabular}{cllcccc}
 \hline \hline
 \noalign{\smallskip}
 WDS & Discoverer & Primary star & $\alpha$ & $\delta$ & \textit{G} & \textit{d} \\
 name & code &  & (J2000) & (J2000) & (mag) & (pc) \\
 \noalign{\smallskip}
 \hline
 \noalign{\smallskip}
00474--7345 & OGL 84 & Gaia DR3 4685766099704705280 & 00:47:25.09 & $-$73:44:42.5 & 17.5 & 1700$\pm$200 \\
00489--7434 & OGL 87 & Gaia DR3 4685482661910629248 & 00:48:55.94 & $-$74:33:46.7 & 19.4 & 504$\pm$54 \\
01121--7400 & OGL 161 & Gaia DR3 4686310976416294528 & 01:12:11.20 & $-$73:59:42.9 & 16.6 & 1870$\pm$150 \\
01235--7356 & OGL 188 & Gaia DR3 4686225867342046848 & 01:23:33.07 & $-$73:55:34.7 & 14.3 & 468.0$\pm$3.1 \\
03074--4655 & TSN 110 & Gaia DR3 4750712533547201920 & 03:07:26.03 & $-$46:54:44.8 & 16.6 & 136.0$\pm$1.0 \\
10181--0130 & TSN 113 & Gaia DR3 3830436797339858176 & 10:18:03.35 & $-$01:30:11.6 & 17.8 & 397$\pm$21 \\
 \noalign{\smallskip}
 \hline
 \end{tabular}
 \label{tab:no_simbad}
\end{table*}
For each primary star of the 504 WDS ultrawide pairs, we collected its main identifier in the Simbad astronomical database \citep{wenger00}. 
Of them, only six do not have a Simbad entry; they are shown in Table~\ref{tab:no_simbad}.
Using {\tt TOPCAT} \citep[][]{taylor05} and the equatorial coordinates tabulated by WDS, we automatically cross-matched every primary with {\it Gaia} DR3.
Next, we visually inspected all cross-matches with the help of the Aladin sky atlas \citep{bonnarel00}, and VizieR \citep{ochsenbein00}.
When there were more than a single {\it Gaia} counterpart per primary within our $\sim$5\,arcsec cross-match radius, we chose the right star by comparing WDS and {\it Gaia} proper motions, magnitudes, and spectral types.

Of the 504 primaries, 44 are redundant (they belong to two or more pairs). 
Only 6 out of the 460 non-redundant primaries had no {\it Gaia} DR3 entry because of their extreme brightness ($\alpha$~Aur --Capella--, $\alpha$~Car --Canopus--, $\alpha$~PsA --Fomalhaut--, $\alpha$~Cen, $\gamma$~Cen, and $\delta$~Vel), for which we took proper motions and parallaxes from \textit{Hipparcos}.
In addition, another 22 primaries are in {\it Gaia} DR3, but do not have a five-parameter astrometric solution.
For 11 of them, namely, those of moderate brightness ($G$ = 2.3--9.5\,mag), we again took proper motions and parallaxes from \textit{Hipparcos}, while for 9 of them, we took the data from {\it Gaia} DR2 \citep{gaiacollaboration18}\footnote{The nine primary stars with {\it Gaia} DR2 data are: LP~295--49, G~202--45, 36~And~A, 4~Sex, HD~111456, HD~125354, HD~340345, BD-12~6174, and HD~213987.}.
The 2 remaining stars are \object{LSPM~J2323+6559} and \object{2MASS~J00202956--1535280}, for which there are only proper motions available from \citet{lepine05d} and \citet{cutri14}, respectively.
Since the 2 later primaries do not have published parallaxes, we discarded the corresponding pairs from the analysis.
Accounting for these two discards, we retained 502 pairs for the next step of the analysis.

\subsection{Search for WDS companions}
\label{sec:search_for_wds_companions}

The WDS catalogue provides the relative positions of the companion stars of the pairs with respect to the primaries through $\rho$ and $\theta$. 
To manually locate the companions to the primaries, we used Aladin.
We loaded different catalogues and services, namely {\it Gaia} DR3, 2MASS \citep{skrutskie06}, Simbad, and WDS, and we used the {\tt dist} tool. 
For the correct identification of the companion, we proceeded to carry out a visual confirmation of the primary cross-match and we chose the {\it Gaia} DR3 candidate companion within 10\,arcsec around the expected location that matched the WDS values of proper motion, magnitude, and spectral type. 
If the companion had not been identified, especially in the widest systems ($\rho >$ 10\,000\,arcsec), we enlarged the search radius in steps up to 120\,arcsec. 
For these outliers, we used all the available information, namely, the WDS remarks and the original publications.
There were only two cases where the companion star was not found by us with the $\rho$ and $\theta$ provided by WDS, even after enlarging the search radius and scouring the literature\footnote{The two WDS pairs with unidentified companion stars are 
WDS~03074--5655 (TSN~110) and 
WDS~03353--4020 (TSN~111).
In a preliminary analysis, there was a third unidentified system, namely
WDS~05463+5627 (LDS~3673), but it suffered from a typographical error in WDS that was corrected afterwards (\citealt{carro21}; B.\,D.~Mason, priv. comm.).
We revise its relative astrometry to $\rho$ = 57.1\,arcsec, $\theta$ = 262.5\,deg, and epoch = J2016.0.}.

As for the primaries, we retrieved Simbad identifiers and \textit{Gaia} DR3 for the corresponding companions. 
Only eight of the companions had not parallaxes (or even proper motions) available in any catalogue, and we also discarded them from the analysis\footnote{The eight companions without parallax are: LSPM J1536+2856, SCR J1900--3939, UCAC3 208--200112, 2MASS J13543510--0607333, 2MASS J14313545--0313117, Gaia DR3 276070675205077632, Gaia DR3 4655216993788228480, and Gaia DR3 601133385210548736.}.
We computed our own $\rho$ and $\theta$ parameters for the 492 ($502-2-8$)
remaining pairs using the standard equations of spherical trigonometry (e.g. \citealt{smolinski06}):

\begin{equation}
\rho= \arccos{[\cos{(\Delta \alpha \cos{\delta_1})}\cos{(\Delta \delta)}]}
\label{eqn:rho_obtention}
,\end{equation}

\noindent and

\begin{equation}
\theta = \frac{\pi}{2} - \arctan\left[\frac{\sin{(\Delta \delta)}}{\cos{(\Delta \delta)}\sin{(\Delta \alpha \cos{\delta_1})}}\right],
\label{eqn:theta_obtention}
\end{equation}

\noindent where $\Delta \alpha$ = $\alpha_2 - \alpha_1$,
$\Delta \delta$ = $\delta_2 - \delta_1$,
and $\alpha_1,\delta_1$ and $\alpha_2,\delta_2$ are the equatorial coordinates of the primary and companion stars, respectively.

We compared the $\rho$ and $\theta$ values we measured  with those tabulated by WDS (in particular, with the latest measurements, i.e. {\tt sep2} and {\tt pa2}).
For the position angle, the standard deviation of the differences between our measurements and those from WDS is 0.84\,deg.
The distribution of the differences in $\theta$ is not Gaussian, with a narrow peak centred at 0\,deg and wide, but shallow, wings at both sides.
Of the 492 identified pairs with parallaxes, only 23 have absolute differences in $\theta$ greater than 1\,deg (and up to 4.2\,deg). Most of the kinds of differences ascribed to uncertainties propagated from inaccurate pre-{\it Gaia} coordinates, especially for the widest systems, such as WDS~23127+6317 (e.g. with Eq.~\ref{eqn:theta_obtention} being highly non-linear).
The distribution of the differences in $\rho$ is similar to that of $\theta$, with a narrow peak centred at 0\,arcsec and a relatively large standard deviation of the differences of 21.0\,arcsec.
This large amount is originated by the difficulty in previous works to measure $\rho$ or even to identify the companion of the widest systems, such as the `outliers' described above and found at more than 10\,arcsec from their expected locations\footnote{For example, \citet{tokovinin12} and we ourselves measured $\rho$ = 1684.2\,arcsec and 1684.49\,arcsec, respectively, for the outlier system \object{HD~45875} + Gaia DR3 1115649542191409664 (TOK~503, WDS~06387+7542~AD), but WDS instead tabulates 1898.64\,arcsec collected in 2015.}.

The distribution of our new values of $\rho$ are plotted with red data points in Fig.~\ref{fig:acumhist}.
Of the 492 identified pairs with parallax, 298 have $\rho$ = 1000--2000\,arcsec, 117 have $\rho$ = 2000--10\,000\,arcsec, and 77 have $\rho>$ 10\,000\,arcsec.
The latter ultrawide pairs come mostly from the works by \cite{probst83} and \cite{shaya11}.
The widest pair has $\rho$ = 66\,094\,arcsec (WDS 02157+6740, SHY~10; \citealt{shaya11}).
As described below, not all of them are physically bound.

\subsection{Pair validation}
\label{sec:pair_validation}

\begin{figure*}
 \centering
    \includegraphics[width=1\linewidth, angle=0]{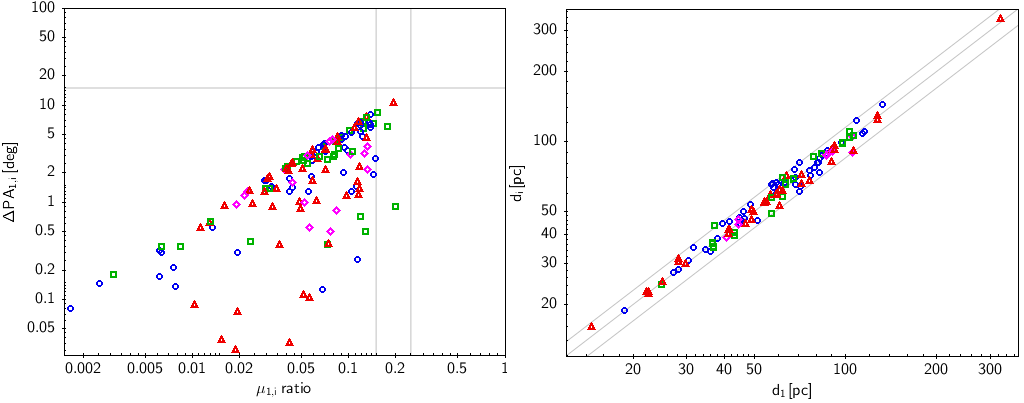}
 \caption{
 Astrometric criteria for pair validation.
 In both diagrams, we plot pairs between primaries and secondaries with blue circles, tertiaries with red triangles, quaternaries with green squares, and higher order companions with purple diamonds. 
 \textit{Left:} $\Delta {\rm PA}_{1,i}$ vs. $\mu_{1,i}$ ratio diagram.
 Vertical and horizontal grey lines mark the $\mu$ ratios of 0.15 and 0.25 $\mu$ and $\Delta$PA of 15\,deg, respectively. 
 Compare with Fig.~2 in \citet{montes18a}.
 \textit{Right:} 
 Distance of companions vs. distance of primaries. 
 Diagonal lines indicate the 1.15:1, 1:1, and 0.85:1 distance relationships.
 The $\alpha$~Cen~AB + Proxima system, at $d \sim$ 1.3\,pc, is not shown.
}
 \label{fig:dpamu_ddd}
\end{figure*}

To validate the 492 pairs, we used the criteria established by \cite{montes18a} to distinguish between physical (bound) and optical (unbound) systems.
For that purpose, we computed two astrometric parameters that quantify the similarity of the proper motions of two stars:

\begin{equation}
\mu\,\mathrm{ratio}=\sqrt{\frac{(\mu_{\alpha} \cos{\delta_1}-\mu_{\alpha} \cos{\delta_2})^2+(\mu_{\delta 1}-\mu_{\delta 2})^2}{(\mu_{\alpha} \cos{\delta_1})^2+(\mu_{\delta 1})^2}}<0.15,
\label{eqn:crit1}
\end{equation}

\noindent and

\begin{equation}
\Delta \text{PA}=\lvert  \text{PA}_1- \text{PA}_2 \rvert<15\,\mathrm{deg},
\label{eqn:crit2}
\end{equation}

\noindent where PA$_i$ are the angles of the proper motion vectors, with $i=1$ for the primary star and $i=2$ for the companion.
We added an extra buffer in the $\mu$ ratio of up to 0.25 to account for projection effects on the celestial sphere of nearby ultrawide systems, as in the case of $\alpha$~Cen~AB + Proxima \citep{innes15,wertheimer06,caballero09}.

At the time of publication by \cite{montes18a}, {\it Gaia} parallaxes were not available except the for 2.5 million stars of the Tycho-Gaia Astrometric Solution \citep{michalik15}.
With the advent of the third {\it Gaia} data release with precise parallaxes for $\sim$700 times more stars, we added one additional condition to our validation. 
\citet{cifuentes21} imposed parallactic distances to agree within 10\%, while \citet{gonzalezpayo21a} did it within 15\%, which is the value we chose to impose. 
In short, our third astrometric criterion was:

\begin{equation}
\Biggl\lvert \frac{\pi_1^{-1}-\pi_2^{-1}}{\pi_1^{-1}} \Biggr\rvert < 0.15,
\label{eqn:crit3}
\end{equation}

\noindent with $\pi_1$ and $\pi_2$ as the parallaxes of both components of the pair (we did not apply any colour correction for computing distances -- \citealt{bailerjones18a,lindegren21}).
In Fig.~\ref{fig:dpamu_ddd}, we show the relations between $\mu$ ratio and $\Delta$PA and between distances of the two components of the 153 pairs that satisfy the three imposed criteria simultaneously (and additionally, the multiple companions obtained in Sect.~\ref{sec:search_additional}). 
Although we refer to them as pairs, in many cases they are actually part of hierarchical multiple (triple, quadruple, quintuple...) systems made of stars at very different angular separations to their primaries.
This is described in detail below.

Except for 2 of them\footnote{The two systems at $d >$ 150\,pc are $\gamma$\,Cas + HD~5408 (188\,pc) and G~143--33 + G~143--27 (324\,pc).}, all the 153 pairs are located at heliocentric distances shorter than 150\,pc, with a distribution peaking at 40--50\,pc.
The distribution of total proper motions is, however, flatter, with only one pair\footnote{The high proper motion pair with $\mu >$ 700\,mas\,a$^{-1}$ is $\alpha$~Cen~AB + Proxima (3710\,mas\,a$^{-1}$).} with a $\mu$ greater than 700\,mas\,a$^{-1}$ and none with a $\mu$ less than 25\,mas\,a$^{-1}$.

We did not keep in our final list of validated pairs an ultrawide system candidate at about 2400\,pc towards the Magellanic Clouds, namely OGL~54 \citep{poleski12}.
It is made of \object{OGLE~SMC-SC1~161-162} and Gaia DR3 4685747717242739328 (``SMC128.7.9551''), which are separated by about 12\,pc.
If truly linked, the pair would be much further and wider than any other system considered here.
Last but not least, we revised the system $\rho$ from 1017\,arcsec to 977\,arcsec, below our boundary at 1000\,arcsec.

\subsection{Additional companions and stellar kinematic groups in \textit{Gaia} DR3}
\label{sec:search_additional}

\begin{figure*}
 \centering
 \includegraphics[width=1\linewidth, angle=0]{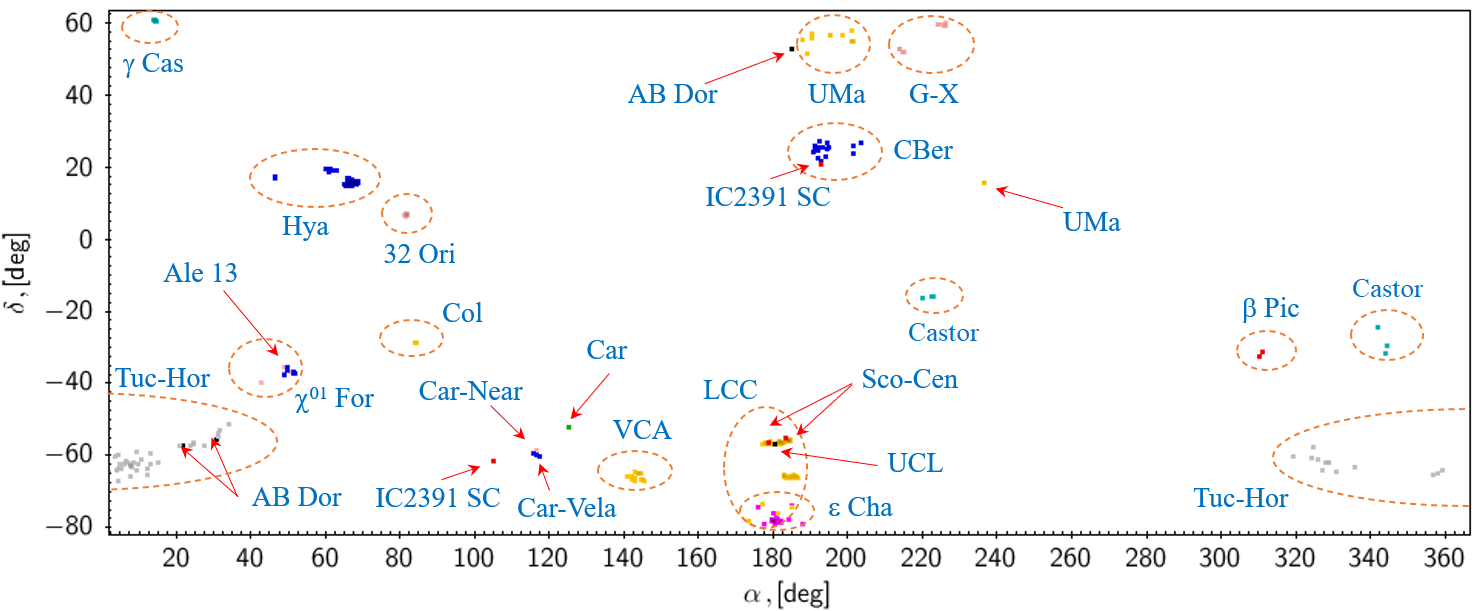}
 \caption{Spatial distribution of the 309 identified young stars in SKGs, associations, and open clusters.
}
\label{fig:spatialdistr}
\end{figure*}

We looked for additional proper motion and parallax companions within 1\,pc around both the primary and the companion of the 153 validated pairs.
We followed the methodology described in Sect.~3 of \cite{gonzalezpayo21a}; however, in our work, apart from {\tt TOPCAT} and a customised code in astronomic data query language \citep{yasuda04}, we used \textit{Gaia} DR3 and the criteria imposed by Eqs.~\ref{eqn:crit1}--\ref{eqn:crit3}.
For a few cases of ultrawide WDS pairs with projected physical separations greater than 1\,pc, we extended the search radius up to the maximum separation between known components.

In our {\it Gaia} search, we identified 349 additional common proper motion and parallax companions that satisfy the astrometric criteria of Eqs.~\ref{eqn:crit1}, \ref{eqn:crit2}, and~\ref{eqn:crit3}.
Of these, 111 additional companions are catalogued by WDS and 239 are not.
The large multiplicity order of some system candidates, made of over a dozen pairs each (i.e. higher than dodecuple), together with the presence of debris discs in some of the components (e.g. \object{$\alpha^{01}$~Lib}, \object{AU~Mic} -- \citealt{kalas04,chen05,mizusawa12,gaspar13,mittal15,plavchan20}), has led us to investigate the membership of all our targets in young stellar kinematic groups (SKGs -- \citealt{eggen65,montes01a,zuckerman04}), stellar associations \citep{ambartsumian49,blaaw91,dezeeuw99}, and even open clusters.

Of the 153 validated pairs in Sect.~\ref{sec:pair_validation}, there are 59 with at least one component (primary, companion, or both) that had previously been considered part of young SKGs, associations, and clusters such as the Tucana-Horologium and Coma Berenices moving groups, the $\epsilon$~Chamaeleontis association, or the Hyades open cluster (e.g.
\citealt{perryman98,murphy13,kraus14,pecaut16,riedel17,gagne18a,tang19}).
Furthermore, of the 239 additional astrometric companions not catalogued by WDS, a total of 199 share proper motion and parallax companions with these young pairs.
Table~\ref{tab:groups_long} shows the name, equatorial coordinates, and $G$-band magnitude of 349 young stars and candidates, together with the corresponding group (SKG, association, or cluster) when available (309 cases), and references.
The full names and acronyms of the 22 considered groups, with ages ranging from 4--8\,Ma (of the Chamaeleon-Scorpius-Centaurus-Crux complex) to 600--800\,Ma (of the Hyades and [TPY2019]~Group-X), are provided in the table notes.
Discoverer codes are given for all components tabulated by WDS (some stars that belong to different WDS systems can have different entries\footnote{For example, \object{HD~1466} is SHY~113~G, SHY~114~G, and CVN~33~G.}), while the 199 additional companions have the string `...' in the discoverer code column. 
The spatial distribution of the 309 young stars and candidates in SKGs, associations, and open clusters is shown in Fig.~\ref{fig:spatialdistr}.

About 80\% of the 199 additional companions had also been ascribed to young groups, but not all.
We report 40 stars, marked with `...' in the group column in Table~\ref{tab:groups_long}, that are new candidate members in young SKGs, associations, and clusters.
Eight of them had actually been considered as previous members, but the most recent works have classified them as ``improbable members'' (e.g. HD~207377~AB in Tucana-Horologium; \citealt{zuckerman01}). 
In any case, some of the 40 stars, because of either their brightness (e.g. \object{HD~71043}, which is probably an A0\,V, $G \approx$ 5.9\,mag, 200--300\,Ma-old star in Carina) or faintness (e.g. 2MASS~J12145318--5519494, which probably is a young brown dwarf in Lower Centaurus-Crux; \citealt{folkes12}), may be interesting to confirm in future works.

One more wide pair, composed by the bright stars \object{$\gamma$~Cas} and \object{HD~5408}, resulted in a nonuple system after our initial astrometric analysis and bibliographic search.
Since $\gamma$~Cas is an extremely young classical Be star (\citealt{poeckert78,white82,henrichs83,stee95}), we also tabulated the resolved pair components in Table~\ref{tab:groups_long}, although it has never been ascribed to any group in particular (but see \citealt{mamajek17}). 
The $\gamma$~Cas system is discussed in further detail in Appendix~\ref{sec:gamCas}.

Despite the far-reaching title of this series of papers, in this particular work, we focus on relatively evolved and old systems in the galactic field.
Common proper motion (and parallax) surveys of resolved companions to bona fide SKG members is indeed a widely used and successful technique for discovering new young stars and brown dwarfs \citep[][and references therein]{alonsofloriano15}.
However, a dedicated work on disentangling actual very wide binaries from unbound components in SKGs with similar galactocentric space velocities is planned.

After removing the 59 pairs with stars in young SKGs, associations, and clusters, we kept 94 wide pairs in the galactic field.
To them, we added the 40 additional astrometric companions found in our {\it Gaia} DR3 search and not catalogued by WDS plus 39 already reported by WDS and separated by less than 1000\,arcsec.
As a result, there were 266 stars\footnote{\object{HD~79392} is catalogued by WDS as the primary of two different systems (WDS 09150+3837/TOK 525 and WDS 09150+3837/DAM1575).} in 243 resolved {\it Gaia} sources and in 94 systems that passed to the next step of our analysis.
All the systems and resolved {\it Gaia} sources are listed in Tables~\ref{tab:systems1} and~\ref{tab:systems2}.

\subsection{New close binary candidates from {\it Gaia} data}
\label{sec:ruwe_vr}

We carried out a cross-matching with WDS and scoured the literature in search of additional companions not identified in our {\it Gaia} DR3 search.
We did not find any additional WDS companion at $\rho <$ 1000\,arcsec that were resolvable by {\it Gaia} and that had not been recovered in our search.
However, WDS also tabulates very close systems ($\rho \lesssim$ 1.3\,arcsec) that were discovered and characterised with micrometers, speckle, lucky imaging, or adaptive optics, and which are unresolvable by {\it Gaia} thanks to the close separation or relatively large magnitude difference between components (e.g. \object{HD~6101}, \object{HD~102590}, \object{HD~186957}).
In addition, there is a number of pair components that are spectroscopic binaries (e.g. \object{HD~120510}; \citealt{pourbaix04}) or triples (e.g. $\delta$~Vel, which is also an eclipsing binary with a close astrometric companion; \citealt{kervella13}), or very close binaries from proper motion anomalies (e.g. \object{HD~125354}; \citealt{kervella19}).
We further consider all this information in Sect.~\ref{sec:results_discussion}.

Three pairs in multiple systems are in the 0.15--0.25 $\mu$ ratio buffer interval in the $\Delta$PA vs. $\mu$ ratio diagram (Fig.~\ref{fig:dpamu_ddd}), namely WDS~09487--2625, WDS~16278--0822, and WDS~23309--5807.
The origin of their large $\mu$ ratio lies on wide amplitude orbital (i.e. proper motion) variations induced by additional components in the systems at 1.83\,arcsec (\object{HD 85043}, 
I~205), $\sim$1.0\,arcsec (\object{$\upsilon$~Oph}, RST~3949), and 1.27\,arcsec (\object{HD 221252}, I~145) to the primaries or companions.
As a result, we also validated the three systems (one triple and two quadruples) in spite of not satisfying our original $\mu$ ratio criterion.

Next, we cross-matched our 243 {\it Gaia} sources in 94 systems with the {\it Hipparcos}-{\it Gaia} catalogues of accelerations of \citet{kervella19} and \citet{brandt21}.
Of them, 54 have a measurable proper motion anomaly (Boolean variable set to unit in \textit{Gaia} DR2, proper motion anomaly binary flag {\tt BinG2} --\citealt{kervella19}--, or {\tt chi2} $>$ 11.8 --\citealt{brandt21}--) that are probably induced by unseen companions.
They are marked with a footnote in Tables~\ref{tab:systems1} and~\ref{tab:systems2}.

In addition, we looked for new very close binary candidates among the 94 systems.
First we used the {\it Gaia} re-normalised unit weight error ({\tt RUWE}), which is a robust indicator of the goodness of a star's astrometric solution \citep{arenou18,lindegren18a}.
Large {\tt RUWE} values correspond to stars with angular separations small enough not to be resolved by {\it Gaia}, but large enough to perturb the astrometric solution.
{\it Gaia} DR3 provides {\tt RUWE} values for 234 {\it Gaia} entries.
The nine sources without {\tt RUWE} values are either very bright stars ($\delta$~Vel, $\alpha$~Cen~A and B, $\upsilon$~Oph) or known close binaries with angular separations $\rho \sim$ 0.2--0.9\,arcsec (e.g. \object{HD~6101}).
There are 14 stars with {\tt RUWE} $> 10$.
They are also marked with a footnote in Tables~\ref{tab:systems1} and ~\ref{tab:systems2}.
The five {\it Gaia} sources with the greatest {\tt RUWE}, of about 20--40, are either already known close binaries below the {\it Gaia} resolution limit (e.g. \object{G~210--44}, $\rho \sim$ 0.1\,arcsec -- HDS~2989 in the Hipparcos Double Stars catalogue) or strong, relatively faint, new binary candidates (e.g. \object{2MASS~J02022892--3849021}, \object{UCAC3~109--11370}, \object{LSPM~J0956+0441}, and \object{HD~59438\,C}). 
Of the other nine {\it Gaia} sources with moderate {\tt RUWE} of about 10--20, some have also been tabulated as candidate binaries, such as \object{HD~75514} and \object{HD~139696}, which were listed in the {\it Hipparcos}–{\it Gaia} catalogue of accelerations \citep{kervella19}, \object{HD~210111}, which is a $\lambda$~Bootis-type spectroscopic binary \citep{paunzen12}, and \object{HD~215243}, which is subgiant spectroscopic binary \citep{gorynya18}. 
The rest of {\it Gaia} sources with {\tt RUWE} $>$ 10 would need an independent confirmation of binarity.  
Being less conservative, we could have extended our analysis down to {\tt RUWE} = 5, which is about three times greater than the critical value of 1.41 of \citet{arenou18}, \citet{lindegren18a}, or \cite{cifuentes20}.
There are only five \textit{Gaia} sources (in double or multiple systems) with $5 \leq {\tt RUWE}\leq 10$. However, as some careful studies of nearby stars indicate, {\tt RUWE} values slightly larger than 1.4 do not necessarily translate into close binarity (\citealt{ramsay22}; Ribas et al. accepted). Since the confirmation of actual close binarity requires a radial-velocity or high-resolution imaging follow-up, we imposed a very conservative {\tt RUWE} limit.
\begin{table*}
 \centering
 \caption[]{Radial-velocity outlier candidates.}
 \label{tab:outliers}
\begin{tabular}{llllc}
    \hline \hline
    \noalign{\smallskip}
WDS & Discoverer & Primary star & Companion star & $\lvert\Delta V_r\rvert$ \\
 & code &  &  & (km\,$\text{s}^{-1}$) \\
    \noalign{\smallskip}
    \hline
    \noalign{\smallskip}
01066+1353 & SHY 396 & HD 6566 & HD 5433$^a$ & 25.5$\pm$0.4 \\
02022--4550 & SHY 410 & HD 12586 & HD 12808 & 30.8$\pm$0.2 \\
02310+0823 & GIC  32 & G 4-24 & G 73-59$^b$ & 63.1$\pm$3.8 \\
02315+0106 & SHY 422 & BD+00 415B & HD 17000 & 4.4$\pm$0.3 \\
02462+0536 & TOK 651 & HD 17250 & HD 17163 & 18.1$\pm$0.6 \\
03503--0131 & SHY 164 & HD 24098A & HD 22584$^a$ & 5.9$\pm$0.2 \\
04346--3539 & TOK 488 & HD 29231 & L 447-2 & 25.5$\pm$0.4 \\
05222+0524 & TOK 497 & HD 35066(A)$^a$ & TYC 109-530-1 & 0.8$\pm$0.3 \\
08211+4021 & TOK 516 & BD+40 2030$^a$ & G 111-70 & 36.9$\pm$0.3 \\
08237--5519 & SHY 526 & HD 71257 & HD 72143$^a$ & 4.9$\pm$0.3 \\
08388--1315 & SHY 201 & HD 73583 & BD-09 2535 & 18.6$\pm$0.3 \\
08480--3115 & SHY 529 & HD 75269$^a$ & HD 75514$^{a,b}$ & 8.5$\pm$1.5 \\
09467+1632 & TOK 531 & BD+17 2130$^a$ & LP 428-36 & 56.6$\pm$2.7 \\
09487--2625 & TOK 532 & HD 85043A$^a$ & PM J09486-2644 & 1.8$\pm$0.3 \\
09568+0415 & TOK 533 & HD 86147 & LSPM J0956+0441$^b$ & 10.7$\pm$0.9 \\
10289+3453 & SHY 215 & HD 90681 & HD 92194 & 4.8$\pm$0.2 \\
10532--3006 & SHY 563 & HD 94375$^a$ & HD 94542$^a$ & 23.6$\pm$0.2 \\
11214+0638 & TOK 544 & HD 98697 & LP 552-34 & 19.4$\pm$3.5 \\
11455+4740 & LEP  45 & HD 102158 & G 122-46 & 19.6$\pm$0.6 \\
13305+2231 & SHY 626 & HD 117528 & BD+22 2587 & 93.2$\pm$0.2 \\
13470+3833 & SHY 633 & HD 120164 & HD 119767 & 21.6$\pm$0.2 \\
15120+0245 & WIS 281 & LP 562-9 & LP 562-10 & 24.5$\pm$3.7 \\
15208+3129 & LEP  74 & HD 136654 & AX CrB & 0.8$\pm$0.2 \\
15318--0204 & SHY 677 & HD 138370 & HD 138159 & 17.3$\pm$0.3 \\
15330--0111 & SHY 678 & 11 Ser & HD 142011 & 12.0$\pm$0.2 \\
15356+7726 & WIS 288 & LSPM J1535+7725 & LP 22-358 & 24.3$\pm$0.3 \\
15408--3252 & SHY 278 & HD 139696$^{a,b}$ & CD-32 10820 & 42.8$\pm$2.9 \\
15590+1820 & SHY 691 & HD 143292$^a$ & HD 142899 & 42.7$\pm$0.3 \\
16278--0822 & SHY 287 & $\upsilon$ Oph$^{a,c}$ & HD 144660 & 11.7$\pm$0.2 \\
17166+0325 & SHY 715 & HD 156287$^a$ & HD 159243 & 8.5$\pm$0.2 \\
18143--4309 & SHY 740 & HD 166793 & HD 166533 & 31.9$\pm$0.4 \\
18496+1313 & SHY 309 & HD 229635 & HD 229830 & 31.0$\pm$0.3 \\
18571+5143 & SHY 749 & HD 176341 & BD+49 2879 & 14.6$\pm$0.2 \\
18597+1615 & TOK 622 & HD 176441$^a$ & LSPM J1858+1613$^b$ & 19.3$\pm$0.5 \\
19290--4952 & SHY 319 & HD 182857 & HD 185112$^a$ & 14.0$\pm$0.3 \\
20084+1503 & LDS1033 & G 143-33 & G 143-27$^c$ & 66.3$\pm$1.7 \\
20371+6122 & SHY 780 & HD 196903 & HD 198662 & 12.4$\pm$0.2 \\
20404--3251 & SHY 781 & HD 196746$^a$ & HD 196189 & 26.2$\pm$0.2 \\
20489--6847 & SHY 782 & HD 197569 & HD 199760 & 7.6$\pm$0.2 \\
21105+2227 & SHY 793 & HD 201670 & HD 198759 & 54.4$\pm$17.8 \\
22175+2335 & GIC 179 & G 127-13$^b$ & G 127-14 & 21.7$\pm$0.9 \\
22220--3431 & SHY 802 & HD 212035 & HD 210111$^c$ & 11.4$\pm$0.5 \\
23506+5412 & SHY 840 & HD 223582$^a$ & HD 223788 & 1.7$\pm$0.2 \\

    \noalign{\smallskip}
    \hline
    \end{tabular}
 \tablefoot{
    \tablefoottext{a}{Stars with proper motion anomaly \citep{kervella19,brandt21}.}
    \tablefoottext{b}{Stars with {\tt RUWE} $> 10$.}
    \tablefoottext{c}{Known spectroscopic binaries.}
}    
\end{table*}

Next, we used the standard deviation of the radial velocities, $V_r$, measured with the {\it Gaia} Radial Velocity Spectrometer, which receives the misleading label {\tt radial\_velocity\_error} \citep{gaiacollaboration22,katz22}.
Of the 243 {\it Gaia} entries, 182 have $V_r$ and its standard deviation, $\sigma_{Vr}$.
The median formal precision of the velocities for the brightest, most stable {\it Gaia} stars lies at about 0.12\,km\,s$^{-1}$ to 0.15\,m\,s$^{-1}$ and smoothly increases for fainter stars \citep{katz22}. 
However, we identified at least six {\it Gaia} sources that have significantly greater $\sigma_{Vr}$ than expected given their magnitudes. 
Being all stars of intermediate ages and spectral types in the main sequence (i.e. no pulsating giants or subgiants, nor very active T~Tauri stars), we adscribed the large $\sigma_{Vr}$ to spectroscopic binarity.
Actually, two of them had already been reported as spectroscopic binaries, namely \object{HD~2000077} (\citealt{konacki10,montes18a} and references therein) and \object{HD~215243} \citep[][which also has a large {\tt RUWE}]{gorynya18}.
A third one, namely \object{HD~75514} \citep{kervella19,brandt21}, has a significant proper motion anomaly.
The other three new spectroscopic binary candidates have a large {\tt RUWE} (8.1; \object{BD+32~2868}), 
moderate {\tt RUWE} and $\sigma_{Vr}$ (2.48, 2.86\,km\,s$^{-1}$), or a small RUWE but a huge $\sigma_{Vr}$ for a bright single star ($G \approx$ 7.7\,mag, 17.76\,km\,s$^{-1}$; \object{HD~201670}).

At this stage, we may wonder why common parallax and proper motion criteria alone were used for system validation, instead of common radial velocities as well, at least for the 99 pairs with data for the two components.
We note that a large difference in radial velocities may be a symptom of long-period spectroscopic binarity of one of the components (by ``long period,'' we mean longer than or of the same order of the 34 months of the {\it Gaia} DR3 radial-velocity coverage).
Nevertheless, the above-mentioned properties of high \texttt{RUWE}, $\sigma_{Vr}$, or, especially, proper motion anomaly do not always indicate unknown close companions, but can also be produced by the already detected close companions. 
Some examples of known pairs with astrometric accelerations and orbital periods of tens to hundreds years are HD~6101, HD~59438, and HD~85043.

In Table~\ref{tab:outliers}, we list 43 pairs of primaries and companions with radial velocity differences larger than three times the quadratic sum of the respective $\sigma_{Vr}$.
Among them, we can find known spectroscopic binaries, components with large {\tt RUWE} values, proper motion anomalies, or a combination of them.
Some of the pairs in Table~\ref{tab:outliers} may be false positives, that is, two unrelated stars with very similar proper motions and parallaxes but very different radial velocities.
However, with the data available to us, it is impossible to disentangle between them and true wide physical systems with one radial-velocity outlier component due to currently unknown long-period spectroscopic binarity.

\subsection{Colour-magnitude diagram}
\label{sec:HR_diagram}

\begin{figure}
 \centering
 \includegraphics[width=1\linewidth, angle=0]{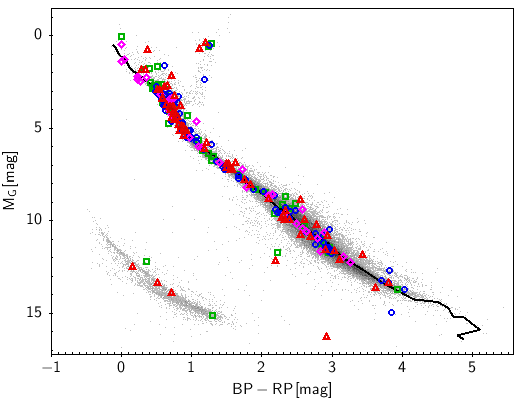}
 \caption{HR diagram of all investigated stars.
 Coloured open symbols stand for stars in double (blue circles), triple (red triangles), quadruple (green squares), and higher-order multiple systems (purple diamonds).
 Grey dots represent selected field stars from \textit{Gaia}.
 The black solid line is the updated main sequence of \citet{pecaut13}.
 The stars outside the main sequence are discussed in the text.
}
 \label{fig:hrdplot}
\end{figure}

Stellar masses are needed to compute gravitational binding energies, while luminosity classes are needed to estimate stellar masses.
Estimated stellar ages are also needed to investigate the evolution of fragile multiple systems, while luminosity classes also shed light on stellar ages, especially outside the main sequence.
The luminosity class of the stars in the 243 {\it Gaia} sources is illustrated by the Hertzprung-Russell (H-R) diagram of Fig.\,\ref{fig:hrdplot}.
We took parallaxes and $G$, $G_{BP}$, and $G_{RP}$ magnitudes from \textit{Gaia} DR3, except for four very brights stars ($\delta$~Vel, $\alpha$~Cen A and B, $\upsilon$~Oph), for which we estimated their magnitudes from their well-determined spectral types, published Johnson $B$, $V$, $R$ photometry, and the main-sequence colour-spectral type relation of \citet{pecaut13}.
We also plot this relation in the diagram, although the main sequence, together with the loci of white dwarfs and giants and subgiants beyond the turnoff point, is clearly marked by 57\,345 field stars with good {\it Gaia} astrometry and photometry following the H-R example of \citet{taylor21}, but with the DR3 data set.

From their position in the H-R diagram, we identified eight giants and subgiants and five white dwarfs (listed in Table\,\ref{tab:giants}).
We confirmed their classification with a comprehensive bibliographic study.
Among the 13 stars, only one is part of a close pair unresolved by {\it Gaia}, namely $\delta$~Vel.
Furthermore, all but one of the giants are so bright that were listed already by \citet{bayer1603} and \citet{flamsteed1725}.

Four of the five white dwarfs have a spectral type determination, with only one presented as a white dwarf candidate by \citet{gentilefusillo19}.
However, all of them are part of multiple systems (i.e. triple or higher).
For example, WDS~01024+0504 is made of two spectroscopic binaries, namely the double, early K dwarf \object{HD~6101} and the double, DA5.9 white dwarf \object{EGGR~7} \citep{giclas59,maxted00,lajoie07,caballero09,gianninas11,toonen17},
while WDS~06536-3956 is made of the early M dwarf \object{L~454--11} \citep{lepine11} and two white dwarfs, \object{WT~201} (DA8.0) and \object{WT~202} (DA7.0) \citep{subasavage08}.
The system may also be quadruple because L~454--11 has a {\tt RUWE} = 18.0.
The other two white dwarfs are in triple and quadruple systems.

\begin{table*}
 \centering
 \caption[]{Giants and white dwarfs in wide double and multiple systems.}
 \begin{tabular}{@{\hspace{1mm}}l@{\hspace{1mm}}l@{\hspace{1mm}}l@{\hspace{2mm}}l@{\hspace{2mm}}c@{\hspace{2mm}}c@{\hspace{2mm}}c@{\hspace{0mm}}c@{\hspace{1mm}}}
 \hline \hline
 \noalign{\smallskip}
 Star & Spectral & WDS & Discoverer & $\alpha$\,(J2000) & $\delta$\,(J2000) & $M$ & Age \\
 & type &  & code & (hh:mm:ss.ss) & (dd:mm:ss.s) & (M$_{\odot}$) & (Ga) \\
 \noalign{\smallskip}
 \hline
 \noalign{\smallskip}
 \multicolumn{8}{c}{\it Giants}\\
 \noalign{\smallskip}
$\delta$ Vel Aa & A2\,IV & 08447--5443 & SHY 49 & 08:44:42.23 & $-$54:42:31.7 & 3.19$\pm$0.03$^a$ & $\sim$0.431$^a$ \\
HD 120164 & K0\,III & 13470+3833 & SHY 633 & 13:46:59.77 & +38:32:33.7 & 2.42$\pm$0.24$^b$ & $\sim$0.7$^b$ \\
$\iota$ Vir & F7\,III & 14190--0636 & SHY 71 & 14:16:00.87 &$-$06:00:02.0 & $\sim$1.81$^c$ & 1.809$\pm$0.001$^d$ \\
11 Ser & K0\,III & 15330--0111 & SHY 678 & 15:32:57.94 & $-$01:11:11.0 & 1.27$\pm$0.35$^e$ & $2.75^{+0.88}_{-0.66}$\,$^e$ \\
64 Aql & K1\,III--IV & 20080--0041 & SHY 325 & 20:08:01.82 & $-$00:40:41.5 & 1.00$\pm$0.27$^e$ & 9.33$\pm$4.17$^e$ \\
$\nu$ Aqr & K0\,III & 21096--1122 & TOK 633 & 21:09:35.64 & $-$11:22:18.1 & $2.01_{-0.11}^{+0.04}$\,$^f$ & $1.26_{-0.19}^{+0.22}$\,$^f$ \\
$\kappa$ Aqr & K1.5\,III & 22378--0414 & TOK 640 & 22:37:45.38 & $-$04:13:41.0 & 2.55$\pm$0.13$^g$ & 2.79$\pm$1.16$^h$ \\
$\iota$ Cep & K1\,III & 22497+6612 & SHY 359 & 22:49:40.81 & +66:12:01.4 & $1.55_{-0.20}^{+0.05}$\,$^f$ & $2.57_{-0.38}^{+0.18}$\,$^f$ \\
 \noalign{\smallskip}
 \hline
 \noalign{\smallskip}
 \multicolumn{8}{c}{\it White dwarfs}\\
 \noalign{\smallskip}
EGGR 7 & DA5.9 & 01024+0504 & WNO 50 & 01:03:49.92 & +05:04:30.6 & $\sim$ 0.77$^i$ &  ... \\
WT 202 & DA7.0 & 06536--3956 & SUB 2 & 06:53:35.44 & $-$39:55:34.8 & 0.64$\pm$0.02$^j$ & $2.4_{-0.1}^{+1.0}$\,$^j$ \\
WT 201 & DA8.0 & 06536--3956 & SUB 2 & 06:53:30.21 & $-$39:54:29.1 & 0.64$\pm$0.02$^j$ & $3.2_{-0.1}^{+1.1}$\,$^j$ \\
Gaia DR3 812109085097488768 & ...$^k$ & 09150+3837 & DAM1575 & 09:14:58.95 & +38:36:58.3 & 0.5$\pm$0.1$^l$ & ... \\
SDSS J230056.41+640815.5 & DC & 22497+6612 & ... & 23:00:56.46 & +64:08:16.0 & 0.5$\pm$0.1$^l$ & ... \\
 \noalign{\smallskip}
 \hline
 \end{tabular}
 \label{tab:giants}
 \tablefoot{
    \tablefoottext{a}{\citet{david15};}
    \tablefoottext{b}{\citet{dasilva15};}
    \tablefoottext{c}{\citet{gontcharov10};}
    \tablefoottext{d}{\citet{eker18};}
    \tablefoottext{e}{\citet{feuillet16};}
    \tablefoottext{f}{\citet{stock18};}
    \tablefoottext{g}{\citet{kervella19};}
    \tablefoottext{h}{\citet{soubiran08};} 
    \tablefoottext{i}{\citet{lajoie07};}
    \tablefoottext{j}{Rebassa-Mansergas (priv. comm.);}
    \tablefoottext{k}{\citet{gentilefusillo19};}
    \tablefoottext{l}{This work.}
}
\end{table*}

Apart from the 13 giants, subgiants, and white dwarfs in Table\,\ref{tab:giants}, there are still some objects lying outside the main sequence in the colour-magnitude diagram of Fig.\,\ref{fig:hrdplot}.
In particular, there are 4 sources apparently below the main sequence.
The origin of this discrepancy lies in the four cases on wrong photometry:
\object{2MASS J02004917--3848535} in the double system WDS~02025--3849 is an $\sim$M7--8 ultracool dwarf with $G_{BP}$ fainter than the {\it Gaia} limit \citep{smart19}; 
\object{Gaia DR3 749786356557791744} in the triple system WDS 10289+3453 is another ultracool dwarf with $G_{BP}$ fainter than the {\it Gaia} limit, but with a spectral type at the M-L boundary; 
\object{Gaia DR3 3923191426460144896} in the quadruple system WDS~11486+1417 is an $\sim$M4--5 late-type dwarf at $\rho \approx$ 10.1\,arcsec of the very bright ($G \approx  5.9$\,mag) A8+G2 binary \object{HD~102590};
and \object{Gaia DR3 1367008242580377216} in the triple system WDS~17415+4924 is an $\sim$M4--5 late-type dwarf with a relatively high value of $G_{BP}/G_{RP}$ excess factor, {\tt E(BP/RP)}, which is an indicator of systematic errors in photometry \citep{riello18}.
Remarkably, 2 of the 4 {\it Gaia} sources with the wrong photometry are the least massive stars in our sample (Sect.~\ref{sec:masses_gravitational_bindings}).
The rest of the {\it Gaia} sources, which are especially redder than the subgiant turnoff point, are reasonably  matched to the main sequence.

\subsection{Masses and gravitational binding energies}
\label{sec:masses_gravitational_bindings}

For stars in the main sequence, we determined stellar masses, $M$, from the $G$-band absolute magnitude, {\it Gaia,} and 2MASS colours, spectral types, and the updated version of Table~4 in \citet{pecaut13}\footnote{\url{https://www.pas.rochester.edu/~emamajek/EEM_dwarf_UBVIJHK_colors_Teff.txt}}.
The match between spectral types derived by us from colours and absolute magnitudes and compiled from the bibliography is excellent (although we estimated spectral types for some {\it Gaia} sources that had previously gone unreported in the literature).
In the case of unresolved (spectroscopic binaries and close WDS astrometric binaries), very bright stars (e.g. $\alpha$~Cen A and B), giants, subgiants, and white dwarfs (Table\,\ref{tab:giants}), we compiled $M$ values from the bibliography \citep[e.g.][]{lajoie07,soubiran08,feuillet16,eker18,stock18,gentilefusillo19}.
If unavailable, we determined $M$ from colours and absolute magnitudes by assuming either two equal-mass stars in double-lined spectroscopic binaries or that the mass of the companion, $M_2$ is much less than the mass of the primary, $M_1$, in single-lined spectroscopic binaries \citep{latham02}.
Because of this na\"ive approach, we established an uncertainty of 10\% for our $M$ values \citep{mann19,schweitzer99}, which may actually be larger in poorly investigated, single-lined spectroscopic binaries. 
In only two cases, namely, of white dwarfs without a public mass determination, we estimated their $M$ as in \citet{rebassamansergas21}.
For the giants, subgiants, and white dwarfs we also compiled ages from the literature, as summarised in the last column of Table\,\ref{tab:giants}; such ages can be extrapolated to their wide companions.
While the masses of the white dwarfs vary between about 0.5 and 0.8\,$M_\odot$ and of the giants between 1.0 and 3.2\,$M_\odot$, the masses of the stars on or near the main sequence vary from about 0.08 to 2.8\,$M_\odot$.
The latter extremes correspond to the new ultracool dwarf Gaia DR3 749786356557791744 at the M-L boundary, which is at 13.7\,arcsec to the solar-like \object{HD~90681} star and the B9.5\,IV \object{HD~188162}, which is the most massive star of a septuple system candidate (Sect.~\ref{sec:results_discussion}).

Next, we computed the projected physical separation, $s$, between every two {\it Gaia}-resolved components in each pair from the angular separation, $\rho$, and the distance, $d$, to the primary.
Given the wide separations considered, instead of using the $s \approx d ~ \rho$ approximation, we used instead the exact definition from the trigonometry:
\begin{equation}
\label{Eq_s}
s = d ~ \sin{\rho}.
\end{equation}
We considered the distance to the primary star (which usually has the smallest parallax uncertainty) as the distance to the whole system.
The determined $s$ vary from $\sim$11\,au in the case of nearby, close astrometric binaries (e.g. HD~6101), to $\sim 2.3 ~ 10^6$\,au (about 11\,pc) in the case of the very widest companions (see below).
The uncertainty in $s$ is underestimated for primaries whose parallaxes may be affected by close binarity.

Finally, we determined reduced binding energies of the wide systems as in \citet{caballero09}:
\begin{equation}
|U^*_g| = G\frac{M_1 M_2}{s},
\label{eqn:binding}
\end{equation}
They are ``reduced'' because we used the projected physical separation for computing $|U^*_g|$ instead of the actual separation or the semi-major axis, $a$, which is unknown.
We did not apply a most probable conversion factor between $a$ and $s$ for easier computation and, especially, comparisons with previous works \citep{close03,burgasser07,radigan09,caballero10,faherty10}.
This conversion factor, resulting from a uniform distribution of tridimensional vectors projected on a bidimensional plane \citep{abt76,fischer92}, would lead to about 26\% larger actual separations and, therefore, 26\% smaller (non-reduced) binding energies\footnote{\citet{fischer92} determined the statistical correction $\overline{a} \approx 1.26 ~ \overline{s}$ between projected separation ($s$)
and true separation ($a$) from Monte Carlo
simulations over a full suite of binary parameters.} \citep{dhital10,oelkers17}.

The resulting $M_1$, $M_2$, $\rho$, $\theta$, $s$, and $|U^*_g|$ are listed in Table~\ref{tab:systems2}. 
We computed $|U^*_g|$ only for systems with double-like hierarchy, that is, actual doubles and multiple systems with
$\rho_{1, \rm wide} \gg \rho_{1,i}$.
Here, `wide' indicates resolved companions at $\rho_{1, \rm wide} > 1000$\,arcsec and `$i$' other components.
As a result, we did not compute $|U^*_g|$ of 14 multiple systems with $\rho_{\rm wide} \sim \rho_{1,i}$, which we called trapezoidal systems or trapezia.

\section{Results and discussion}
\label{sec:results_discussion}

\begin{table}
 \centering
 \caption[]{Multiplicity order rates of ultrawide galactic systems.}
 \begin{tabular}{lcc}
 \hline \hline
 \noalign{\smallskip}
System type & Minimum  & Estimated  \\
 & rate$^a$ (\%) & rate$^b$(\%) \\
 \noalign{\smallskip}
 \hline
 \noalign{\smallskip}
 Double & 51.6$\pm$14.6 &  32.3$\pm$11.5 \\
 Triple & 25.8$\pm$10.3 &  23.7$\pm$9.9 \\
 Quadruple & 15.1$\pm$7.9 &  21.5$\pm$9.4 \\
 Quintuple or higher & 7.5$\pm$5.6 &  22.6$\pm$9.7 \\
  \noalign{\smallskip}
 \hline
 \noalign{\smallskip}
 \end{tabular}
 \label{tab:multiplicity}
 \tablefoot{
    \tablefoottext{a}{Multiplicity order rate including only systems resolved by {\it Gaia} or tabulated by WDS.}
    \tablefoottext{b}{Multiplicity order rate including also close binary candidates with large {\tt RUWE}, $\sigma_{Vr}$, or proper motion anomaly.}
}
\end{table}

Among the 155\,159 pairs contained in the WDS catalogue at the time of our analysis, 153 pairs with common-parallax, common-proper motion, ultrawide components at $\rho >$ 1000\,arcsec passed our astrometric criteria in Sect.~\ref{sec:pair_validation}, of which 59 (38.6$\pm$9.8\%) are part of young SKGs, associations, or open clusters (Table~\ref{tab:groups_long}), and 95 (61.4$\pm$12.4\%) are ultrawide pairs in 94 galactic systems --one triple is made of two pairs with $\rho >$ 1000\,arcsec and different WDS entries (see Sect.~\ref{sec:search_additional}), which makes 95 WDS pairs.
Because of the small sample size, we used the Wald interval \citep{agresti98} with a 95\% of confidence to calculate the ratio uncertainties\footnote{Wald 95\% confidence interval is $(\lambda - 1.96\sqrt{\lambda/n}, \lambda + 1.96\sqrt{\lambda/n})$, where $\lambda$ is the number of successes in $n$ trials.}. 
To the galactic systems, we added 39 companions from the literature and separated by $\rho <$ 1000\,arcsec.
In our {\it Gaia} DR3 search, we also found 39 additional astrometric companions not catalogued by WDS; that is, we found new companions in about a quarter of the investigated systems. 
In contrast, WDS tabulated a number of additional companion candidates with accurate {\it Gaia} DR3 data that did not pass our conservative astrometric criteria (Sect.\,\ref{sec:pair_validation}).
Most, but not all, of them are flagged by WDS with `{\tt U}' (`proper motion or other technique indicates that this pair is non-physical').

The 94 galactic field systems and their components are listed in Tables\,\ref{tab:systems1} (basic astrometry and photometry) and~\ref{tab:systems2} (stellar masses, angular and projected physical separations, position angles, and binding energies).
We remark that we reclassified the stars in Tables\,\ref{tab:systems1} and~\ref{tab:systems2} as primaries, secondaries, tertiaries, and so on, according to their $G$-band magnitudes. As a result, the WDS nomenclature ``A'', ``B'', ``C'' (etc.) does not always match our re-ordering.

Among the 94 galactic field systems, there are 48 double, 24 triple, 14 quadruple, 2 quintuple, 3 sextuples, and 2 septuples.
The corresponding minimum multiplicity order rates are displayed in Table~\ref{tab:multiplicity}.
The estimated multiplicity order rates and, therefore, the number of multiple systems increase significantly at the expense of the number of doubles if the new candidate companions with large {\tt RUWE}, $\sigma_{Vr}$, or proper motion anomaly are included (Sect.~\ref{sec:ruwe_vr}).
When these close binary candidates are taken into account, most of the ultrawide systems (67.8\%) become multiple: 23.7\% are triple, 21.5\% are quadruple, and 22.6\% have a higher multiplicity order.
These rates are far greater than what is found in less separated multiple systems in the field \citep{chaname04,duchene13,tokovinin97}.
Such a higher-than-usual multiplicity order implies a larger total mass, which, in turn, implies a larger binding energy.

\begin{figure*}
 \centering
 \includegraphics[width=1\linewidth, angle=0]{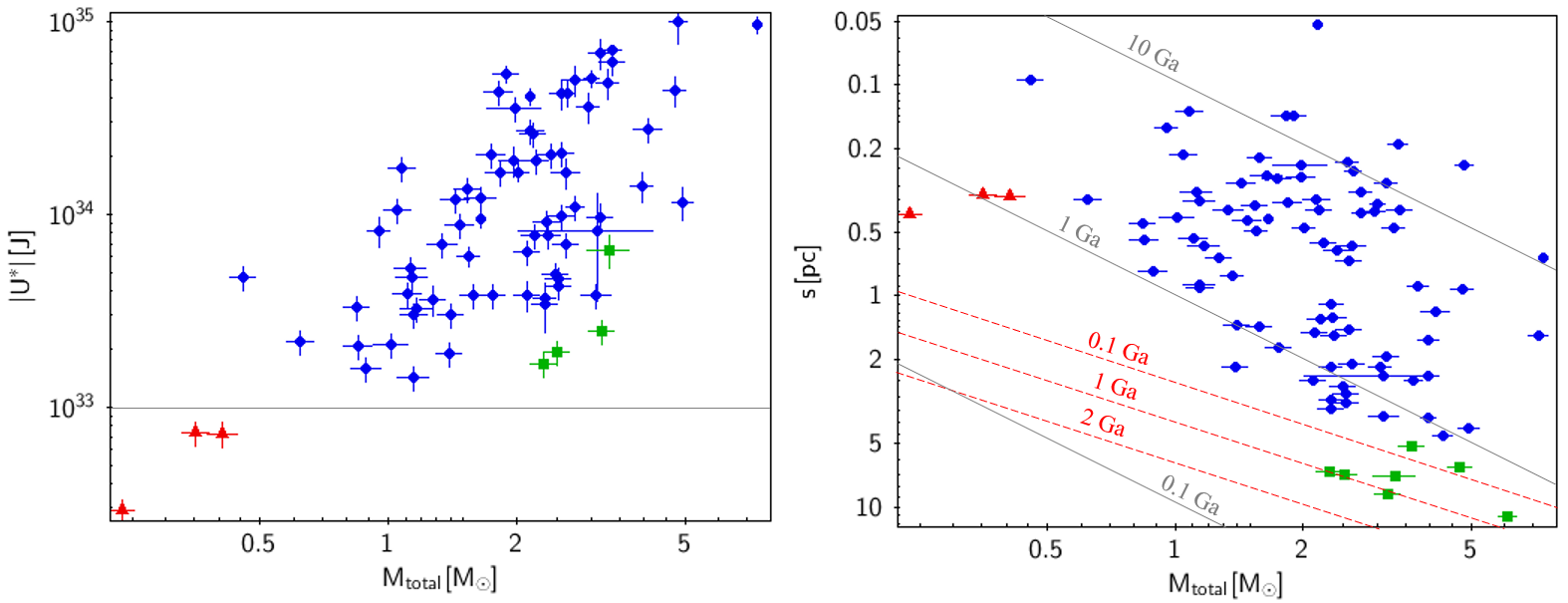}
 \caption{Reduced binding energy ({\it left}) and projected physical separation ({\it right}) as functions of ultrawide system total mass.
 In both panels, the three most fragile systems (top of Table~\ref{tab:fragile}) and the most separated systems (bottom of  Table~\ref{tab:fragile}) are plotted in red triangles and green squares, respectively, while the rest of investigated ultrawide systems are plotted in blue circles.
 The error bars in $s$ are smaller than the used symbols.
 In the {\it left} panel, the horizontal line marks the limit of $|U^*_g|$ at $10^{33}$\,J.
 In the {\it right} panel, the grey solid diagonal lines mark the statistical maximum ages of 0.1, 1, and 10\,Ga at which the systems are likely bound (computed with Eq.~\ref{eqn:seplimit}), while the red dashed diagonal lines mark the corresponding orbital periods of 0.1, 1, and 2\,Ga (computed with Kepler's third law). 
 In both cases we used the correction $a \approx 1.26~s$ \citep{fischer92}.
}
\label{fig:massenergysep}
\end{figure*}

\begin{table*}
 \centering
 \caption[]{Most fragile ($|U^*_g|<10^{33}$\,J) and the most separated ($s\geq 5$\,pc) systems.}
 \begin{tabular}{cllcccc}
 \hline \hline
 \noalign{\smallskip}
WDS & Discoverer & Star & $M$ & \textit{d}$^a$ & \textit{s}$^b$ & $|U^*_g|$ \\
 & code &  & ($M_{\odot}$) & (pc) & ($10^3$\,au) & ($10^{33}$ J) \\
 \noalign{\smallskip}
 \hline
 \noalign{\smallskip}
 \multicolumn{7}{c}{\textit{The most fragile systems}}\\
 \noalign{\smallskip}
00016--0102 &  & 2MASS J00013688-0101441 & 0.22$\pm$0.02 & \multirow{2}{*}{60.33$\pm$0.13} &  \multirow{2}{*}{68.5$\pm$0.2} & \multirow{2}{*}{0.74$\pm$0.10}  \\
 & WIS 1 & SIPS J0000-0112 & 0.13$\pm 0.01$ & & & \\
 \noalign{\smallskip}  
02025--3849 &  & 2MASS J02022892-3849021$^c$ & 0.32$\pm$0.03 & \multirow{2}{*}{59.46$\pm$2.29} & \multirow{2}{*}{69.3$\pm$2.7} & \multirow{2}{*}{0.72$\pm$0.11} \\
 & WIS 248 & 2MASS J02004917-3848535 & 0.09$\pm$0.01 & & & \\
 \noalign{\smallskip}
15488+4929 & & LSPM J1548+4928 & 0.12$\pm$0.01 & \multirow{2}{*}{76.81$\pm$0.63} & \multirow{2}{*}{85.0$\pm$0.7} & \multirow{2}{*}{0.29$\pm$0.04} \\
 & WIS 295 & LSPM J1550+4921 & 0.11$\pm$0.01 & & & \\
 \noalign{\smallskip}
 \hline
 \noalign{\smallskip}
 \multicolumn{7}{c}{\textit{The most separated systems}}\\  
 \noalign{\smallskip} 
02315+0106 &  & HD 15695 & 1.75$\pm$0.18 & \multirow{6}{*}{105.52$\pm$0.33} & \multirow{6}{*}{2284.7$\pm$7.2} & \multirow{6}{*}{...} \\
 & STF 274 & BD+00 415B & 1.63$\pm$0.16 &  &  &  \\
 & SHY 422 & HD 17000 & 1.14$\pm$0.11 &  &  &  \\
 & ... & HD 16985 & 1.07$\pm$0.11 &  &  &  \\
 & ... & Gaia DR3 2497835645142616192$^d$ & 0.30$\pm$0.03 &  &  &  \\
 & ... & Gaia DR3 2514005200579732608 & 0.20$\pm$0.02 &  &  &  \\
\noalign{\smallskip} 
07166--2319 &  & HD 56578$^e$ & 2.42$\pm$0.24 & \multirow{3}{*}{106.32$\pm$0.47} & \multirow{3}{*}{1340.9$\pm$6.0} & \multirow{3}{*}{...} \\
 & SHY 508 & HD 57527$^e$ & 1.92$\pm$0.19 &  &  &  \\
 & ... & Gaia DR3 5613164850183516544 & 0.35$\pm$0.03 &  &  &  \\
\noalign{\smallskip} 
10532--3006 &  & HD 94375$^e$ & 1.32$\pm$0.13 & \multirow{2}{*}{82.17$\pm$0.16} & \multirow{2}{*}{1439.6$\pm$2.8} & \multirow{2}{*}{1.91$\pm$0.27} \\
 & SHY 563 & HD 94542$^e$ & 1.19$\pm$0.12 &  &  &  \\
\noalign{\smallskip} 
15330--0111 &  & 11 Ser & 1.27$\pm$0.35 & \multirow{4}{*}{83.61$\pm$0.42} & \multirow{4}{*}{1483.0$\pm$7.4} & \multirow{4}{*}{3.08$\pm$0.62} \\
 & SHY 678 & HD 142011 & 1.21$\pm$0.12 &  &  &  \\
 & ... & Gaia DR3 4403070145373483392 & 0.43$\pm$0.04 &  &  &  \\
 & ... & Gaia DR3 4403070149671286272$^d$ & 0.40$\pm$0.04 &  &  &  \\
\noalign{\smallskip} 
17166+0325 &  & HD 156287$^e$ & 1.24$\pm$0.12 & \multirow{2}{*}{82.15$\pm$0.17} & \multirow{2}{*}{1407.6$\pm$2.8} & \multirow{2}{*}{1.68$\pm$0.24} \\
 & SHY 715 & HD 159243 & 1.08$\pm$0.11 &  &  &  \\
\noalign{\smallskip} 
21105+2227 &  & HD 201670$^d$ & 1.74$\pm$0.17 & \multirow{2}{*}{113.81$\pm$0.97} & \multirow{2}{*}{1783.3$\pm$15.2} & \multirow{2}{*}{2.49$\pm$0.35} \\
 & SHY 793 & HD 198759 & 1.45$\pm$0.15 &  &  &  \\
\noalign{\smallskip} 
22497+6612 &  & $\iota$ Cep & 1.55$\pm$0.20 & \multirow{4}{*}{36.65$\pm$0.18} & \multirow{4}{*}{1061.3$\pm$5.1} & \multirow{4}{*}{...} \\
 & SHY 359 & HD 215588 & 1.23$\pm$0.12 &  &  &  \\
 & ... & UCAC3 297-187960  & 0.35$\pm$0.03 &  &  &  \\
 & ... & SDSS J230056.41+640815.5 & 0.50$\pm$0.10 &  &  &  \\
 \noalign{\smallskip}
 \hline
 \noalign{\smallskip}
 \end{tabular}
 \label{tab:fragile}
 \tablefoot{
    \tablefoottext{a}{Distance of the primary star.}
    \tablefoottext{b}{Maximum separation between stars inside the system.}
    \tablefoottext{c}{{\tt RUWE}$>10$.}
    \tablefoottext{d}{Large $\sigma_{Vr}$ for its $G$ magnitude.}
    \tablefoottext{e}{Proper motion anomaly measured by \citet{kervella19}, \citet{brandt21} or both.}
}
\end{table*}

In the left panel of Fig.\,\ref{fig:massenergysep}, we display the minimum reduced gravitational binding energy of the 80 systems for which we were able to compute their $|U^*_g|$ as a function of the total mass in the system, $M_{\rm total} = \sum M_i$, $i=1:7$ (i.e. all except for the 14 trapezia).
This diagram would be complete only by adding systems with angular separations $\rho < 1000$\,arcsec but with very low masses \citep[e.g.][]{caballero07a,caballero07b,artigau07,rica12}.
It is complete, however, at the highest total masses and lowest binding energies.
Actual total masses and binding energies, when close binary candidates are taken into account, are larger.

There are three systems with $|U^*_g| < 10 ^{33}$\,J, significantly lower that those of the other 77 systems.
They are listed at the top of Table~\ref{tab:fragile} with their WDS identifiers, discoverer codes (i.e. Wide-field Infrared Survey Explorer, WIS, \citealt{kirkpatrick16}), Simbad names, stellar masses, distances, projected physical separations, and gravitational binding energies.
The three systems are doubles composed of M3--6\,V primaries and M5--9\,V secondaries.
These spectral types were estimated by us from $M_G$ from the relations of \citet{pecaut13} and \citet{cifuentes20}, except for the secondary star in WDS~15488+4929, namely, \object{LSPM~J1550+4921}, whose spectral type M7.0\,V was determined by \citet{west11} from low-resolution spectroscopy. 
With a mass of about 0.09\,$M_\odot$, the secondary in the system WDS~02025--3849, namely, \object{2MASS J02004917--3848535} ($\sim$M7--8\,V), is the second-least massive star in our whole sample.
The low masses of the system components and the wide separations, of about 68--85\,10$^3$\,au (six to eight times wider than $\alpha$~Cen + Proxima), explain the very low $|U^*_g|$.
Actual binding energies may be larger, as the primary in WDS~02025--3849 has a {\tt RUWE} = 40.7; 
assuming an equal-mass binary, the corrected binding energy would double.
None of the three systems have radial-velocity determinations (from {\it Gaia} DR3 and \citealt{west11}) for the two resolved components. 
Given their relatively large $\mu$ ratios and $\Delta$PA (but within our boundary conditions), a dedicated radial-velocity follow-up would be necessary to ascertain whether the three fragile binaries are actually triples. 

Even if each of the three systems had an additional component and, therefore, higher total masses and binding energies than estimated above, there seems to be a lower boundary of $|U^*_g|$ for the most fragile systems at about $10^{33}$\,J (first mentioned by \citealt{caballero10}).
This lower limit may be a consequence of the tidal disruption of wide systems by the galactic gravitational potential, via energy and momentum exchange in encounters with other stars or even the interstellar medium \citep{heggie75,draine80,bahcall81,dhital10,jiang10}.
Actually, during the lifetime of a stellar system, the continuous small and dissipative encounters with other stars are much more disruptive than occasional single catastrophic encounters \citep{retterer82,weinberg87}.
As a result of these interactions, the initial distribution of separations of stellar systems change (increase) over time until eventual disruption.
Using the Fokker–Planck coefficients to describe the effects produced on the orbital binding energies due to those small encounters over time, \citet{weinberg87} estimated the average lifetime of a binary as:

\begin{equation}
\begin{aligned}
t_*(a) \simeq 18\,\text{Ga} ~
\left(\frac{n_*}{0.05\,\text{pc}^{-3}}\right)^{-1}
\left(\frac{M_*}{M_\odot}\right)^{-2}
\left(\frac{M_{\rm tot}}{M_\odot}\right)\\
\left(\frac{V_{\rm rel}}{20\,\text{km\,s}^{-1}}\right)
\left(\frac{a}{0.1\,\text{pc}}\right)^{-1}
\text{ln}^{-1}\Lambda,
\end{aligned}
\label{eqn:weinberg}
\end{equation}

\noindent where $n_*$ and $M_*$ are the number density and average mass of the perturber objects, $V_{\rm rel}$ is the relative velocity between the binary system and the perturber, $M_{\rm tot}$ and $a$ are the total mass and semi-major axis of the binary system, and ln\,$\Lambda$ is the Coulomb logarithm.
The calculation was simplified by \citet{dhital10} by setting the values $n_* = 0.1\,M_\odot$\,pc$^{-3}$, $M_* = 0.7\,M_\odot$, $V_{\rm rel} = 20$\,km\,s$^{-1}$, and $\ln{\Lambda} = 1$ \citep{close07}, and produced an equation that describes in a statistical way the maximum separation of a surviving stellar system at a given age:

\begin{equation}
a \simeq 1.212\,\frac{M_{\text{total}}}{t_*},
\label{eqn:seplimit}
\end{equation}

\noindent where the total mass is in $M_{\odot}$, the average lifetime in Ga, and the semi-major axis in~pc. 

We plot the projected physical separation $s$ as a function of the total mass $M_{\odot}$ of the 94 ultrawide systems in the right panel of Fig.~\ref{fig:massenergysep}.
Overplotted on them, we display the physical separations corresponding to 0.1, 1.0, and 10\,Ga and the orbital periods for 0.1, 1.0, and 2\,Ga.
The three most fragile systems may have survived in their current configuration by about 1\,Ga or slightly less in the case of WDS~15488+4929.
However, there are other systems that are less fragile (i.e. have higher reduced binding energies, comparable to those of well-recognised systems) but that can be disrupted in a few hundred million years.
As we may expect, they are among the most separated systems.

There are seven system candidates with $s$ = 1.1--2.3~$10^6$\,au (5.1--11.1\,pc), listed at the bottom of Table~\ref{tab:fragile}. 
These refer to the kinds of systems that lend their name to the topic of this work (Reaching the boundary between stellar kinematic groups and very wide binaries).
In the spherical volume of radius 10\,pc centred on the Sun, according to the exhaustive compendium by \citet{reyle21}, there are 339 systems containing stars, brown dwarfs, and exoplanets.
As a result, regardless of their (unknown) age, the ultrawide binary and multiple systems may be at the last stages of disruption and follow the formation-evolution-dissolution sequence described by \citet{close03}, who predicted an overabundance of very low-mass binaries far from the centre of the original `minicluster.'
This is the case of the most separated components in the systems WDS~02315+0106 and WDS~15330--0111, which have low masses and large $\sigma_{Vr}$ for their $G$ magnitudes.
The seven system candidates, all of them identified by \citet{shaya11}, may be the remnants of previous SKGs that are being dissolved in the Milky Way and that are older than the ones identified in Sect.~\ref{sec:search_additional}.
Three system candidates, including the sextuple (perhaps septuple) WDS~02315+0106 system, are trapezia and, therefore, their reduced binding energies were not computed.
The other four systems consist of three very wide binaries made of two bright Henry Draper stars \citep{cannon18} and one double-like hierarchical quadruple (perhaps quintuple) system.
The latter, namely WDS~15330--01110, is made of the K0 giant 11~Ser (Table~\ref{tab:giants}), the G1 dwarf \object{HD~142011}, and two anonymous early M dwarfs, one of which has a large $\sigma_{Vr}$ for their $G$ magnitude.
These two (or three) M dwarfs are very close to each other ($s \sim$ 92\,au) and to the Sun-like star ($s \sim$ 630--690\,au), which allowed us to compute $|U^*_g|$.
However, the K0 giant and the G1+M+M triple are separated by about 1.5~$10^6$\,au (7.2\,pc).
Between them,  dozens of unrelated stars with similar parallaxes must exist, but with very different proper motions that exert smooth `continuous small and dissipative' gravitational thrusts. 

\begin{figure}
 \centering
 \includegraphics[width=1\linewidth, angle=0]{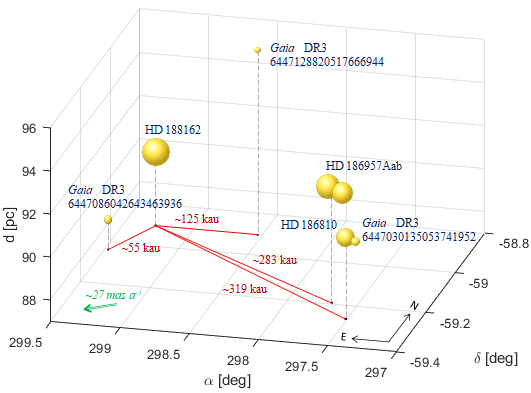}
 \caption{Spatial distribution of the septuple system WDS 19507--5912. 
 The size of the spheres representing every star, labelled in blue, are proportional to their brightness. 
 The projected physical separations from the primary star, HD 188162, in red, are in $10^3$\,au. 
 The overall proper motion, in green, is in mas\,a$^{-1}$.}
 \label{fig:septuple}
\end{figure}

Something similar may occur in the case of the 14 ultrawide trapezia, which tend to have a high multiplicity hierarchy: 1 of the 2 septuple systems, the 3 sextuples, 7 of the 15 quadruples, and 3 of the 25 triples are trapezia.
In particular, the trapezoidal septuple system WDS~19507--5912 is sketched in Fig.~\ref{fig:septuple}.
It consists of three late-B-to-early-A stars (one is a spectroscopic binary) and three intermediate-to-late M dwarfs (one is close to an A star).
Given the early spectral types of the most massive stars, this system may approximately be the age of the Hyades  (600--700\,Ma; \citealt{perryman98,gossage18,martin18}) and, therefore, stand as an unidentified sparse young SKG\footnote{If confirmed in the future, we propose naming the SKG following the discovery name of the brightest, earliest star: \object{HD~188162}.}.
All these results are in accordance with the suggestions by \citet{basri06} and \citet{caballero07a}, who proposed a major prevalence of wide triples over wide binaries.
We also confirm that the individual components of systems at very wide separations are often multiple systems themselves, as stated by
\citet{cifuentes21}.

Five of the seven ultrawidest systems have orbital periods of the order of 1\,Ga, even older than the Hyades.
Such long orbital periods stand as a challenge to the `binary' definition itself, namely: a system of two stars that are gravitationally bound to and in orbit around each other. 
As a result, the ultrawidest pairs may have not completed one revolution either because of their young age or because they were recently disrupted by passing stars and, therefore, should not be called `binaries'.
This statement should also be extrapolated to the system candidates in young stellar kinematic groups (Section~\ref{sec:search_additional}), including less separated but also less massive, pairs.
Furthermore, \citet{caballero09} already claimed that the \object{AU~Mic}+\object{AT~Mic} system in the $\beta$~Pictoris moving group has only completed at most two orbital periods since its formation.
All of this makes the boundary between stellar kinematic groups and very wide binaries blurrier and blurrier.

Finally, we used the accurate {\it Gaia} astrometry to measure the relative transverse velocity, $\Delta V$, as a function of the projected physical separation of the 48 widest systems with $s >$ 0.1\,pc, and compared it with the maximum velocity allowed for a bound binary, as done by some other authors (e.g. \citealt{el_badry19b,el-badry21}).
This comparison may interpret several observational studies that have reported that the difference in the proper motions or radial velocities of the components of nearby wide binaries appear larger than predicted by Kepler’s laws, indicating a potential breakdown of general relativity at low accelerations.
However, our data, which are relatively scarce compared to extensive simulations (cf. \citealt{el_badry19b}), do not even show projection effects.
Furthermore, inner subsystems, which are frequent, disturb our $\Delta V$ estimates.
As a result, the actual bound nature of our most fragile system is hardly verifiable.

To sum up, wide pairs with very low-mass components and $|U^*_g| \sim 10^{33}$\,J (e.g. ultracool dwarf binaries with late-M and early-L spectral types and projected physical separations of a few thousand astronomical units -- \citealt{caballero07a,caballero07b,artigau07}) are perhaps more relevant for investigating the disruption of fragile systems by the galactic gravitational potential rather than ultrawide systems of gargantuan projected physical separations (much larger than those proposed by \citealt{tolbert64}, \citealt{bahcall81}, or \citealt{retterer82}) caught in the act of destruction and, that probably are the leftover of past SKGs.

\section{Summary}
\label{sec:summary}

Thanks to the {\it Gaia} DR3 \citep{gaiacollaboration22} and a number of parallax and proper motion searches in the previous decade \citep[e.g.][]{caballero10, shaya11, tokovinin12, kirkpatrick16}, we present a leap forward with respect to the first item of this series of papers, on the Washington double stars with the widest angular separations \citep{caballero09}.
Accordingly, we increase by over an order of magnitude the sample size and the astrometric precision of systems with angular separations $\rho >$ 1000\,arcsec.
Among other results of our analysis, we present:
($i$) 40 additional astrometric companions not catalogued yet by WDS, including several ultracool dwarfs at the M-L boundary and one hot white dwarf, not counting several dozens close binary candidates from large {\it Gaia} DR3 {\tt RUWE} and $\sigma_{\rm RV}$; 
($ii$) a general confusion in the literature between actual, physically bound, ultrawide pairs and components in young SKGs, associations, and even clusters with identical galactocentric space velocities;
($iii$) three very fragile systems discovered by \citet{kirkpatrick16} that are made of intermediate and late M dwarfs with large projected physical separations of 0.33--0.41\,pc and small reduced binding energies $|U_g^*| \lesssim 10^{33}$\,J, which is probably the smallest value found among gravitationally bound systems;
($iv$) the individual components of systems at very wide separations are often multiple systems themselves \citep{cifuentes21}, which implies an overabundance of high-order multiples (triples, quadruples, quintuples, and more) among the widest systems, and larger binding energies and total masses than systems with comparable separations but lower multiplicity order;  
and ($v$) an additional observational confirmation of classical theoretical predictions \citep[e.g.][]{weinberg87} of disruption of binary systems by the galactic gravitational potential, which destroys the ultrawidest systems with total masses below 10\,$M_\odot$ in less that 600--700\,Ma (the age of the Hyades cluster).
As incidental results, we report 40 new candidate stars in known young SKGs and at least one new young stellar association around the bright star $\gamma$~Cas, although some of our highest-order multiple systems, such as the septuple around the B9.5\,IV star HD~188162, may also be association remnants.

In conclusion, the total mass, binding energy, and probability that an ultrawide system is actually bound increase as the stellar multiplicity order also increases.
Many systems reported here have overwhelmingly large projected physical separations, but have instead total masses large enough for the binding energies being comparable to those of less separated, less massive systems that are widely accepted as physical.
However, none of the ultrawide systems will survive for another few hundred million years.
In a sense, the widest multiple systems of today, which are now being torn apart by the Galaxy, will be the single stars of tomorrow.

%
%
\begin{acknowledgements}
We thank the anonymous reviewer for their constructive suggestions and comments, Jes\'us Ma\'iz Apell\'aniz for discussion on the $\gamma$~Cas system, Alberto Rebassa-Mansergas for estimating masses of two white dwarfs,
Brian~D. Mason for help with five unidentified wide WDS companions,
and Andrei Tokovinin for providing us the $\rho$ of a wide pair and very useful discussion on hierarchical multiplicity.
We acknowledge financial support from the Agencia Estatal de Investigaci\'on 10.13039/501100011033 of the Ministerio de Ciencia e Innovaci\'on and the ERDF ``A way of making Europe'' through projects PID2019-109522GB-C51 and PID2020-112949GB-I00, and the Centre of Excellence ``Mar\'ia de Maeztu'' award to the Centro de Astrobiolog\'ia (MDM-2017-0737), 
and from the European Commission Framework Programme Horizon 2020 Research and Innovation through the ESCAPE project under grant agreement no.~824064.
This research made use of the Washington Double Star Catalog maintained at the U.S. Naval Observatory,
NASA’s Astrophysics Data System Bibliographic Services, the Simbad database, VizieR catalogue access tool, and Aladin sky atlas at the CDS, Strasbourg (France), and the TOPCAT tool.
\end{acknowledgements}
%

%
%
\bibliographystyle{aa.bst}
\bibliography{mnemonic,biblio}

\begin{thebibliography}{207}
\expandafter\ifx\csname natexlab\endcsname\relax\def\natexlab#1{#1}\fi

\bibitem[{{Abt}(1988)}]{abt88}
{Abt}, H.~A. 1988, \apj, 331, 922

\bibitem[{{Abt} \& {Levy}(1976)}]{abt76}
{Abt}, H.~A. \& {Levy}, S.~G. 1976, \apjs, 30, 273

\bibitem[{{Agresti} \& {Coull}(1998)}]{agresti98}
{Agresti}, A. \& {Coull}, B. 1998, Am. Stat., 52, 119

\bibitem[{{Allen} {et~al.}(1998){Allen}, {Herrera}, \& {Poveda}}]{allen98}
{Allen}, C., {Herrera}, M.~A., \& {Poveda}, A. 1998, in IX Latin American
  Regional IAU Meeting, ``Focal Points in Latin American Astronomy'', ed.
  A.~{Aguilar} \& A.~{Carraminana}, 21

\bibitem[{{Allen} {et~al.}(1997){Allen}, {Poveda}, \& {Herrera}}]{allen97}
{Allen}, C., {Poveda}, A., \& {Herrera}, M.~A. 1997, {The Distribution of
  Separations of Wide Binaries}, Vol. 223 (Springer), 133

\bibitem[{{Allen} {et~al.}(2000){Allen}, {Poveda}, \& {Herrera}}]{allen00}
{Allen}, C., {Poveda}, A., \& {Herrera}, M.~A. 2000, A\&A, 356, 529

\bibitem[{{Allen}(1899)}]{allen1899}
{Allen}, R.~H. 1899, {Star-names and their meanings}

\bibitem[{{Alonso-Floriano} {et~al.}(2015){Alonso-Floriano}, {Caballero},
  {Cort{\'e}s-Contreras}, {Solano}, \& {Montes}}]{alonsofloriano15}
{Alonso-Floriano}, F.~J., {Caballero}, J.~A., {Cort{\'e}s-Contreras}, M.,
  {Solano}, E., \& {Montes}, D. 2015, \aap, 583, A85

\bibitem[{{Ambartsumian}(1949)}]{ambartsumian49}
{Ambartsumian}, V.~A. 1949, Sov. Astron., 26, 3

\bibitem[{{Arenou} {et~al.}(2018){Arenou}, {Luri}, {Babusiaux}, {Fabricius},
  {Helmi}, {Muraveva}, {Robin}, {Spoto}, {Vallenari}, {Antoja},
  {Cantat-Gaudin}, {Jordi}, {Leclerc}, {Reyl{\'e}}, {Romero-G{\'o}mez}, {Shih},
  {Soria}, {Barache}, {Bossini}, {Bragaglia}, {Breddels}, {Fabrizio},
  {Lambert}, {Marrese}, {Massari}, {Moitinho}, {Robichon}, {Ruiz-Dern},
  {Sordo}, {Veljanoski}, {Eyer}, {Jasniewicz}, {Pancino}, {Soubiran}, {Spagna},
  {Tanga}, {Turon}, \& {Zurbach}}]{arenou18}
{Arenou}, F., {Luri}, X., {Babusiaux}, C., {et~al.} 2018, \aap, 616, A17

\bibitem[{{Argelander}(1903)}]{argelander03}
{Argelander}, F. W.~A. 1903, Eds Marcus and Weber's Verlag, 0

\bibitem[{{Artigau} {et~al.}(2007){Artigau}, {Lafreni{\`e}re}, {Doyon},
  {Albert}, {Nadeau}, \& {Robert}}]{artigau07}
{Artigau}, {\'E}., {Lafreni{\`e}re}, D., {Doyon}, R., {et~al.} 2007, \apjl,
  659, L49

\bibitem[{{Bahcall} \& {Soneira}(1981)}]{bahcall81}
{Bahcall}, J.~N. \& {Soneira}, R.~M. 1981, \apj, 246, 122

\bibitem[{{Bailer-Jones} {et~al.}(2018){Bailer-Jones}, {Rybizki}, {Fouesneau},
  {Mantelet}, \& {Andrae}}]{bailerjones18a}
{Bailer-Jones}, C.~A.~L., {Rybizki}, J., {Fouesneau}, M., {Mantelet}, G., \&
  {Andrae}, R. 2018, AJ, 156, 58

\bibitem[{{Basri} \& {Reiners}(2006)}]{basri06}
{Basri}, G. \& {Reiners}, A. 2006, \aj, 132, 663

\bibitem[{{Batten}(1973)}]{batten73}
{Batten}, A.~H. 1973, {Binary and multiple systems of stars: International
  Series of Monographs in Natural Philosophy} (Pergamon)

\bibitem[{{Bayer}(1603)}]{bayer1603}
{Bayer}, J. 1603, {Uranometria omnium asterismorum continens schemata, nova
  methodo delineata aereis laminis expressa}

\bibitem[{{Bell} {et~al.}(2017){Bell}, {Murphy}, \& {Mamajek}}]{bell17}
{Bell}, C. P.~M., {Murphy}, S.~J., \& {Mamajek}, E.~E. 2017, \mnras, 468, 1198

\bibitem[{{Blaauw}(1991)}]{blaaw91}
{Blaauw}, A. 1991, in NATO Advanced Study Institute (ASI) Series C, Vol. 342,
  The Physics of Star Formation and Early Stellar Evolution, ed. C.~J. {Lada}
  \& N.~D. {Kylafis}, 125

\bibitem[{{Bohn} {et~al.}(2022){Bohn}, {Ginski}, {Kenworthy}, {Mamajek},
  {Meshkat}, {Pecaut}, {Reggiani}, {Seay}, {Brown}, {Cugno}, {Henning},
  {Launhardt}, {Quirrenbach}, {Rickman}, \& {S{\'e}gransan}}]{bohn22}
{Bohn}, A.~J., {Ginski}, C., {Kenworthy}, M.~A., {et~al.} 2022, \aap, 657, A53

\bibitem[{{Bonnarel} {et~al.}(2000){Bonnarel}, {Fernique}, {Bienaym{\'e}},
  {Egret}, {Genova}, {Louys}, {Ochsenbein}, {Wenger}, \&
  {Bartlett}}]{bonnarel00}
{Bonnarel}, F., {Fernique}, P., {Bienaym{\'e}}, O., {et~al.} 2000, A\&AS, 143,
  33

\bibitem[{{Bouvier} {et~al.}(2008){Bouvier}, {Kendall}, {Meeus}, {Testi},
  {Moraux}, {Stauffer}, {James}, {Cuillandre}, {Irwin}, {McCaughrean},
  {Baraffe}, \& {Bertin}}]{bouvier08}
{Bouvier}, J., {Kendall}, T., {Meeus}, G., {et~al.} 2008, \aap, 481, 661

\bibitem[{{Brandt}(2021)}]{brandt21}
{Brandt}, T.~D. 2021, \apjs, 254, 42

\bibitem[{{Bressan} {et~al.}(2012){Bressan}, {Marigo}, {Girardi}, {Salasnich},
  {Dal Cero}, {Rubele}, \& {Nanni}}]{bressan12}
{Bressan}, A., {Marigo}, P., {Girardi}, L., {et~al.} 2012, \mnras, 427, 127

\bibitem[{{Burgasser} {et~al.}(2007){Burgasser}, {Reid}, {Siegler}, {Close},
  {Allen}, {Lowrance}, \& {Gizis}}]{burgasser07}
{Burgasser}, A.~J., {Reid}, I.~N., {Siegler}, N., {et~al.} 2007, in Protostars
  and Planets V, ed. B.~{Reipurth}, D.~{Jewitt}, \& K.~{Keil}, 427

\bibitem[{{Burnham}(1978)}]{burnham78}
{Burnham}, Robert, J. 1978, {Burnham's Celestial Handbook: An Observer's Guide
  to the Universe Beyond the Solar System, in three volumes.}

\bibitem[{{Caballero}(2007{\natexlab{a}})}]{caballero07b}
{Caballero}, J.~A. 2007{\natexlab{a}}, \apj, 667, 520

\bibitem[{{Caballero}(2007{\natexlab{b}})}]{caballero07a}
{Caballero}, J.~A. 2007{\natexlab{b}}, \aap, 462, L61

\bibitem[{{Caballero}(2008)}]{caballero08}
{Caballero}, J.~A. 2008, \mnras, 383, 375

\bibitem[{{Caballero}(2009)}]{caballero09}
{Caballero}, J.~A. 2009, A\&A, 507, 251

\bibitem[{{Caballero}(2010)}]{caballero10}
{Caballero}, J.~A. 2010, A\&A, 514, A98

\bibitem[{{Caballero} {et~al.}(2013){Caballero}, {Genebriera}, {Tobal},
  {Miret}, {Rica}, {Cairol}, {Miret}, {Novalbos}, {Montes}, \&
  {Klutsch}}]{caballero13}
{Caballero}, J.~A., {Genebriera}, J., {Tobal}, T., {et~al.} 2013, in Highlights
  of Spanish Astrophysics VII, ed. J.~C. {Guirado}, L.~M. {Lara}, V.~{Quilis},
  \& J.~{Gorgas}, 971--976

\bibitem[{{Cannon} \& {Pickering}(1918)}]{cannon18}
{Cannon}, A.~J. \& {Pickering}, E.~C. 1918, Annals of Harvard College
  Observatory, 91, 1

\bibitem[{{Cantat-Gaudin} {et~al.}(2018){Cantat-Gaudin}, {Jordi}, {Vallenari},
  {Bragaglia}, {Balaguer-N{\'u}{\~n}ez}, {Soubiran}, {Bossini}, {Moitinho},
  {Castro-Ginard}, {Krone-Martins}, {Casamiquela}, {Sordo}, \&
  {Carrera}}]{cantat-gaudin18}
{Cantat-Gaudin}, T., {Jordi}, C., {Vallenari}, A., {et~al.} 2018, \aap, 618,
  A93

\bibitem[{{Carro}(2021)}]{carro21}
{Carro}, J.~M. 2021, Il Bollettino delle Stelle Doppie, 33, 12

\bibitem[{{Chabrier}(2003)}]{chabrier03}
{Chabrier}, G. 2003, \pasp, 115, 763

\bibitem[{{Chanam{\'e}} \& {Gould}(2004)}]{chaname04}
{Chanam{\'e}}, J. \& {Gould}, A. 2004, ApJ, 601, 289

\bibitem[{{Chen} {et~al.}(2005){Chen}, {Patten}, {Werner}, {Dowell},
  {Stapelfeldt}, {Song}, {Stauffer}, {Blaylock}, {Gordon}, \&
  {Krause}}]{chen05}
{Chen}, C.~H., {Patten}, B.~M., {Werner}, M.~W., {et~al.} 2005, \apj, 634, 1372

\bibitem[{{Christy} \& {Walker}(1969)}]{christy69}
{Christy}, J.~W. \& {Walker}, R.~L., J. 1969, \pasp, 81, 643

\bibitem[{{Cifuentes} {et~al.}(2021){Cifuentes}, {Caballero}, \&
  {Agust{\'\i}}}]{cifuentes21}
{Cifuentes}, C., {Caballero}, J.~A., \& {Agust{\'\i}}, S. 2021, RNAAS, 5, 129

\bibitem[{{Cifuentes} {et~al.}(2020){Cifuentes}, {Caballero},
  {Cort{\'e}s-Contreras}, {Montes}, {Abell{\'a}n}, {Dorda}, {Holgado},
  {Zapatero Osorio}, {Morales}, {Amado}, {Passegger}, {Quirrenbach}, {Reiners},
  {Ribas}, {Sanz-Forcada}, {Schweitzer}, {Seifert}, \& {Solano}}]{cifuentes20}
{Cifuentes}, C., {Caballero}, J.~A., {Cort{\'e}s-Contreras}, M., {et~al.} 2020,
  A\&A, 642, A115

\bibitem[{{Close} {et~al.}(1990){Close}, {Richer}, \& {Crabtree}}]{close90}
{Close}, L.~M., {Richer}, H.~B., \& {Crabtree}, D.~R. 1990, \aj, 100, 1968

\bibitem[{{Close} {et~al.}(2003){Close}, {Siegler}, \& {Freed}}]{close03}
{Close}, L.~M., {Siegler}, N., \& {Freed}, M. 2003, in Brown Dwarfs, ed.
  E.~{Mart{\'\i}n}, Vol. 211, 249

\bibitem[{{Close} {et~al.}(2007){Close}, {Zuckerman}, {Song}, {Barman},
  {Marois}, {Rice}, {Siegler}, {Macintosh}, {Becklin}, {Campbell}, {Lyke},
  {Conrad}, \& {Le Mignant}}]{close07}
{Close}, L.~M., {Zuckerman}, B., {Song}, I., {et~al.} 2007, \apj, 660, 1492

\bibitem[{{Cutri} {et~al.}(2014){Cutri}, {Wright}, {Conrow}, {Fowler},
  {Eisenhardt}, {Grillmair}, {Kirkpatrick}, {Masci}, {McCallon}, {Wheelock},
  {Fajardo-Acosta}, {Yan}, {Benford}, {Harbut}, {Jarrett}, {Lake}, {Leisawitz},
  {Ressler}, {Stanford}, {Tsai}, {Liu}, {Helou}, {Mainzer}, {Gettngs},
  {Gonzalez}, {Hoffman}, {Marsh}, {Padgett}, {Skrutskie}, {Beck}, {Papin}, \&
  {Wittman}}]{cutri14}
{Cutri}, R.~M., {Wright}, E.~L., {Conrow}, T., {et~al.} 2014, VizieR Online
  Data Catalog, 2328, 0

\bibitem[{{da Silva} {et~al.}(2015){da Silva}, {Milone}, \&
  {Rocha-Pinto}}]{dasilva15}
{da Silva}, R., {Milone}, A. d.~C., \& {Rocha-Pinto}, H.~J. 2015, \aap, 580,
  A24

\bibitem[{{David} \& {Hillenbrand}(2015)}]{david15}
{David}, T.~J. \& {Hillenbrand}, L.~A. 2015, \apj, 804, 146

\bibitem[{{de Zeeuw} {et~al.}(1999){de Zeeuw}, {Hoogerwerf}, {de Bruijne},
  {Brown}, \& {Blaauw}}]{dezeeuw99}
{de Zeeuw}, P.~T., {Hoogerwerf}, R., {de Bruijne}, J.~H.~J., {Brown}, A.~G.~A.,
  \& {Blaauw}, A. 1999, \aj, 117, 354

\bibitem[{{Dhital} {et~al.}(2010){Dhital}, {West}, {Stassun}, \&
  {Bochanski}}]{dhital10}
{Dhital}, S., {West}, A.~A., {Stassun}, K.~G., \& {Bochanski}, J.~J. 2010, AJ,
  139, 2566

\bibitem[{{Dopcke} {et~al.}(2019){Dopcke}, {Porto de Mello}, \&
  {Sneden}}]{dopcke19}
{Dopcke}, G., {Porto de Mello}, G.~F., \& {Sneden}, C. 2019, \mnras, 485, 4375

\bibitem[{{Draine}(1980)}]{draine80}
{Draine}, B.~T. 1980, \apj, 241, 1021

\bibitem[{{Duch{\^e}ne} \& {Kraus}(2013)}]{duchene13}
{Duch{\^e}ne}, G. \& {Kraus}, A. 2013, ARA\&A, 51, 269

\bibitem[{{Duquennoy} \& {Mayor}(1991)}]{duquennoy91}
{Duquennoy}, A. \& {Mayor}, M. 1991, \aap, 248, 485

\bibitem[{{Dzib} {et~al.}(2018){Dzib}, {Loinard}, {Ortiz-Le{\'o}n},
  {Rodr{\'\i}guez}, \& {Galli}}]{dzib18}
{Dzib}, S.~A., {Loinard}, L., {Ortiz-Le{\'o}n}, G.~N., {Rodr{\'\i}guez}, L.~F.,
  \& {Galli}, P. A.~B. 2018, \apj, 867, 151

\bibitem[{{Eggen}(1965)}]{eggen65}
{Eggen}, O.~J. 1965, in Galactic structure. Edited by Adriaan Blaauw and
  Maarten Schmidt. Published by the University of Chicago Press, 111

\bibitem[{{Eggleton} \& {Tokovinin}(2008)}]{eggleton08}
{Eggleton}, P.~P. \& {Tokovinin}, A.~A. 2008, \mnras, 389, 869

\bibitem[{{Eker} {et~al.}(2018){Eker}, {Bak{\i}{\c{s}}}, {Bilir}, {Soydugan},
  {Steer}, {Soydugan}, {Bak{\i}{\c{s}}}, {Ali{\c{c}}avu{\c{s}}}, {Aslan}, \&
  {Alpsoy}}]{eker18}
{Eker}, Z., {Bak{\i}{\c{s}}}, V., {Bilir}, S., {et~al.} 2018, \mnras, 479, 5491

\bibitem[{{El-Badry}(2019)}]{el_badry19b}
{El-Badry}, K. 2019, \mnras, 482, 5018

\bibitem[{{El-Badry} {et~al.}(2021){El-Badry}, {Rix}, \& {Heintz}}]{el-badry21}
{El-Badry}, K., {Rix}, H.-W., \& {Heintz}, T.~M. 2021, \mnras, 506, 2269

\bibitem[{{Faherty} {et~al.}(2010){Faherty}, {Burgasser}, {West}, {Bochanski},
  {Cruz}, {Shara}, \& {Walter}}]{faherty10}
{Faherty}, J.~K., {Burgasser}, A.~J., {West}, A.~A., {et~al.} 2010, \aj, 139,
  176

\bibitem[{{Fekel}(1979)}]{fekel79}
{Fekel}, F.~C., J. 1979, PhD thesis, University of Texas, Austin

\bibitem[{{Feuillet} {et~al.}(2016){Feuillet}, {Bovy}, {Holtzman}, {Girardi},
  {MacDonald}, {Majewski}, \& {Nidever}}]{feuillet16}
{Feuillet}, D.~K., {Bovy}, J., {Holtzman}, J., {et~al.} 2016, VizieR Online
  Data Catalog, J/ApJ/817/40

\bibitem[{{Fischer} \& {Marcy}(1992)}]{fischer92}
{Fischer}, D.~A. \& {Marcy}, G.~W. 1992, ApJ, 396, 178

\bibitem[{{Flamsteed}(1725)}]{flamsteed1725}
{Flamsteed}, J. 1725, {Historia Coelestis Britannicae, tribus Voluminibus
  contenta (1675-1689), (1689-1720), vol. 1, 2, 3}

\bibitem[{{Folkes} {et~al.}(2012){Folkes}, {Pinfield}, {Jones}, {Kurtev},
  {Zhang}, {G{\'a}lvez-Ortiz}, {Marocco}, {Day-Jones}, \& {Clarke}}]{folkes12}
{Folkes}, S.~L., {Pinfield}, D.~J., {Jones}, H.~R.~A., {et~al.} 2012, \mnras,
  427, 3280

\bibitem[{{Fracastoro}(1988)}]{fracastoro88}
{Fracastoro}, M.~G. 1988, Ap\&SS, 142, 11

\bibitem[{{Freund} {et~al.}(2020){Freund}, {Robrade}, {Schneider}, \&
  {Schmitt}}]{freund20}
{Freund}, S., {Robrade}, J., {Schneider}, P.~C., \& {Schmitt}, J.~H.~M.~M.
  2020, \aap, 640, A66

\bibitem[{{F{\"u}rnkranz} {et~al.}(2019){F{\"u}rnkranz}, {Meingast}, \&
  {Alves}}]{furnkranz19}
{F{\"u}rnkranz}, V., {Meingast}, S., \& {Alves}, J. 2019, \aap, 624, L11

\bibitem[{{Gagn{\'e}} \& {Faherty}(2018)}]{gagne18c}
{Gagn{\'e}}, J. \& {Faherty}, J.~K. 2018, \apj, 862, 138

\bibitem[{{Gagn{\'e}} {et~al.}(2015){Gagn{\'e}}, {Lafreni{\`e}re}, {Doyon},
  {Malo}, \& {Artigau}}]{gagne15}
{Gagn{\'e}}, J., {Lafreni{\`e}re}, D., {Doyon}, R., {Malo}, L., \& {Artigau},
  {\'E}. 2015, \apj, 798, 73

\bibitem[{{Gagn{\'e}} {et~al.}(2018{\natexlab{a}}){Gagn{\'e}}, {Mamajek},
  {Malo}, {Riedel}, {Rodriguez}, {Lafreni{\`e}re}, {Faherty}, {Roy-Loubier},
  {Pueyo}, {Robin}, \& {Doyon}}]{gagne18a}
{Gagn{\'e}}, J., {Mamajek}, E.~E., {Malo}, L., {et~al.} 2018{\natexlab{a}},
  \apj, 856, 23

\bibitem[{{Gagn{\'e}} {et~al.}(2018{\natexlab{b}}){Gagn{\'e}}, {Roy-Loubier},
  {Faherty}, {Doyon}, \& {Malo}}]{gagne18b}
{Gagn{\'e}}, J., {Roy-Loubier}, O., {Faherty}, J.~K., {Doyon}, R., \& {Malo},
  L. 2018{\natexlab{b}}, \apj, 860, 43

\bibitem[{{Gaia Collaboration} {et~al.}(2018){Gaia Collaboration}, {Brown},
  {Vallenari}, {Prusti}, {de Bruijne}, {Babusiaux}, {Bailer-Jones}, {Biermann},
  {Evans}, {Eyer}, {Jansen}, {Jordi}, {Klioner}, {Lammers}, {Lindegren},
  {Luri}, {Mignard}, {Panem}, {Pourbaix}, {Randich}, {Sartoretti}, {Siddiqui},
  {Soubiran}, {van Leeuwen}, {Walton}, {Arenou}, {Bastian}, {Cropper},
  {Drimmel}, {Katz}, {Lattanzi}, {Bakker}, {Cacciari}, {Casta{\~n}eda},
  {Chaoul}, {Cheek}, {De Angeli}, {Fabricius}, {Guerra}, {Holl}, {Masana},
  {Messineo}, {Mowlavi}, {Nienartowicz}, {Panuzzo}, {Portell}, {Riello},
  {Seabroke}, {Tanga}, {Th{\'e}venin}, {Gracia-Abril}, {Comoretto},
  {Garcia-Reinaldos}, {Teyssier}, {Altmann}, {Andrae}, {Audard},
  {Bellas-Velidis}, {Benson}, {Berthier}, {Blomme}, {Burgess}, {Busso},
  {Carry}, {Cellino}, {Clementini}, {Clotet}, {Creevey}, {Davidson}, {De
  Ridder}, {Delchambre}, {Dell'Oro}, {Ducourant},
  {Fern{\'a}ndez-Hern{\'a}ndez}, {Fouesneau}, {Fr{\'e}mat}, {Galluccio},
  {Garc{\'\i}a-Torres}, {Gonz{\'a}lez-N{\'u}{\~n}ez}, {Gonz{\'a}lez-Vidal},
  {Gosset}, {Guy}, {Halbwachs}, {Hambly}, {Harrison}, {Hern{\'a}ndez},
  {Hestroffer}, {Hodgkin}, {Hutton}, {Jasniewicz}, {Jean-Antoine-Piccolo},
  {Jordan}, {Korn}, {Krone-Martins}, {Lanzafame}, {Lebzelter}, {L{\"o}ffler},
  {Manteiga}, {Marrese}, {Mart{\'\i}n-Fleitas}, {Moitinho}, {Mora}, {Muinonen},
  {Osinde}, {Pancino}, {Pauwels}, {Petit}, {Recio-Blanco}, {Richards},
  {Rimoldini}, {Robin}, {Sarro}, {Siopis}, {Smith}, {Sozzetti}, {S{\"u}veges},
  {Torra}, {van Reeven}, {Abbas}, {Abreu Aramburu}, {Accart}, {Aerts},
  {Altavilla}, {{\'A}lvarez}, {Alvarez}, {Alves}, {Anderson}, {Andrei},
  {Anglada Varela}, {Antiche}, {Antoja}, {Arcay}, {Astraatmadja}, {Bach},
  {Baker}, {Balaguer-N{\'u}{\~n}ez}, {Balm}, {Barache}, {Barata}, {Barbato},
  {Barblan}, {Barklem}, {Barrado}, {Barros}, {Barstow}, {Bartholom{\'e}
  Mu{\~n}oz}, {Bassilana}, {Becciani}, {Bellazzini}, {Berihuete}, {Bertone},
  {Bianchi}, {Bienaym{\'e}}, {Blanco-Cuaresma}, {Boch}, {Boeche}, {Bombrun},
  {Borrachero}, {Bossini}, {Bouquillon}, {Bourda}, {Bragaglia}, {Bramante},
  {Breddels}, {Bressan}, {Brouillet}, {Br{\"u}semeister}, {Brugaletta},
  {Bucciarelli}, {Burlacu}, {Busonero}, {Butkevich}, {Buzzi}, {Caffau},
  {Cancelliere}, {Cannizzaro}, {Cantat-Gaudin}, {Carballo}, {Carlucci},
  {Carrasco}, {Casamiquela}, {Castellani}, {Castro-Ginard}, {Charlot},
  {Chemin}, {Chiavassa}, {Cocozza}, {Costigan}, {Cowell}, {Crifo}, {Crosta},
  {Crowley}, {Cuypers}, {Dafonte}, {Damerdji}, {Dapergolas}, {David}, {David},
  {de Laverny}, {De Luise}, {De March}, {de Martino}, {de Souza}, {de Torres},
  {Debosscher}, {del Pozo}, {Delbo}, {Delgado}, {Delgado}, {Di Matteo},
  {Diakite}, {Diener}, {Distefano}, {Dolding}, {Drazinos}, {Dur{\'a}n},
  {Edvardsson}, {Enke}, {Eriksson}, {Esquej}, {Eynard Bontemps}, {Fabre},
  {Fabrizio}, {Faigler}, {Falc{\~a}o}, {Farr{\`a}s Casas}, {Federici},
  {Fedorets}, {Fernique}, {Figueras}, {Filippi}, {Findeisen}, {Fonti},
  {Fraile}, {Fraser}, {Fr{\'e}zouls}, {Gai}, {Galleti}, {Garabato},
  {Garc{\'\i}a-Sedano}, {Garofalo}, {Garralda}, {Gavel}, {Gavras}, {Gerssen},
  {Geyer}, {Giacobbe}, {Gilmore}, {Girona}, {Giuffrida}, {Glass}, {Gomes},
  {Granvik}, {Gueguen}, {Guerrier}, {Guiraud}, {Guti{\'e}rrez-S{\'a}nchez},
  {Haigron}, {Hatzidimitriou}, {Hauser}, {Haywood}, {Heiter}, {Helmi}, {Heu},
  {Hilger}, {Hobbs}, {Hofmann}, {Holland}, {Huckle}, {Hypki}, {Icardi},
  {Jan{\ss}en}, {Jevardat de Fombelle}, {Jonker}, {Juh{\'a}sz}, {Julbe},
  {Karampelas}, {Kewley}, {Klar}, {Kochoska}, {Kohley}, {Kolenberg},
  {Kontizas}, {Kontizas}, {Koposov}, {Kordopatis}, {Kostrzewa-Rutkowska},
  {Koubsky}, {Lambert}, {Lanza}, {Lasne}, {Lavigne}, {Le Fustec}, {Le
  Poncin-Lafitte}, {Lebreton}, {Leccia}, {Leclerc}, {Lecoeur-Taibi},
  {Lenhardt}, {Leroux}, {Liao}, {Licata}, {Lindstr{\o}m}, {Lister}, {Livanou},
  {Lobel}, {L{\'o}pez}, {Managau}, {Mann}, {Mantelet}, {Marchal}, {Marchant},
  {Marconi}, {Marinoni}, {Marschalk{\'o}}, {Marshall}, {Martino}, {Marton},
  {Mary}, {Massari}, {Matijevi{\v{c}}}, {Mazeh}, {McMillan}, {Messina},
  {Michalik}, {Millar}, {Molina}, {Molinaro}, {Moln{\'a}r}, {Montegriffo},
  {Mor}, {Morbidelli}, {Morel}, {Morris}, {Mulone}, {Muraveva}, {Musella},
  {Nelemans}, {Nicastro}, {Noval}, {O'Mullane}, {Ord{\'e}novic},
  {Ord{\'o}{\~n}ez-Blanco}, {Osborne}, {Pagani}, {Pagano}, {Pailler},
  {Palacin}, {Palaversa}, {Panahi}, {Pawlak}, {Piersimoni}, {Pineau}, {Plachy},
  {Plum}, {Poggio}, {Poujoulet}, {Pr{\v{s}}a}, {Pulone}, {Racero}, {Ragaini},
  {Rambaux}, {Ramos-Lerate}, {Regibo}, {Reyl{\'e}}, {Riclet}, {Ripepi}, {Riva},
  {Rivard}, {Rixon}, {Roegiers}, {Roelens}, {Romero-G{\'o}mez}, {Rowell},
  {Royer}, {Ruiz-Dern}, {Sadowski}, {Sagrist{\`a} Sell{\'e}s}, {Sahlmann},
  {Salgado}, {Salguero}, {Sanna}, {Santana-Ros}, {Sarasso}, {Savietto},
  {Schultheis}, {Sciacca}, {Segol}, {Segovia}, {S{\'e}gransan}, {Shih},
  {Siltala}, {Silva}, {Smart}, {Smith}, {Solano}, {Solitro}, {Sordo}, {Soria
  Nieto}, {Souchay}, {Spagna}, {Spoto}, {Stampa}, {Steele},
  {Steidelm{\"u}ller}, {Stephenson}, {Stoev}, {Suess}, {Surdej}, {Szabados},
  {Szegedi-Elek}, {Tapiador}, {Taris}, {Tauran}, {Taylor}, {Teixeira},
  {Terrett}, {Teyssandier}, {Thuillot}, {Titarenko}, {Torra Clotet}, {Turon},
  {Ulla}, {Utrilla}, {Uzzi}, {Vaillant}, {Valentini}, {Valette}, {van Elteren},
  {Van Hemelryck}, {van Leeuwen}, {Vaschetto}, {Vecchiato}, {Veljanoski},
  {Viala}, {Vicente}, {Vogt}, {von Essen}, {Voss}, {Votruba}, {Voutsinas},
  {Walmsley}, {Weiler}, {Wertz}, {Wevers}, {Wyrzykowski}, {Yoldas},
  {{\v{Z}}erjal}, {Ziaeepour}, {Zorec}, {Zschocke}, {Zucker}, {Zurbach}, \&
  {Zwitter}}]{gaiacollaboration18}
{Gaia Collaboration}, {Brown}, A.~G.~A., {Vallenari}, A., {et~al.} 2018, A\&A,
  616, A1

\bibitem[{{Gaia Collaboration} {et~al.}(2021){Gaia Collaboration}, {Brown},
  {Vallenari}, {Prusti}, {de Bruijne}, {Babusiaux}, {Biermann}, {Creevey},
  {Evans}, {Eyer}, {Hutton}, {Jansen}, {Jordi}, {Klioner}, {Lammers},
  {Lindegren}, {Luri}, {Mignard}, {Panem}, {Pourbaix}, {Randich}, {Sartoretti},
  {Soubiran}, {Walton}, {Arenou}, {Bailer-Jones}, {Bastian}, {Cropper},
  {Drimmel}, {Katz}, {Lattanzi}, {van Leeuwen}, {Bakker}, {Cacciari},
  {Casta{\~n}eda}, {De Angeli}, {Ducourant}, {Fabricius}, {Fouesneau},
  {Fr{\'e}mat}, {Guerra}, {Guerrier}, {Guiraud}, {Jean-Antoine Piccolo},
  {Masana}, {Messineo}, {Mowlavi}, {Nicolas}, {Nienartowicz}, {Pailler},
  {Panuzzo}, {Riclet}, {Roux}, {Seabroke}, {Sordo}, {Tanga}, {Th{\'e}venin},
  {Gracia-Abril}, {Portell}, {Teyssier}, {Altmann}, {Andrae}, {Bellas-Velidis},
  {Benson}, {Berthier}, {Blomme}, {Brugaletta}, {Burgess}, {Busso}, {Carry},
  {Cellino}, {Cheek}, {Clementini}, {Damerdji}, {Davidson}, {Delchambre},
  {Dell'Oro}, {Fern{\'a}ndez-Hern{\'a}ndez}, {Galluccio}, {Garc{\'\i}a-Lario},
  {Garcia-Reinaldos}, {Gonz{\'a}lez-N{\'u}{\~n}ez}, {Gosset}, {Haigron},
  {Halbwachs}, {Hambly}, {Harrison}, {Hatzidimitriou}, {Heiter},
  {Hern{\'a}ndez}, {Hestroffer}, {Hodgkin}, {Holl}, {Jan{\ss}en}, {Jevardat de
  Fombelle}, {Jordan}, {Krone-Martins}, {Lanzafame}, {L{\"o}ffler}, {Lorca},
  {Manteiga}, {Marchal}, {Marrese}, {Moitinho}, {Mora}, {Muinonen}, {Osborne},
  {Pancino}, {Pauwels}, {Petit}, {Recio-Blanco}, {Richards}, {Riello},
  {Rimoldini}, {Robin}, {Roegiers}, {Rybizki}, {Sarro}, {Siopis}, {Smith},
  {Sozzetti}, {Ulla}, {Utrilla}, {van Leeuwen}, {van Reeven}, {Abbas}, {Abreu
  Aramburu}, {Accart}, {Aerts}, {Aguado}, {Ajaj}, {Altavilla}, {{\'A}lvarez},
  {{\'A}lvarez Cid-Fuentes}, {Alves}, {Anderson}, {Anglada Varela}, {Antoja},
  {Audard}, {Baines}, {Baker}, {Balaguer-N{\'u}{\~n}ez}, {Balbinot}, {Balog},
  {Barache}, {Barbato}, {Barros}, {Barstow}, {Bartolom{\'e}}, {Bassilana},
  {Bauchet}, {Baudesson-Stella}, {Becciani}, {Bellazzini}, {Bernet}, {Bertone},
  {Bianchi}, {Blanco-Cuaresma}, {Boch}, {Bombrun}, {Bossini}, {Bouquillon},
  {Bragaglia}, {Bramante}, {Breedt}, {Bressan}, {Brouillet}, {Bucciarelli},
  {Burlacu}, {Busonero}, {Butkevich}, {Buzzi}, {Caffau}, {Cancelliere},
  {C{\'a}novas}, {Cantat-Gaudin}, {Carballo}, {Carlucci}, {Carnerero},
  {Carrasco}, {Casamiquela}, {Castellani}, {Castro-Ginard}, {Castro Sampol},
  {Chaoul}, {Charlot}, {Chemin}, {Chiavassa}, {Cioni}, {Comoretto}, {Cooper},
  {Cornez}, {Cowell}, {Crifo}, {Crosta}, {Crowley}, {Dafonte}, {Dapergolas},
  {David}, {David}, {de Laverny}, {De Luise}, {De March}, {De Ridder}, {de
  Souza}, {de Teodoro}, {de Torres}, {del Peloso}, {del Pozo}, {Delbo},
  {Delgado}, {Delgado}, {Delisle}, {Di Matteo}, {Diakite}, {Diener},
  {Distefano}, {Dolding}, {Eappachen}, {Edvardsson}, {Enke}, {Esquej}, {Fabre},
  {Fabrizio}, {Faigler}, {Fedorets}, {Fernique}, {Fienga}, {Figueras},
  {Fouron}, {Fragkoudi}, {Fraile}, {Franke}, {Gai}, {Garabato},
  {Garcia-Gutierrez}, {Garc{\'\i}a-Torres}, {Garofalo}, {Gavras}, {Gerlach},
  {Geyer}, {Giacobbe}, {Gilmore}, {Girona}, {Giuffrida}, {Gomel}, {Gomez},
  {Gonzalez-Santamaria}, {Gonz{\'a}lez-Vidal}, {Granvik},
  {Guti{\'e}rrez-S{\'a}nchez}, {Guy}, {Hauser}, {Haywood}, {Helmi}, {Hidalgo},
  {Hilger}, {H{\l}adczuk}, {Hobbs}, {Holland}, {Huckle}, {Jasniewicz},
  {Jonker}, {Juaristi Campillo}, {Julbe}, {Karbevska}, {Kervella}, {Khanna},
  {Kochoska}, {Kontizas}, {Kordopatis}, {Korn}, {Kostrzewa-Rutkowska},
  {Kruszy{\'n}ska}, {Lambert}, {Lanza}, {Lasne}, {Le Campion}, {Le Fustec},
  {Lebreton}, {Lebzelter}, {Leccia}, {Leclerc}, {Lecoeur-Taibi}, {Liao},
  {Licata}, {Lindstr{\o}m}, {Lister}, {Livanou}, {Lobel}, {Madrero Pardo},
  {Managau}, {Mann}, {Marchant}, {Marconi}, {Marcos Santos}, {Marinoni},
  {Marocco}, {Marshall}, {Martin Polo}, {Mart{\'\i}n-Fleitas}, {Masip},
  {Massari}, {Mastrobuono-Battisti}, {Mazeh}, {McMillan}, {Messina},
  {Michalik}, {Millar}, {Mints}, {Molina}, {Molinaro}, {Moln{\'a}r},
  {Montegriffo}, {Mor}, {Morbidelli}, {Morel}, {Morris}, {Mulone}, {Munoz},
  {Muraveva}, {Murphy}, {Musella}, {Noval}, {Ord{\'e}novic}, {Orr{\`u}},
  {Osinde}, {Pagani}, {Pagano}, {Palaversa}, {Palicio}, {Panahi}, {Pawlak},
  {Pe{\~n}alosa Esteller}, {Penttil{\"a}}, {Piersimoni}, {Pineau}, {Plachy},
  {Plum}, {Poggio}, {Poretti}, {Poujoulet}, {Pr{\v{s}}a}, {Pulone}, {Racero},
  {Ragaini}, {Rainer}, {Raiteri}, {Rambaux}, {Ramos}, {Ramos-Lerate}, {Re
  Fiorentin}, {Regibo}, {Reyl{\'e}}, {Ripepi}, {Riva}, {Rixon}, {Robichon},
  {Robin}, {Roelens}, {Rohrbasser}, {Romero-G{\'o}mez}, {Rowell}, {Royer},
  {Rybicki}, {Sadowski}, {Sagrist{\`a} Sell{\'e}s}, {Sahlmann}, {Salgado},
  {Salguero}, {Samaras}, {Sanchez Gimenez}, {Sanna}, {Santove{\~n}a},
  {Sarasso}, {Schultheis}, {Sciacca}, {Segol}, {Segovia}, {S{\'e}gransan},
  {Semeux}, {Shahaf}, {Siddiqui}, {Siebert}, {Siltala}, {Slezak}, {Smart},
  {Solano}, {Solitro}, {Souami}, {Souchay}, {Spagna}, {Spoto}, {Steele},
  {Steidelm{\"u}ller}, {Stephenson}, {S{\"u}veges}, {Szabados}, {Szegedi-Elek},
  {Taris}, {Tauran}, {Taylor}, {Teixeira}, {Thuillot}, {Tonello}, {Torra},
  {Torra}, {Turon}, {Unger}, {Vaillant}, {van Dillen}, {Vanel}, {Vecchiato},
  {Viala}, {Vicente}, {Voutsinas}, {Weiler}, {Wevers}, {Wyrzykowski}, {Yoldas},
  {Yvard}, {Zhao}, {Zorec}, {Zucker}, {Zurbach}, \&
  {Zwitter}}]{gaiacollaboration21}
{Gaia Collaboration}, {Brown}, A.~G.~A., {Vallenari}, A., {et~al.} 2021, \aap,
  649, A1

\bibitem[{{Gaia Collaboration} {et~al.}(2016){Gaia Collaboration}, {Brown},
  {Vallenari}, {Prusti}, {de Bruijne}, {Mignard}, {Drimmel}, {Babusiaux},
  {Bailer-Jones}, {Bastian}, {Biermann}, {Evans}, {Eyer}, {Jansen}, {Jordi},
  {Katz}, {Klioner}, {Lammers}, {Lindegren}, {Luri}, {O'Mullane}, {Panem},
  {Pourbaix}, {Randich}, {Sartoretti}, {Siddiqui}, {Soubiran}, {Valette}, {van
  Leeuwen}, {Walton}, {Aerts}, {Arenou}, {Cropper}, {H{\o}g}, {Lattanzi},
  {Grebel}, {Holland}, {Huc}, {Passot}, {Perryman}, {Bramante}, {Cacciari},
  {Casta{\~n}eda}, {Chaoul}, {Cheek}, {De Angeli}, {Fabricius}, {Guerra},
  {Hern{\'a}ndez}, {Jean-Antoine-Piccolo}, {Masana}, {Messineo}, {Mowlavi},
  {Nienartowicz}, {Ord{\'o}{\~n}ez-Blanco}, {Panuzzo}, {Portell}, {Richards},
  {Riello}, {Seabroke}, {Tanga}, {Th{\'e}venin}, {Torra}, {Els},
  {Gracia-Abril}, {Comoretto}, {Garcia-Reinaldos}, {Lock}, {Mercier},
  {Altmann}, {Andrae}, {Astraatmadja}, {Bellas-Velidis}, {Benson}, {Berthier},
  {Blomme}, {Busso}, {Carry}, {Cellino}, {Clementini}, {Cowell}, {Creevey},
  {Cuypers}, {Davidson}, {De Ridder}, {de Torres}, {Delchambre}, {Dell'Oro},
  {Ducourant}, {Fr{\'e}mat}, {Garc{\'\i}a-Torres}, {Gosset}, {Halbwachs},
  {Hambly}, {Harrison}, {Hauser}, {Hestroffer}, {Hodgkin}, {Huckle}, {Hutton},
  {Jasniewicz}, {Jordan}, {Kontizas}, {Korn}, {Lanzafame}, {Manteiga},
  {Moitinho}, {Muinonen}, {Osinde}, {Pancino}, {Pauwels}, {Petit},
  {Recio-Blanco}, {Robin}, {Sarro}, {Siopis}, {Smith}, {Smith}, {Sozzetti},
  {Thuillot}, {van Reeven}, {Viala}, {Abbas}, {Abreu Aramburu}, {Accart},
  {Aguado}, {Allan}, {Allasia}, {Altavilla}, {{\'A}lvarez}, {Alves},
  {Anderson}, {Andrei}, {Anglada Varela}, {Antiche}, {Antoja}, {Ant{\'o}n},
  {Arcay}, {Bach}, {Baker}, {Balaguer-N{\'u}{\~n}ez}, {Barache}, {Barata},
  {Barbier}, {Barblan}, {Barrado y Navascu{\'e}s}, {Barros}, {Barstow},
  {Becciani}, {Bellazzini}, {Bello Garc{\'\i}a}, {Belokurov}, {Bendjoya},
  {Berihuete}, {Bianchi}, {Bienaym{\'e}}, {Billebaud}, {Blagorodnova},
  {Blanco-Cuaresma}, {Boch}, {Bombrun}, {Borrachero}, {Bouquillon}, {Bourda},
  {Bouy}, {Bragaglia}, {Breddels}, {Brouillet}, {Br{\"u}semeister},
  {Bucciarelli}, {Burgess}, {Burgon}, {Burlacu}, {Busonero}, {Buzzi}, {Caffau},
  {Cambras}, {Campbell}, {Cancelliere}, {Cantat-Gaudin}, {Carlucci},
  {Carrasco}, {Castellani}, {Charlot}, {Charnas}, {Chiavassa}, {Clotet},
  {Cocozza}, {Collins}, {Costigan}, {Crifo}, {Cross}, {Crosta}, {Crowley},
  {Dafonte}, {Damerdji}, {Dapergolas}, {David}, {David}, {De Cat}, {de Felice},
  {de Laverny}, {De Luise}, {De March}, {de Martino}, {de Souza}, {Debosscher},
  {del Pozo}, {Delbo}, {Delgado}, {Delgado}, {Di Matteo}, {Diakite},
  {Distefano}, {Dolding}, {Dos Anjos}, {Drazinos}, {Duran}, {Dzigan},
  {Edvardsson}, {Enke}, {Evans}, {Eynard Bontemps}, {Fabre}, {Fabrizio},
  {Faigler}, {Falc{\~a}o}, {Farr{\`a}s Casas}, {Federici}, {Fedorets},
  {Fern{\'a}ndez-Hern{\'a}ndez}, {Fernique}, {Fienga}, {Figueras}, {Filippi},
  {Findeisen}, {Fonti}, {Fouesneau}, {Fraile}, {Fraser}, {Fuchs}, {Gai},
  {Galleti}, {Galluccio}, {Garabato}, {Garc{\'\i}a-Sedano}, {Garofalo},
  {Garralda}, {Gavras}, {Gerssen}, {Geyer}, {Gilmore}, {Girona}, {Giuffrida},
  {Gomes}, {Gonz{\'a}lez-Marcos}, {Gonz{\'a}lez-N{\'u}{\~n}ez},
  {Gonz{\'a}lez-Vidal}, {Granvik}, {Guerrier}, {Guillout}, {Guiraud},
  {G{\'u}rpide}, {Guti{\'e}rrez-S{\'a}nchez}, {Guy}, {Haigron},
  {Hatzidimitriou}, {Haywood}, {Heiter}, {Helmi}, {Hobbs}, {Hofmann}, {Holl},
  {Holland}, {Hunt}, {Hypki}, {Icardi}, {Irwin}, {Jevardat de Fombelle},
  {Jofr{\'e}}, {Jonker}, {Jorissen}, {Julbe}, {Karampelas}, {Kochoska},
  {Kohley}, {Kolenberg}, {Kontizas}, {Koposov}, {Kordopatis}, {Koubsky},
  {Krone-Martins}, {Kudryashova}, {Kull}, {Bachchan}, {Lacoste-Seris}, {Lanza},
  {Lavigne}, {Le Poncin-Lafitte}, {Lebreton}, {Lebzelter}, {Leccia}, {Leclerc},
  {Lecoeur-Taibi}, {Lemaitre}, {Lenhardt}, {Leroux}, {Liao}, {Licata},
  {Lindstr{\o}m}, {Lister}, {Livanou}, {Lobel}, {L{\"o}ffler}, {L{\'o}pez},
  {Lorenz}, {MacDonald}, {Magalh{\~a}es Fernandes}, {Managau}, {Mann},
  {Mantelet}, {Marchal}, {Marchant}, {Marconi}, {Marinoni}, {Marrese},
  {Marschalk{\'o}}, {Marshall}, {Mart{\'\i}n-Fleitas}, {Martino}, {Mary},
  {Matijevi{\v{c}}}, {Mazeh}, {McMillan}, {Messina}, {Michalik}, {Millar},
  {Miranda}, {Molina}, {Molinaro}, {Molinaro}, {Moln{\'a}r}, {Moniez},
  {Montegriffo}, {Mor}, {Mora}, {Morbidelli}, {Morel}, {Morgenthaler},
  {Morris}, {Mulone}, {Muraveva}, {Musella}, {Narbonne}, {Nelemans},
  {Nicastro}, {Noval}, {Ord{\'e}novic}, {Ordieres-Mer{\'e}}, {Osborne},
  {Pagani}, {Pagano}, {Pailler}, {Palacin}, {Palaversa}, {Parsons}, {Pecoraro},
  {Pedrosa}, {Pentik{\"a}inen}, {Pichon}, {Piersimoni}, {Pineau}, {Plachy},
  {Plum}, {Poujoulet}, {Pr{\v{s}}a}, {Pulone}, {Ragaini}, {Rago}, {Rambaux},
  {Ramos-Lerate}, {Ranalli}, {Rauw}, {Read}, {Regibo}, {Reyl{\'e}}, {Ribeiro},
  {Rimoldini}, {Ripepi}, {Riva}, {Rixon}, {Roelens}, {Romero-G{\'o}mez},
  {Rowell}, {Royer}, {Ruiz-Dern}, {Sadowski}, {Sagrist{\`a} Sell{\'e}s},
  {Sahlmann}, {Salgado}, {Salguero}, {Sarasso}, {Savietto}, {Schultheis},
  {Sciacca}, {Segol}, {Segovia}, {Segransan}, {Shih}, {Smareglia}, {Smart},
  {Solano}, {Solitro}, {Sordo}, {Soria Nieto}, {Souchay}, {Spagna}, {Spoto},
  {Stampa}, {Steele}, {Steidelm{\"u}ller}, {Stephenson}, {Stoev}, {Suess},
  {S{\"u}veges}, {Surdej}, {Szabados}, {Szegedi-Elek}, {Tapiador}, {Taris},
  {Tauran}, {Taylor}, {Teixeira}, {Terrett}, {Tingley}, {Trager}, {Turon},
  {Ulla}, {Utrilla}, {Valentini}, {van Elteren}, {Van Hemelryck}, {van
  Leeuwen}, {Varadi}, {Vecchiato}, {Veljanoski}, {Via}, {Vicente}, {Vogt},
  {Voss}, {Votruba}, {Voutsinas}, {Walmsley}, {Weiler}, {Weingrill}, {Wevers},
  {Wyrzykowski}, {Yoldas}, {{\v{Z}}erjal}, {Zucker}, {Zurbach}, {Zwitter},
  {Alecu}, {Allen}, {Allende Prieto}, {Amorim}, {Anglada-Escud{\'e}},
  {Arsenijevic}, {Azaz}, {Balm}, {Beck}, {Bernstein}, {Bigot}, {Bijaoui},
  {Blasco}, {Bonfigli}, {Bono}, {Boudreault}, {Bressan}, {Brown}, {Brunet},
  {Bunclark}, {Buonanno}, {Butkevich}, {Carret}, {Carrion}, {Chemin},
  {Ch{\'e}reau}, {Corcione}, {Darmigny}, {de Boer}, {de Teodoro}, {de Zeeuw},
  {Delle Luche}, {Domingues}, {Dubath}, {Fodor}, {Fr{\'e}zouls}, {Fries},
  {Fustes}, {Fyfe}, {Gallardo}, {Gallegos}, {Gardiol}, {Gebran}, {Gomboc},
  {G{\'o}mez}, {Grux}, {Gueguen}, {Heyrovsky}, {Hoar}, {Iannicola}, {Isasi
  Parache}, {Janotto}, {Joliet}, {Jonckheere}, {Keil}, {Kim}, {Klagyivik},
  {Klar}, {Knude}, {Kochukhov}, {Kolka}, {Kos}, {Kutka}, {Lainey}, {LeBouquin},
  {Liu}, {Loreggia}, {Makarov}, {Marseille}, {Martayan}, {Martinez-Rubi},
  {Massart}, {Meynadier}, {Mignot}, {Munari}, {Nguyen}, {Nordlander}, {Ocvirk},
  {O'Flaherty}, {Olias Sanz}, {Ortiz}, {Osorio}, {Oszkiewicz}, {Ouzounis},
  {Palmer}, {Park}, {Pasquato}, {Peltzer}, {Peralta}, {P{\'e}turaud},
  {Pieniluoma}, {Pigozzi}, {Poels}, {Prat}, {Prod'homme}, {Raison}, {Rebordao},
  {Risquez}, {Rocca-Volmerange}, {Rosen}, {Ruiz-Fuertes}, {Russo}, {Sembay},
  {Serraller Vizcaino}, {Short}, {Siebert}, {Silva}, {Sinachopoulos}, {Slezak},
  {Soffel}, {Sosnowska}, {Strai{\v{z}}ys}, {ter Linden}, {Terrell}, {Theil},
  {Tiede}, {Troisi}, {Tsalmantza}, {Tur}, {Vaccari}, {Vachier}, {Valles}, {Van
  Hamme}, {Veltz}, {Virtanen}, {Wallut}, {Wichmann}, {Wilkinson}, {Ziaeepour},
  \& {Zschocke}}]{gaiacollaboration16}
{Gaia Collaboration}, {Brown}, A.~G.~A., {Vallenari}, A., {et~al.} 2016, \aap,
  595, A2

\bibitem[{{Gaia Collaboration} {et~al.}(2022){Gaia Collaboration}, {Vallenari,
  A.}, {Brown, A.G.A.}, {Prusti, T.}, \& {et al.}}]{gaiacollaboration22}
{Gaia Collaboration}, {Vallenari, A.}, {Brown, A.G.A.}, {Prusti, T.}, \& {et
  al.} 2022, A\&A

\bibitem[{{Garnavich}(1993)}]{garnavich93}
{Garnavich}, P.~M. 1993, \pasp, 105, 321

\bibitem[{{G{\'a}sp{\'a}r} {et~al.}(2013){G{\'a}sp{\'a}r}, {Rieke}, \&
  {Balog}}]{gaspar13}
{G{\'a}sp{\'a}r}, A., {Rieke}, G.~H., \& {Balog}, Z. 2013, \apj, 768, 25

\bibitem[{{Gentile Fusillo} {et~al.}(2019){Gentile Fusillo}, {Tremblay},
  {G{\"a}nsicke}, {Manser}, {Cunningham}, {Cukanovaite}, {Hollands}, {Marsh},
  {Raddi}, {Jordan}, {Toonen}, {Geier}, {Barstow}, \&
  {Cummings}}]{gentilefusillo19}
{Gentile Fusillo}, N.~P., {Tremblay}, P.-E., {G{\"a}nsicke}, B.~T., {et~al.}
  2019, \mnras, 482, 4570

\bibitem[{{Gianninas} {et~al.}(2011){Gianninas}, {Bergeron}, \&
  {Ruiz}}]{gianninas11}
{Gianninas}, A., {Bergeron}, P., \& {Ruiz}, M.~T. 2011, \apj, 743, 138

\bibitem[{{Giclas} {et~al.}(1959){Giclas}, {Slaughter}, \&
  {Burnham}}]{giclas59}
{Giclas}, H.~L., {Slaughter}, C.~D., \& {Burnham}, R. 1959, Lowell Observatory
  Bulletin, 4, 136

\bibitem[{{Goldman} {et~al.}(2018){Goldman}, {R{\"o}ser}, {Schilbach},
  {Mo{\'o}r}, \& {Henning}}]{goldman18}
{Goldman}, B., {R{\"o}ser}, S., {Schilbach}, E., {Mo{\'o}r}, A.~C., \&
  {Henning}, T. 2018, \apj, 868, 32

\bibitem[{Gontcharov \& Kiyaeva(2010)}]{gontcharov10}
Gontcharov, G.~A. \& Kiyaeva, O.~V. 2010, New Astron., 15, 324

\bibitem[{{Gonz{\'a}lez-Payo} {et~al.}(2021){Gonz{\'a}lez-Payo},
  {Cort{\'e}s-Contreras}, {Lodieu}, {Solano}, {Zhang}, \&
  {G{\'a}lvez-Ortiz}}]{gonzalezpayo21a}
{Gonz{\'a}lez-Payo}, J., {Cort{\'e}s-Contreras}, M., {Lodieu}, N., {et~al.}
  2021, \aap, 650, A190

\bibitem[{{Gorynya} \& {Tokovinin}(2018)}]{gorynya18}
{Gorynya}, N.~A. \& {Tokovinin}, A. 2018, \mnras, 475, 1375

\bibitem[{{Gossage} {et~al.}(2018){Gossage}, {Conroy}, {Dotter}, {Choi},
  {Rosenfield}, {Cargile}, \& {Dolphin}}]{gossage18}
{Gossage}, S., {Conroy}, C., {Dotter}, A., {et~al.} 2018, \apj, 863, 67

\bibitem[{{Guenther} {et~al.}(2005){Guenther}, {Paulson}, {Cochran},
  {Patience}, {Hatzes}, \& {Macintosh}}]{guenther05}
{Guenther}, E.~W., {Paulson}, D.~B., {Cochran}, W.~D., {et~al.} 2005, \aap,
  442, 1031

\bibitem[{{Heggie}(1975)}]{heggie75}
{Heggie}, D.~C. 1975, \mnras, 173, 729

\bibitem[{{Henrichs} {et~al.}(1983){Henrichs}, {Hammerschlag-Hensberge},
  {Howarth}, \& {Barr}}]{henrichs83}
{Henrichs}, H.~F., {Hammerschlag-Hensberge}, G., {Howarth}, I.~D., \& {Barr},
  P. 1983, \apj, 268, 807

\bibitem[{{Herschel}(1802)}]{herschel1802}
{Herschel}, W. 1802, Philos. T. R. Soc. Lond., 92, 213

\bibitem[{{Hirshfeld}(2001)}]{hirshfeld01}
{Hirshfeld}, A.~W. 2001, {Parallax: The Race to Measure the Cosmos}

\bibitem[{{Hoogerwerf}(2000)}]{hoogerwerf00}
{Hoogerwerf}, R. 2000, \mnras, 313, 43

\bibitem[{{Hubble}(1922)}]{hubble22}
{Hubble}, E.~P. 1922, \apj, 56, 162

\bibitem[{{Hutter} {et~al.}(2021){Hutter}, {Tycner}, {Zavala}, {Benson},
  {Hummel}, \& {Zirm}}]{hutter21}
{Hutter}, D.~J., {Tycner}, C., {Zavala}, R.~T., {et~al.} 2021, \apjs, 257, 69

\bibitem[{{Innes}(1915)}]{innes15}
{Innes}, R.~T.~A. 1915, Circular of the Union Observatory Johannesburg, 30, 235

\bibitem[{{Jansen} {et~al.}(1994){Jansen}, {van Dishoeck}, \&
  {Black}}]{jansen94}
{Jansen}, D.~J., {van Dishoeck}, E.~F., \& {Black}, J.~H. 1994, \aap, 282, 605

\bibitem[{{Jensen} {et~al.}(1993){Jensen}, {Mathieu}, \& {Fuller}}]{jensen93}
{Jensen}, E.~L., {Mathieu}, R.~D., \& {Fuller}, G.~A. 1993, in American
  Astronomical Society Meeting Abstracts, Vol. 182, American Astronomical
  Society Meeting Abstracts \#182, 62.21

\bibitem[{{Jiang} \& {Tremaine}(2010)}]{jiang10}
{Jiang}, Y.-F. \& {Tremaine}, S. 2010, \mnras, 401, 977

\bibitem[{{Kalas} {et~al.}(2004){Kalas}, {Liu}, \& {Matthews}}]{kalas04}
{Kalas}, P., {Liu}, M.~C., \& {Matthews}, B.~C. 2004, Science, 303, 1990

\bibitem[{{Karr} {et~al.}(2005){Karr}, {Noriega-Crespo}, \& {Martin}}]{karr05}
{Karr}, J.~L., {Noriega-Crespo}, A., \& {Martin}, P.~G. 2005, \aj, 129, 954

\bibitem[{{Katz} {et~al.}(2022){Katz}, {Sartoretti}, {Guerrier}, {Panuzzo},
  {Seabroke}, {Th{\'e}venin}, {Cropper}, {Benson}, {Blomme}, {Haigron},
  {Marchal}, {Smith}, {Baker}, {Chemin}, {Damerdji}, {David}, {Dolding},
  {Fr{\'e}mat}, {Gosset}, {Jan{\ss}en}, {Jasniewicz}, {Lobel}, {Plum},
  {Samaras}, {Snaith}, {Soubiran}, {Vanel}, {Zwitter}, {Antoja}, {Arenou},
  {Babusiaux}, {Brouillet}, {Caffau}, {Di Matteo}, {Fabre}, {Fabricius},
  {Frakgoudi}, {Haywood}, {Huckle}, {Hottier}, {Lasne}, {Leclerc},
  {Mastrobuono-Battisti}, {Royer}, {Teyssier}, {Zorec}, {Crifo}, {Jean-Antoine
  Piccolo}, {Turon}, \& {Viala}}]{katz22}
{Katz}, D., {Sartoretti}, P., {Guerrier}, A., {et~al.} 2022, arXiv e-prints,
  arXiv:2206.05902

\bibitem[{{Kervella} {et~al.}(2019){Kervella}, {Arenou}, {Mignard}, \&
  {Th{\'e}venin}}]{kervella19}
{Kervella}, P., {Arenou}, F., {Mignard}, F., \& {Th{\'e}venin}, F. 2019, A\&A,
  623, A72

\bibitem[{{Kervella} {et~al.}(2022){Kervella}, {Arenou}, \&
  {Th{\'e}venin}}]{kervella22}
{Kervella}, P., {Arenou}, F., \& {Th{\'e}venin}, F. 2022, \aap, 657, A7

\bibitem[{{Kervella} {et~al.}(2013){Kervella}, {M{\'e}rand}, {Petr-Gotzens},
  {Pribulla}, \& {Th{\'e}venin}}]{kervella13}
{Kervella}, P., {M{\'e}rand}, A., {Petr-Gotzens}, M.~G., {Pribulla}, T., \&
  {Th{\'e}venin}, F. 2013, \aap, 552, A18

\bibitem[{{Kirkpatrick} {et~al.}(2016){Kirkpatrick}, {Kellogg}, {Schneider},
  {Fajardo-Acosta}, {Cushing}, {Greco}, {Mace}, {Gelino}, {Wright},
  {Eisenhardt}, {Stern}, {Faherty}, {Sheppard}, {Lansbury}, {Logsdon},
  {Martin}, {McLean}, {Schurr}, {Cutri}, \& {Conrow}}]{kirkpatrick16}
{Kirkpatrick}, J.~D., {Kellogg}, K., {Schneider}, A.~C., {et~al.} 2016, \apjs,
  224, 36

\bibitem[{{Konacki} {et~al.}(2010){Konacki}, {Muterspaugh}, {Kulkarni}, \&
  {He{\l}miniak}}]{konacki10}
{Konacki}, M., {Muterspaugh}, M.~W., {Kulkarni}, S.~R., \& {He{\l}miniak},
  K.~G. 2010, \apj, 719, 1293

\bibitem[{{Kopytova} {et~al.}(2016){Kopytova}, {Brandner}, {Tognelli}, {Prada
  Moroni}, {Da Rio}, {R{\"o}ser}, \& {Schilbach}}]{kopytova16}
{Kopytova}, T.~G., {Brandner}, W., {Tognelli}, E., {et~al.} 2016, \aap, 585, A7

\bibitem[{{Kouwenhoven} {et~al.}(2010){Kouwenhoven}, {Goodwin}, {Parker},
  {Davies}, {Malmberg}, \& {Kroupa}}]{kouwenhoven10}
{Kouwenhoven}, M.~B.~N., {Goodwin}, S.~P., {Parker}, R.~J., {et~al.} 2010,
  \mnras, 404, 1835

\bibitem[{{Kraicheva} {et~al.}(1985){Kraicheva}, {Popova}, {Tutukov}, \&
  {Iungelson}}]{kraicheva85}
{Kraicheva}, Z.~T., {Popova}, E.~I., {Tutukov}, A.~V., \& {Iungelson}, L.~R.
  1985, Astrofizika, 22, 105

\bibitem[{{Kraus} \& {Hillenbrand}(2009)}]{kraus09}
{Kraus}, A.~L. \& {Hillenbrand}, L.~A. 2009, \apj, 703, 1511

\bibitem[{{Kraus} {et~al.}(2014){Kraus}, {Shkolnik}, {Allers}, \&
  {Liu}}]{kraus14}
{Kraus}, A.~L., {Shkolnik}, E.~L., {Allers}, K.~N., \& {Liu}, M.~C. 2014, \aj,
  147, 146

\bibitem[{{Kroupa}(2001)}]{kroupa01}
{Kroupa}, P. 2001, \mnras, 322, 231

\bibitem[{{Lajoie} \& {Bergeron}(2007)}]{lajoie07}
{Lajoie}, C.~P. \& {Bergeron}, P. 2007, \apj, 667, 1126

\bibitem[{{Latham} {et~al.}(1991){Latham}, {Mazeh}, {Davis}, {Stefanik}, \&
  {Abt}}]{latham91}
{Latham}, D.~W., {Mazeh}, T., {Davis}, R.~J., {Stefanik}, R.~P., \& {Abt},
  H.~A. 1991, \aj, 101, 625

\bibitem[{{Latham} {et~al.}(2002){Latham}, {Stefanik}, {Torres}, {Davis},
  {Mazeh}, {Carney}, {Laird}, \& {Morse}}]{latham02}
{Latham}, D.~W., {Stefanik}, R.~P., {Torres}, G., {et~al.} 2002, AJ, 124, 1144

\bibitem[{{Lee} {et~al.}(2017){Lee}, {Lee}, {Dunham}, {Tatematsu}, {Choi},
  {Bergin}, \& {Evans}}]{lee17}
{Lee}, J.-E., {Lee}, S., {Dunham}, M.~M., {et~al.} 2017, Nature Astronomy, 1,
  0172

\bibitem[{{L{\'e}pine} \& {Bongiorno}(2007)}]{lepine07a}
{L{\'e}pine}, S. \& {Bongiorno}, B. 2007, \aj, 133, 889

\bibitem[{{L{\'e}pine} \& {Gaidos}(2011)}]{lepine11}
{L{\'e}pine}, S. \& {Gaidos}, E. 2011, \aj, 142, 138

\bibitem[{{L{\'e}pine} \& {Shara}(2005)}]{lepine05d}
{L{\'e}pine}, S. \& {Shara}, M.~M. 2005, AJ, 129, 1483

\bibitem[{{Lindegren} {et~al.}(2021){Lindegren}, {Bastian}, {Biermann},
  {Bombrun}, {de Torres}, {Gerlach}, {Geyer}, {Hern{\'a}ndez}, {Hilger},
  {Hobbs}, {Klioner}, {Lammers}, {McMillan}, {Ramos-Lerate},
  {Steidelm{\"u}ller}, {Stephenson}, \& {van Leeuwen}}]{lindegren21}
{Lindegren}, L., {Bastian}, U., {Biermann}, M., {et~al.} 2021, \aap, 649, A4

\bibitem[{{Lindegren} {et~al.}(2018){Lindegren}, {Hern{\'a}ndez}, {Bombrun},
  {Klioner}, {Bastian}, {Ramos-Lerate}, {de Torres}, {Steidelm{\"u}ller},
  {Stephenson}, {Hobbs}, \& {80 co-authors}}]{lindegren18a}
{Lindegren}, L., {Hern{\'a}ndez}, J., {Bombrun}, A., {et~al.} 2018, A\&A, 616,
  A2

\bibitem[{{Malkov} {et~al.}(2012){Malkov}, {Tamazian}, {Docobo}, \&
  {Chulkov}}]{malkov12}
{Malkov}, O.~Y., {Tamazian}, V.~S., {Docobo}, J.~A., \& {Chulkov}, D.~A. 2012,
  \aap, 546, A69

\bibitem[{{Mamajek}(2017)}]{mamajek17}
{Mamajek}, E. 2017, JDSO, 13, 264

\bibitem[{{Mann} {et~al.}(2019){Mann}, {Dupuy}, {Kraus}, {Gaidos}, {Ansdell},
  {Ireland}, {Rizzuto}, {Hung}, {Dittmann}, {Factor}, {Feiden}, {Martinez},
  {Ru{\'\i}z-Rodr{\'\i}guez}, \& {Thao}}]{mann19}
{Mann}, A.~W., {Dupuy}, T., {Kraus}, A.~L., {et~al.} 2019, \apj, 871, 63

\bibitem[{{Mart{\'\i}n} {et~al.}(2018){Mart{\'\i}n}, {Lodieu}, {Pavlenko}, \&
  {B{\'e}jar}}]{martin18}
{Mart{\'\i}n}, E.~L., {Lodieu}, N., {Pavlenko}, Y., \& {B{\'e}jar}, V. J.~S.
  2018, \apj, 856, 40

\bibitem[{{Mason} {et~al.}(2001){Mason}, {Wycoff}, {Hartkopf}, {Douglass}, \&
  {Worley}}]{mason01}
{Mason}, B.~D., {Wycoff}, G.~L., {Hartkopf}, W.~I., {Douglass}, G.~G., \&
  {Worley}, C.~E. 2001, AJ, 122, 3466

\bibitem[{{Maxted} {et~al.}(2000){Maxted}, {Marsh}, \& {Moran}}]{maxted00}
{Maxted}, P.~F.~L., {Marsh}, T.~R., \& {Moran}, C.~K.~J. 2000, \mnras, 319, 305

\bibitem[{{Michalik} {et~al.}(2015){Michalik}, {Lindegren}, \&
  {Hobbs}}]{michalik15}
{Michalik}, D., {Lindegren}, L., \& {Hobbs}, D. 2015, \aap, 574, A115

\bibitem[{{Mitrofanova} {et~al.}(2021){Mitrofanova}, {Dyachenko}, {Beskakotov},
  {Balega}, {Maksimov}, \& {Rastegaev}}]{mitrofanova21}
{Mitrofanova}, A., {Dyachenko}, V., {Beskakotov}, A., {et~al.} 2021, \aj, 162,
  156

\bibitem[{{Mittal} {et~al.}(2015){Mittal}, {Chen}, {Jang-Condell}, {Manoj},
  {Sargent}, {Watson}, \& {Lisse}}]{mittal15}
{Mittal}, T., {Chen}, C.~H., {Jang-Condell}, H., {et~al.} 2015, \apj, 798, 87

\bibitem[{{Mizusawa} {et~al.}(2012){Mizusawa}, {Rebull}, {Stauffer}, {Bryden},
  {Meyer}, \& {Song}}]{mizusawa12}
{Mizusawa}, T.~F., {Rebull}, L.~M., {Stauffer}, J.~R., {et~al.} 2012, \aj, 144,
  135

\bibitem[{{Montes} {et~al.}(2018){Montes}, {Gonz{\'a}lez-Peinado}, {Tabernero},
  {Caballero}, {Marfil}, {Alonso-Floriano}, {Cort{\'e}s-Contreras},
  {Gonz{\'a}lez Hern{\'a}ndez}, {Klutsch}, \& {Moreno-J{\'o}dar}}]{montes18a}
{Montes}, D., {Gonz{\'a}lez-Peinado}, R., {Tabernero}, H.~M., {et~al.} 2018,
  \mnras, 479, 1332

\bibitem[{{Montes} {et~al.}(2001){Montes}, {L{\'o}pez-Santiago}, {G{\'a}lvez},
  {Fern{\'a}ndez-Figueroa}, {De Castro}, \& {Cornide}}]{montes01a}
{Montes}, D., {L{\'o}pez-Santiago}, J., {G{\'a}lvez}, M.~C., {et~al.} 2001,
  MNRAS, 328, 45

\bibitem[{{Morgan} {et~al.}(1943){Morgan}, {Keenan}, \& {Kellman}}]{morgan43}
{Morgan}, W.~W., {Keenan}, P.~C., \& {Kellman}, E. 1943, {An atlas of stellar
  spectra, with an outline of spectral classification} (Chicago, Ill., The
  University of Chicago press)

\bibitem[{{Murphy} {et~al.}(2013){Murphy}, {Lawson}, \& {Bessell}}]{murphy13}
{Murphy}, S.~J., {Lawson}, W.~A., \& {Bessell}, M.~S. 2013, \mnras, 435, 1325

\bibitem[{{Naz{\'e}} {et~al.}(2022){Naz{\'e}}, {Rauw}, {Czesla}, {Smith}, \&
  {Robrade}}]{naze22}
{Naz{\'e}}, Y., {Rauw}, G., {Czesla}, S., {Smith}, M.~A., \& {Robrade}, J.
  2022, \mnras, 510, 2286

\bibitem[{{Nemravov{\'a}} {et~al.}(2012){Nemravov{\'a}}, {Harmanec},
  {Koubsk{\'y}}, {Miroshnichenko}, {Yang}, {{\v{S}}lechta}, {Buil},
  {Kor{\v{c}}{\'a}kov{\'a}}, \& {Votruba}}]{nemravova12}
{Nemravov{\'a}}, J., {Harmanec}, P., {Koubsk{\'y}}, P., {et~al.} 2012, \aap,
  537, A59

\bibitem[{{Nesci} {et~al.}(2018){Nesci}, {Tuvikene}, {Rossi}, {Gaudenzi},
  {Galleti}, {Ochner}, \& {Enke}}]{nesci18}
{Nesci}, R., {Tuvikene}, T., {Rossi}, C., {et~al.} 2018, \rmxaa, 54, 341

\bibitem[{{Niemela}(2001)}]{niemela01}
{Niemela}, V. 2001, in Rev. Mex. Astron. Astrofis. Conf. Ser., Vol.~11, Rev.
  Mex. Astron. Astrofis. Conf. Ser., 23--26

\bibitem[{{Ochsenbein} {et~al.}(2000){Ochsenbein}, {Bauer}, \&
  {Marcout}}]{ochsenbein00}
{Ochsenbein}, F., {Bauer}, P., \& {Marcout}, J. 2000, A\&AS, 143, 23

\bibitem[{{Oelkers} {et~al.}(2017){Oelkers}, {Stassun}, \&
  {Dhital}}]{oelkers17}
{Oelkers}, R.~J., {Stassun}, K.~G., \& {Dhital}, S. 2017, \aj, 153, 259

\bibitem[{{{\"O}pik}(1924)}]{opik24}
{{\"O}pik}, E. 1924, Tartu Obs. Publ., 25

\bibitem[{{Osterbrock}(1957)}]{osterbrock57}
{Osterbrock}, D.~E. 1957, \apj, 125, 622

\bibitem[{{Patience} {et~al.}(2002){Patience}, {Ghez}, {Reid}, \&
  {Matthews}}]{patience02}
{Patience}, J., {Ghez}, A.~M., {Reid}, I.~N., \& {Matthews}, K. 2002, \aj, 123,
  1570

\bibitem[{{Paunzen} {et~al.}(2012){Paunzen}, {Heiter}, {Fraga}, \&
  {Pintado}}]{paunzen12}
{Paunzen}, E., {Heiter}, U., {Fraga}, L., \& {Pintado}, O. 2012, \mnras, 419,
  3604

\bibitem[{{Pe{\~n}a Ram{\'\i}rez} {et~al.}(2012){Pe{\~n}a Ram{\'\i}rez},
  {B{\'e}jar}, {Zapatero Osorio}, {Petr-Gotzens}, \&
  {Mart{\'\i}n}}]{penaramirez12}
{Pe{\~n}a Ram{\'\i}rez}, K., {B{\'e}jar}, V.~J.~S., {Zapatero Osorio}, M.~R.,
  {Petr-Gotzens}, M.~G., \& {Mart{\'\i}n}, E.~L. 2012, \apj, 754, 30

\bibitem[{{Pecaut} \& {Mamajek}(2013)}]{pecaut13}
{Pecaut}, M.~J. \& {Mamajek}, E.~E. 2013, ApJS, 208, 9

\bibitem[{{Pecaut} \& {Mamajek}(2016)}]{pecaut16}
{Pecaut}, M.~J. \& {Mamajek}, E.~E. 2016, \mnras, 461, 794

\bibitem[{{Perryman} {et~al.}(1998){Perryman}, {Brown}, {Lebreton}, {Gomez},
  {Turon}, {Cayrel de Strobel}, {Mermilliod}, {Robichon}, {Kovalevsky}, \&
  {Crifo}}]{perryman98}
{Perryman}, M.~A.~C., {Brown}, A.~G.~A., {Lebreton}, Y., {et~al.} 1998, \aap,
  331, 81

\bibitem[{{Perryman} {et~al.}(1997){Perryman}, {Lindegren}, {Kovalevsky},
  {Hog}, {Bastian}, {Bernacca}, {Creze}, {Donati}, {Grenon}, {Grewing}, {van
  Leeuwen}, {van der Marel}, {Mignard}, {Murray}, {Le Poole}, {Schrijver},
  {Turon}, {Arenou}, {Froeschle}, \& {Petersen}}]{perryman97}
{Perryman}, M.~A.~C., {Lindegren}, L., {Kovalevsky}, J., {et~al.} 1997, \aap,
  500, 501

\bibitem[{{Plavchan} {et~al.}(2020){Plavchan}, {Barclay}, {Gagn{\'e}}, {Gao},
  {Cale}, {Matzko}, {Dragomir}, {Quinn}, {Feliz}, {Stassun}, {Crossfield},
  {Berardo}, {Latham}, {Tieu}, {Anglada-Escud{\'e}}, {Ricker}, {Vanderspek},
  {Seager}, {Winn}, {Jenkins}, {Rinehart}, {Krishnamurthy}, {Dynes}, {Doty},
  {Adams}, {Afanasev}, {Beichman}, {Bottom}, {Bowler}, {Brinkworth}, {Brown},
  {Cancino}, {Ciardi}, {Clampin}, {Clark}, {Collins}, {Davison},
  {Foreman-Mackey}, {Furlan}, {Gaidos}, {Geneser}, {Giddens}, {Gilbert},
  {Hall}, {Hellier}, {Henry}, {Horner}, {Howard}, {Huang}, {Huber}, {Kane},
  {Kenworthy}, {Kielkopf}, {Kipping}, {Klenke}, {Kruse}, {Latouf}, {Lowrance},
  {Mennesson}, {Mengel}, {Mills}, {Morton}, {Narita}, {Newton}, {Nishimoto},
  {Okumura}, {Palle}, {Pepper}, {Quintana}, {Roberge}, {Roccatagliata},
  {Schlieder}, {Tanner}, {Teske}, {Tinney}, {Vanderburg}, {von Braun}, {Walp},
  {Wang}, {Wang}, {Weigand}, {White}, {Wittenmyer}, {Wright}, {Youngblood},
  {Zhang}, \& {Zilberman}}]{plavchan20}
{Plavchan}, P., {Barclay}, T., {Gagn{\'e}}, J., {et~al.} 2020, \nat, 582, 497

\bibitem[{{Poeckert} \& {Marlborough}(1978)}]{poeckert78}
{Poeckert}, R. \& {Marlborough}, J.~M. 1978, \apj, 220, 940

\bibitem[{{Poleski} {et~al.}(2012){Poleski}, {Soszy{\'n}ski}, {Udalski},
  {Szyma{\'n}ski}, {Kubiak}, {Pietrzy{\'n}ski}, {Wyrzykowski}, \&
  {Ulaczyk}}]{poleski12}
{Poleski}, R., {Soszy{\'n}ski}, I., {Udalski}, A., {et~al.} 2012, \actaa, 62, 1

\bibitem[{{Pourbaix} \& {Boffin}(2016)}]{pourbaix16}
{Pourbaix}, D. \& {Boffin}, H. M.~J. 2016, \aap, 586, A90

\bibitem[{{Pourbaix} {et~al.}(2004){Pourbaix}, {Tokovinin}, {Batten}, {Fekel},
  {Hartkopf}, {Levato}, {Morrell}, {Torres}, \& {Udry}}]{pourbaix04}
{Pourbaix}, D., {Tokovinin}, A.~A., {Batten}, A.~H., {et~al.} 2004, \aap, 424,
  727

\bibitem[{{Probst}(1983)}]{probst83}
{Probst}, R.~G. 1983, \apjs, 53, 335

\bibitem[{{Radigan} {et~al.}(2009){Radigan}, {Lafreni{\`e}re}, {Jayawardhana},
  \& {Doyon}}]{radigan09}
{Radigan}, J., {Lafreni{\`e}re}, D., {Jayawardhana}, R., \& {Doyon}, R. 2009,
  \apj, 698, 405

\bibitem[{{Raghavan} {et~al.}(2010){Raghavan}, {McAlister}, {Henry}, {Latham},
  {Marcy}, {Mason}, {Gies}, {White}, \& {ten Brummelaar}}]{raghavan10}
{Raghavan}, D., {McAlister}, H.~A., {Henry}, T.~J., {et~al.} 2010, ApJS, 190, 1

\bibitem[{{Ramsay} {et~al.}(2022){Ramsay}, {Hakala}, {Doyle}, {Doyle}, \&
  {Bagnulo}}]{ramsay22}
{Ramsay}, G., {Hakala}, P., {Doyle}, J.~G., {Doyle}, L., \& {Bagnulo}, S. 2022,
  \mnras, 511, 2755

\bibitem[{{Rebassa-Mansergas} {et~al.}(2021){Rebassa-Mansergas}, {Maldonado},
  {Raddi}, {Knowles}, {Torres}, {Hoskin}, {Cunningham}, {Hollands}, {Ren},
  {G{\"a}nsicke}, {Tremblay}, {Castro-Rodr{\'\i}guez}, {Camisassa}, \&
  {Koester}}]{rebassamansergas21}
{Rebassa-Mansergas}, A., {Maldonado}, J., {Raddi}, R., {et~al.} 2021, \mnras,
  505, 3165

\bibitem[{{Reid}(1993)}]{reid93}
{Reid}, N. 1993, \mnras, 265, 785

\bibitem[{{Reino} {et~al.}(2018){Reino}, {de Bruijne}, {Zari}, {d'Antona}, \&
  {Ventura}}]{reino18}
{Reino}, S., {de Bruijne}, J., {Zari}, E., {d'Antona}, F., \& {Ventura}, P.
  2018, \mnras, 477, 3197

\bibitem[{{Reipurth} \& {Mikkola}(2012)}]{reipurth12}
{Reipurth}, B. \& {Mikkola}, S. 2012, \nat, 492, 221

\bibitem[{{Retterer} \& {King}(1982)}]{retterer82}
{Retterer}, J.~M. \& {King}, I.~R. 1982, \apj, 254, 214

\bibitem[{{Reyl{\'e}} {et~al.}(2021){Reyl{\'e}}, {Jardine}, {Fouqu{\'e}},
  {Caballero}, {Smart}, \& {Sozzetti}}]{reyle21}
{Reyl{\'e}}, C., {Jardine}, K., {Fouqu{\'e}}, P., {et~al.} 2021, \aap, 650,
  A201

\bibitem[{{Rica} \& {Caballero}(2012)}]{rica12}
{Rica}, F.~M. \& {Caballero}, J.~A. 2012, The Observatory, 132, 305

\bibitem[{{Riedel} {et~al.}(2017){Riedel}, {Blunt}, {Lambrides}, {Rice},
  {Cruz}, \& {Faherty}}]{riedel17}
{Riedel}, A.~R., {Blunt}, S.~C., {Lambrides}, E.~L., {et~al.} 2017, \aj, 153,
  95

\bibitem[{{Riello} {et~al.}(2018){Riello}, {De Angeli}, {Evans}, {Busso},
  {Hambly}, {Davidson}, {Burgess}, {Montegriffo}, {Osborne}, {Kewley},
  {Carrasco}, {Fabricius}, {Jordi}, {Cacciari}, {van Leeuwen}, \&
  {Holland}}]{riello18}
{Riello}, M., {De Angeli}, F., {Evans}, D.~W., {et~al.} 2018, \aap, 616, A3

\bibitem[{{Rizzuto} {et~al.}(2011){Rizzuto}, {Ireland}, \&
  {Robertson}}]{rizzuto11}
{Rizzuto}, A.~C., {Ireland}, M.~J., \& {Robertson}, J.~G. 2011, \mnras, 416,
  3108

\bibitem[{{R{\"o}ser} {et~al.}(2011){R{\"o}ser}, {Schilbach}, {Piskunov},
  {Kharchenko}, \& {Scholz}}]{roser11}
{R{\"o}ser}, S., {Schilbach}, E., {Piskunov}, A.~E., {Kharchenko}, N.~V., \&
  {Scholz}, R.~D. 2011, \aap, 531, A92

\bibitem[{{Sarro} {et~al.}(2022){Sarro}, {Berihuete}, {Smart}, {Reyl{\'e}},
  {Barrado}, {Garc{\'\i}a-Torres}, {Cooper}, {Jones}, {Marocco}, {Creevey},
  {Sordo}, {Bailer-Jones}, {Montegriffo}, {Carballo}, {Andrae}, {Fouesneau},
  {Lanzafame}, {Pailler}, {Th{\'e}venin}, {Lobel}, {Delchambre}, {Korn},
  {Recio-Blanco}, {Schultheis}, {De Angeli}, {Brouillet}, {Casamiquela},
  {Contursi}, {de Laverny}, {Garc{\'\i}a-Lario}, {Kordopatis}, {Lebreton},
  {Livanou}, {Lorca}, {Palicio}, {Slezak-Oreshina}, {Soubiran}, {Ulla}, \&
  {Zhao}}]{sarro22}
{Sarro}, L.~M., {Berihuete}, A., {Smart}, R.~L., {et~al.} 2022, arXiv e-prints,
  arXiv:2211.03641

\bibitem[{{Sch{\"o}nfeld}(1886)}]{schonfeld1886}
{Sch{\"o}nfeld}, E. 1886, Eds Marcus and Weber's Verlag, 0

\bibitem[{{Schweitzer} {et~al.}(1999){Schweitzer}, {Scholz}, {Stauffer},
  {Irwin}, \& {McCaughrean}}]{schweitzer99}
{Schweitzer}, A., {Scholz}, R.-D., {Stauffer}, J., {Irwin}, M., \&
  {McCaughrean}, M.~J. 1999, A\&A, 350, L62

\bibitem[{{Sharpless}(1959)}]{sharpless59}
{Sharpless}, S. 1959, \apjs, 4, 257

\bibitem[{{Shaya} \& {Olling}(2011)}]{shaya11}
{Shaya}, E.~J. \& {Olling}, R.~P. 2011, ApJS, 192, 2

\bibitem[{{Skrutskie} {et~al.}(2006){Skrutskie}, {Cutri}, {Stiening},
  {Weinberg}, {Schneider}, {Carpenter}, {Beichman}, {Capps}, {Chester},
  {Elias}, {Huchra}, {Liebert}, {Lonsdale}, {Monet}, {Price}, {Seitzer},
  {Jarrett}, {Kirkpatrick}, {Gizis}, {Howard}, {Evans}, {Fowler}, {Fullmer},
  {Hurt}, {Light}, {Kopan}, {Marsh}, {McCallon}, {Tam}, {Van Dyk}, \&
  {Wheelock}}]{skrutskie06}
{Skrutskie}, M.~F., {Cutri}, R.~M., {Stiening}, R., {et~al.} 2006, \aj, 131,
  1163

\bibitem[{{Smart} {et~al.}(2019){Smart}, {Marocco}, {Sarro}, {Barrado},
  {Beam{\'\i}n}, {Caballero}, \& {Jones}}]{smart19}
{Smart}, R.~L., {Marocco}, F., {Sarro}, L.~M., {et~al.} 2019, \mnras, 485, 4423

\bibitem[{{Smolinski} \& {Osborn}(2006)}]{smolinski06}
{Smolinski}, J. \& {Osborn}, W. 2006, in Rev. Mex. Astron. Astrofis. Conf.
  Ser., Vol.~25, Rev. Mex. Astron. Astrofis. Conf. Ser., ed. C.~{Abad},
  A.~{Bongiovanni}, \& Y.~{Guillen}, 65--68

\bibitem[{{Soubiran} {et~al.}(2008){Soubiran}, {Bienaym{\'e}}, {Mishenina}, \&
  {Kovtyukh}}]{soubiran08}
{Soubiran}, C., {Bienaym{\'e}}, O., {Mishenina}, T.~V., \& {Kovtyukh}, V.~V.
  2008, \aap, 480, 91

\bibitem[{{Stauffer} {et~al.}(2021){Stauffer}, {Rebull}, {Jardine}, {Collier
  Cameron}, {Cody}, {Hillenbrand}, {Barrado}, {Kruse}, \&
  {Powell}}]{stauffer21}
{Stauffer}, J., {Rebull}, L.~M., {Jardine}, M., {et~al.} 2021, \aj, 161, 60

\bibitem[{{Stee} {et~al.}(1995){Stee}, {de Araujo}, {Vakili}, {Mourard},
  {Arnold}, {Bonneau}, {Morand}, \& {Tallon-Bosc}}]{stee95}
{Stee}, P., {de Araujo}, F.~X., {Vakili}, F., {et~al.} 1995, \aap, 300, 219

\bibitem[{{Stock} {et~al.}(2018){Stock}, {Reffert}, \& {Quirrenbach}}]{stock18}
{Stock}, S., {Reffert}, S., \& {Quirrenbach}, A. 2018, \aap, 616, A33

\bibitem[{{Subasavage} {et~al.}(2008){Subasavage}, {Henry}, {Bergeron},
  {Dufour}, \& {Hambly}}]{subasavage08}
{Subasavage}, J.~P., {Henry}, T.~J., {Bergeron}, P., {Dufour}, P., \& {Hambly},
  N.~C. 2008, \aj, 136, 899

\bibitem[{{Tang} {et~al.}(2019){Tang}, {Pang}, {Yuan}, {Chen}, {Hong},
  {Goldman}, {Just}, {Shukirgaliyev}, \& {Lin}}]{tang19}
{Tang}, S.-Y., {Pang}, X., {Yuan}, Z., {et~al.} 2019, \apj, 877, 12

\bibitem[{{Taylor}(2005)}]{taylor05}
{Taylor}, M.~B. 2005, in ASPCS, Vol. 347, Astronomical Data Analysis Software
  and Systems XIV, ed. P.~{Shopbell}, M.~{Britton}, \& R.~{Ebert}, 29

\bibitem[{{Taylor}(2021)}]{taylor21}
{Taylor}, M.~B. 2021, {Tutorial: Exploring Gaia data with TOPCAT and STILTS},
  University of Bristol, Bristol UK

\bibitem[{{Tokovinin}(2008)}]{tokovinin08}
{Tokovinin}, A. 2008, \mnras, 389, 925

\bibitem[{{Tokovinin}(2014)}]{tokovinin14b}
{Tokovinin}, A. 2014, \aj, 147, 87

\bibitem[{{Tokovinin}(2017)}]{tokovinin17}
{Tokovinin}, A. 2017, \mnras, 468, 3461

\bibitem[{{Tokovinin}(2018)}]{tokovinin18}
{Tokovinin}, A. 2018, \apjs, 235, 6

\bibitem[{{Tokovinin}(2021)}]{tokovinin21}
{Tokovinin}, A. 2021, \aj, 161, 144

\bibitem[{{Tokovinin} \& {L{\'e}pine}(2012)}]{tokovinin12}
{Tokovinin}, A. \& {L{\'e}pine}, S. 2012, \aj, 144, 102

\bibitem[{{Tokovinin}(1997)}]{tokovinin97}
{Tokovinin}, A.~A. 1997, \aaps, 124, 75

\bibitem[{{Tolbert}(1964)}]{tolbert64}
{Tolbert}, C.~R. 1964, \apj, 139, 1105

\bibitem[{{Toonen} {et~al.}(2017){Toonen}, {Hollands}, {G{\"a}nsicke}, \&
  {Boekholt}}]{toonen17}
{Toonen}, S., {Hollands}, M., {G{\"a}nsicke}, B.~T., \& {Boekholt}, T. 2017,
  \aap, 602, A16

\bibitem[{{Wasserman} \& {Weinberg}(1991)}]{wasserman91}
{Wasserman}, I. \& {Weinberg}, M.~D. 1991, \apj, 382, 149

\bibitem[{{Weinberg} {et~al.}(1987){Weinberg}, {Shapiro}, \&
  {Wasserman}}]{weinberg87}
{Weinberg}, M.~D., {Shapiro}, S.~L., \& {Wasserman}, I. 1987, \apj, 312, 367

\bibitem[{{Weinberg} \& {Wasserman}(1988)}]{weinberg88}
{Weinberg}, M.~D. \& {Wasserman}, I. 1988, \apj, 329, 253

\bibitem[{{Wenger} {et~al.}(2000){Wenger}, {Ochsenbein}, {Egret}, {Dubois},
  {Bonnarel}, {Borde}, {Genova}, {Jasniewicz}, {Lalo{\"e}}, {Lesteven}, \&
  {Monier}}]{wenger00}
{Wenger}, M., {Ochsenbein}, F., {Egret}, D., {et~al.} 2000, \aaps, 143, 9

\bibitem[{{Wertheimer} \& {Laughlin}(2006)}]{wertheimer06}
{Wertheimer}, J.~G. \& {Laughlin}, G. 2006, \aj, 132, 1995

\bibitem[{{West} {et~al.}(2011){West}, {Morgan}, {Bochanski}, {Andersen},
  {Bell}, {Kowalski}, {Davenport}, {Hawley}, {Schmidt}, {Bernat}, {Hilton},
  {Muirhead}, {Covey}, {Rojas-Ayala}, {Schlawin}, {Gooding}, {Schluns},
  {Dhital}, {Pineda}, \& {Jones}}]{west11}
{West}, A.~A., {Morgan}, D.~P., {Bochanski}, J.~J., {et~al.} 2011, \aj, 141, 97

\bibitem[{{White} {et~al.}(1982){White}, {Swank}, {Holt}, \&
  {Parmar}}]{white82}
{White}, N.~E., {Swank}, J.~H., {Holt}, S.~S., \& {Parmar}, A.~N. 1982, \apj,
  263, 277

\bibitem[{{Yasuda} {et~al.}(2004){Yasuda}, {Mizumoto}, {Ohishi}, {O'Mullane},
  {Budav{\'a}ri}, {Haridas}, {Li}, {Malik}, {Szalay}, {Hill}, {Linde}, {Mann},
  \& {Page}}]{yasuda04}
{Yasuda}, N., {Mizumoto}, Y., {Ohishi}, M., {et~al.} 2004, in ASPCS, Vol. 314,
  Astronomical Data Analysis Software and Systems (ADASS) XIII, ed.
  F.~{Ochsenbein}, M.~G. {Allen}, \& D.~{Egret}, 293

\bibitem[{{Zapatero Osorio} \& {Mart{\'\i}n}(2005)}]{zapatero05}
{Zapatero Osorio}, M.~R. \& {Mart{\'\i}n}, E.~L. 2005, in Rev. Mex. Astron.
  Astrofis. Conf. Ser., Vol.~24, Rev. Mex. Astron. Astrofis. Conf. Ser., ed.
  A.~M. {Hidalgo-G{\'a}mez}, J.~J. {Gonz{\'a}lez}, J.~M. {Rodr{\'\i}guez
  Espinosa}, \& S.~{Torres-Peimbert}, 192--197

\bibitem[{{Zorec} {et~al.}(2005){Zorec}, {Fr{\'e}mat}, \& {Cidale}}]{zorec05}
{Zorec}, J., {Fr{\'e}mat}, Y., \& {Cidale}, L. 2005, \aap, 441, 235

\bibitem[{{Zuckerman} \& {Song}(2004)}]{zuckerman04}
{Zuckerman}, B. \& {Song}, I. 2004, \araa, 42, 685

\bibitem[{{Zuckerman} {et~al.}(2001){Zuckerman}, {Song}, \&
  {Webb}}]{zuckerman01}
{Zuckerman}, B., {Song}, I., \& {Webb}, R.~A. 2001, \apj, 559, 388

\end{thebibliography}

\appendix

\section{The $\gamma$~Cas association}
\label{sec:gamCas}

\begin{figure*}
 \centering
 \includegraphics[width=1\linewidth,angle=0]{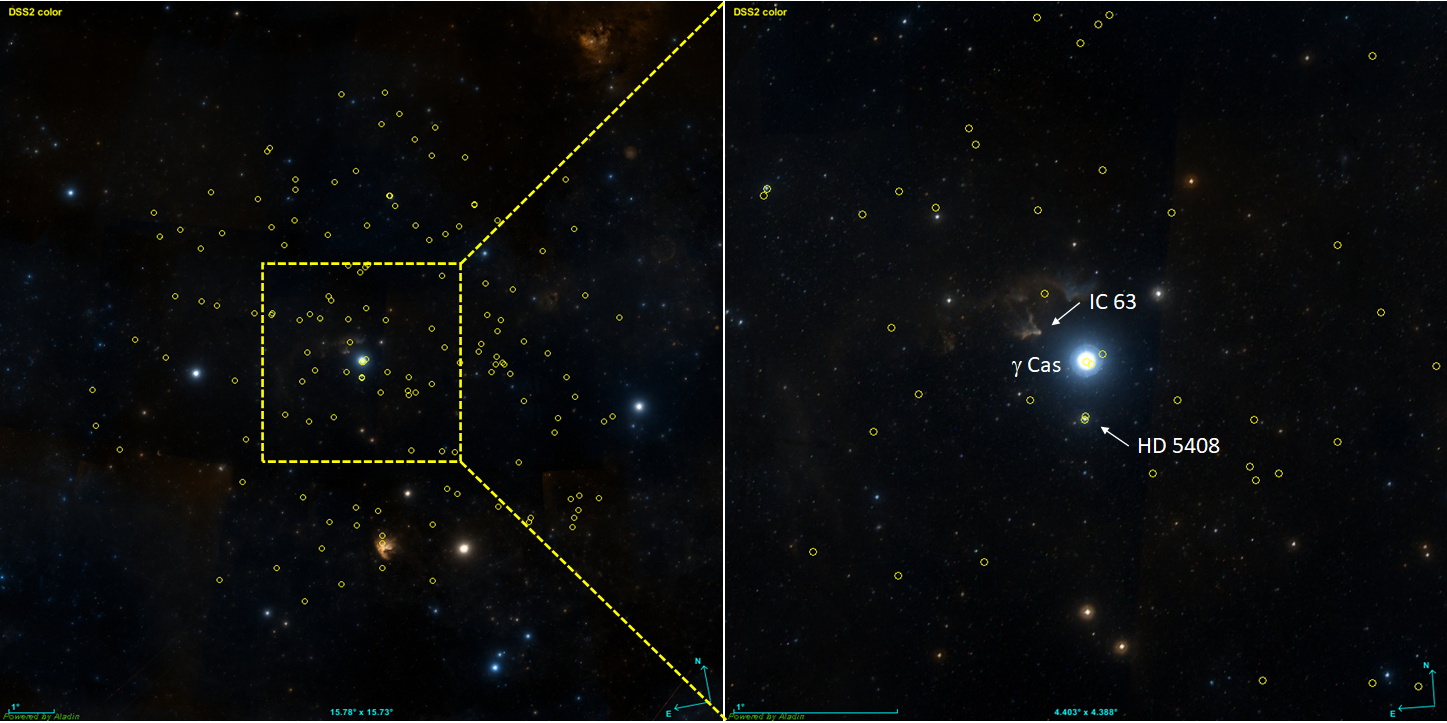}
 \caption{Spatial distribution of candidate young stars (open yellow circles) at less than 6\,deg ({\it left}) and 2\,deg ({\it right}) to $\gamma$~Cas.
 In the right panel we also highlight HD~5408 (the most massive star after $\gamma$~Cas) and the IC~63 emission nebula. 
 The images were created with the Aladin sky atlas and blue, red, and infrared Digitised Sky Survey data.}
 \label{fig:spatial_dist}
 \end{figure*}

Figure~\ref{fig:spatial_dist} shows the spatial distribution around $\gamma$ Cas in two panels. The left panel shows the 145 stars in Table~\ref{tab:gamCas_candidates} forming a circle area with a radius of 6\,degrees. The right panel reduce the radius to about 2\,degrees to have only 30 stars including $\gamma$ Cas. The IC 63 nebula emission, also named the `Phantom nebula', is easily observed in the right panel.

The stars $\gamma$~Cas and HD~5408, separated by 1274.5\,arcsec, constitute the wide pair MAM~20~AD \citep{mamajek17}.
They actually form a quintuple system, as they are reported to be close double (B0.5\,IV + F6\,V)\footnote{BU~499~AC is an optical pair, with ``ADS 782 C'' at 53\,arcsec to $\gamma$~Cas being a background star.} and triple (B7\,V + B9\,V + A1\,V) stars, respectively \citep{morgan43,osterbrock57,christy69,fekel79,nemravova12,hutter21}.

With such an early spectral type and an age of only about 8\,Ma \citep{zorec05}, $\gamma$~Cas is the ionising source of the nearby ($\rho \sim$ 1200\,arcsec) reflection nebulae \object{IC~63} (The Ghost of Cassiopeia) and \object{IC~59}
\citep{hubble22,sharpless59,jansen94}, as well as of an irregular, $\sim$3\,deg-diameter, H~{\sc ii} region \citep{karr05}.
With a mass of about 19\,$M_\odot$ and a surrounding disc, it is also the prototype of the $\gamma$~Cas type of stars (\citealt{poeckert78,stee95,naze22}).

Our astrometric search for common proper motion and parallax with the criteria in Sect.~\ref{sec:pair_validation} resulted in four additional stars not tabulated by WDS.
Of them, only one, namely \object{UCAC4~752--011208} (M4\,Ve), had been catalogued in the literature \citep{nesci18}.
Together with the five components of the $\gamma$~Cas+HD~5408 system, they made an agglomerate of nine stars of 8\,Ma at about 188\,pc.
Since 19\,$M_\odot$-mass stars do not form in isolation \citep{kroupa01,chabrier03,penaramirez12}, we extended our astrometric search and found additional stars that satisfy our criteria.
In particular, we enlarged our search radius centred on $\gamma$~Cas in consecutive steps and, besides $\gamma$~Cas and HD~5408, we found 10, 30, 51, 85, 115, and 143  {\it Gaia} DR3 stars with a 2MASS counterpart, and that satisfy our astrometric criteria, up to 1, 2, 3, 4, 5, and 6\,deg, respectively (Fig.~\ref{fig:spatial_dist}).
Some of these stars are in turn spectroscopic binaries, so their total number is larger.
Most selected stars follow the 10\,Ma theoretical isochrones of  PARSEC\footnote{\url{http://stev.oapd.inaf.it/cgi-bin/cmd}} \citep[][version 1.2S with the default values]{bressan12} at 188\,pc in {\it Gaia}-2MASS colour-magnitude diagrams.
Besides, the mass function computed from masses derived from the $J$-band absolute magnitude and PARSEC models do not deviate too much from Salpeter's.
However, we did not find a clustering of stars towards the most massive stars, as it is usually observed in open clusters of similar age (e.g. \citealt{caballero08}).
The hypothetical stars of an open cluster or, more likely, a stellar association around $\gamma$~Cas \citep{mamajek17} overlaps with the extended and also young population of the Cas-Tau OB1 association \citep{blaaw91,dezeeuw99}.
As a result, additional work is necessary to disentangle the stars that were born together with $\gamma$~Cas and HD~5408.

\onecolumn
\section{Additional tables}
\label{sec:additional_tables}

\centering
\begin{longtable}{l@{\hspace{3mm}}l@{\hspace{3mm}}c@{\hspace{2mm}}c@{\hspace{3mm}}c@{\hspace{2mm}}c@{\hspace{2mm}}c@{\hspace{2mm}}}
\caption{Stars in young stellar kinematic groups.}\\
\hline 
\hline
\noalign{\smallskip}

Star & Discoverer & $\alpha$\,(J2000) & $\delta$\,(J2000) & \textit{G} & Group$^{b}$ & Refs.$^{c}$ \\
 & code$^{a}$ & (hh:mm:ss.ss) & (dd:mm:ss.s) & (mag) & & \\
\noalign{\smallskip}
\noalign{\smallskip}
\hline
\noalign{\smallskip}
\noalign{\smallskip}
\endfirsthead
\caption{Stars in young stellar kinematic groups (continued).}\\
\hline \hline
\noalign{\smallskip} 
Star & Discoverer & $\alpha$\,(J2000) & $\delta$\,(J2000) & \textit{G} & Group$^{b}$ & Refs.$^{c}$ \\
 & code$^{a}$ & (hh:mm:ss.ss) & (dd:mm:ss.s) & (mag) & & \\
\noalign{\smallskip}
\noalign{\smallskip}
\hline
\noalign{\smallskip}
\noalign{\smallskip}
\endhead
\noalign{\smallskip}
\noalign{\smallskip}
\hline
\noalign{\smallskip}
\endfoot 
DENIS J000657.9-643654 & ... & 00:06:57.93 & $-$64:36:54.2 & 18.2 & Tuc-Hor & 1, 2 \\
UPM J0014-6003 & ... & 00:14:47.68 & $-$60:03:47.8 & 12.7 & Tuc-Hor & 1, 3 \\
UCAC3 52-533 & ... & 00:15:27.52 & $-$64:14:54.8 & 11.9 & Tuc-Hor & 1, 3 \\
CD-60 31 & ... & 00:17:30.50 & $-$59:57:04.4 & 10.6 & Tuc-Hor & 3 \\
HD 1466 & SHY 113 G, & 00:18:26.12 & $-$63:28:39.0 & 7.3 & Tuc-Hor & 1, 4 \\
 & SHY 114 G, & & & & & \\
 & CVN 33 G & & & & & \\
2MASS J00182834-6703130 & ... & 00:18:28.33 & $-$67:03:13.0 & 20.6 & Tuc-Hor & 2 \\
\end{longtable}
 \tablefoot{
    \tablefoottext{a}{Marked with `...' are new young star candidates.}
    \tablefoottext{b}{32 Ori: 32 Orionis; AB Dor: AB Doradus; Ale13: Alessi 13; $\beta$ Pic: $\beta$ Pictoris; Car: Carina; Car-Near: Carina-Near; Car-Vela: Carina-Vela; $\gamma$ Cas: $\gamma$ Cassiopeia; Castor: Castor ($\alpha$ Geminorum); CBer: Coma Berenice; $\chi^{\,01}$ For: $\chi^{\,01}$ Fornacis; Col: Columba; $\epsilon$ Cha: $\epsilon$ Chamaelontis; G-X: [TPY2019] Group-X; Hya: Hyades; IC2391: IC2391 Supercluster; LCC: Lower Centarus Crux; SCO: Scorpio-Centaurus; Tuc-Hor: Tucana-Horologium; UCL: Upper Centaurus-Lupus; UMa: Ursa Major; VCA: Volans-Carina.}
    \tablefoottext{c}{1. \citet{riedel17}; 2. \citet{gagne15}; 3. \citet{kraus14}; 4. \citet{gagne18a}; 5. This work; 6. \citet{gagne18c}; 7. \citet{gagne18b}; 8. \citet{freund20}; 9. \citet{kopytova16}; 10. \citet{roser11}; 11. \citet{cantat-gaudin18}; 12. \citet{reino18}; 13. \citet{reid93}; 14. \citet{bouvier08}; 15. \citet{guenther05}; 16. \citet{bell17}; 17. \citet{goldman18}; 18. \citet{murphy13}; 19. \citet{hoogerwerf00}; 20. \citet{dezeeuw99}; 21. \citet{rizzuto11}; 22. \citet{pecaut16}; 23. \citet{stauffer21}; 24. \citet{bohn22}; 25. \citet{dopcke19}; 26. \citet{dzib18}; 27. \citet{furnkranz19}; 28. \citet{tang19}.}
    \tablefoottext{d}{Not available in \textit{Gaia}.}
    }

\newpage

\centering
\begin{longtable}{lllcccc}
\caption{Basic data of the 94 galactic field systems.}\\
\noalign{\smallskip}
\hline \hline
\noalign{\smallskip}
WDS & Discoverer & Star & $\alpha$ (J2000) & $\delta$ (J2000) & $d$ & $G^a$ \\
 & code & & (hh:mm:ss.ss) & (dd:mm:ss.s) & (pc) & (mag) \\
\noalign{\smallskip}
\hline
\noalign{\smallskip}
\endfirsthead
\caption{Basic data of the 94 galactic field systems (continued).}\\
\hline \hline
\noalign{\smallskip} 
WDS & Discoverer & Star & $\alpha$ (J2000) & $\delta$ (J2000) & $d$ & $G^a$ \\
 & code & & (hh:mm:ss.ss) & (dd:mm:ss.s) & (pc) & (mag) \\
\noalign{\smallskip}
\hline
\noalign{\smallskip}
\noalign{\smallskip}
\endhead
\noalign{\smallskip}
\hline
\noalign{\smallskip}
\endfoot 
\multicolumn{7}{c}{\it Double systems}\\
\noalign{\smallskip}
 
00016--0102 &  & 2MASS J00013688-0101441 & 00:01:36.89 & $-$01:01:44.2 & 60.33$\pm$0.13 & 15.2 \\
 & WIS   1 & SIPS J0000-0112 & 00:00:35.39 & $-$01:12:46.3 & 65.41$\pm$0.64 & 17.3\\
 \noalign{\smallskip}
01005--1923 &  & HD 5911 & 01:00:28.01 & $-$19:23:21.8 & 81.63$\pm$0.17 & 7.9 \\
 & SHY 392 & HD 6103 & 01:02:06.90 & $-$19:40:10.8 & 81.59$\pm$0.16 & 8.1 \\
 \noalign{\smallskip}
01066+1353 &  & HD 6566 & 01:06:36.52 & +13:53:04.6 & 58.08$\pm$0.10 & 7.1 \\
 & SHY 396 & HD 5433$^b$ & 00:56:13.48 & +15:39:22.2 & 63.44$\pm$0.43 & 8.5 \\
 \noalign{\smallskip}
01134--3932 &  & HD 7382 & 01:13:26.52 & $-$39:32:24.3 & 59.18$\pm$0.11 & 7.2 \\
\end{longtable}
\normalsize
\tablefoot{
    \tablefoottext{a}{The maximum error in $G$ provided by \textit{Gaia} DR3 is less than 0.005\,mag;}
    \tablefoottext{b}{Proper motion anomaly measured by \citet{kervella19}, \citet{brandt21} or both;}
    \tablefoottext{c}{{\tt RUWE} $> 10$;}
    \tablefoottext{d}{Large $\sigma_{Vr}$ for its $G$ magnitude;}       
    \tablefoottext{e}{HD~35066(Aab) was recently resolved at SOAR with $\rho=0.14$\,arcsec (anonymous reviewer, priv.comm);}
    \tablefoottext{f}{Value from \textit{Gaia} DR2;}
    \tablefoottext{g}{Very shiny star, not resolved by \textit{Gaia} DR3. Astrometric data from \textit{Hipparcos};} 
    \tablefoottext{h}{\citet{pourbaix16};} 
    \tablefoottext{i}{\citet{pecaut13}.}   
}

\newpage

\centering
\begin{longtable}{l@{\hspace{2mm}}l@{\hspace{2mm}}l@{\hspace{2mm}}c@{\hspace{1mm}}c@{\hspace{1mm}}c@{\hspace{1mm}}c@{\hspace{1mm}}c}
\caption{Masses, $\rho$, $\theta$, separations and gravitational binding energy of the systems identified in the sample of work.}\\
\noalign{\smallskip}
\hline \hline
\noalign{\smallskip}
WDS & Discoverer & Star & \textit{M} & $\rho$ & $\theta$ & \textit{s} & $|U^*_g|$ \\
 & code &  & (M$_\odot$) & (arcsec) & (deg) & ($10^3$\,au) & ($10^{33}$\,J) \\
\noalign{\smallskip}
\hline
\noalign{\smallskip}
\endfirsthead
\caption{Masses, $\rho$, $\theta$, separations, and gravitational binding energy of the systems identified in the sample of work (continued).}\\
\hline 
\hline
\noalign{\smallskip} 
WDS & Discoverer & Star & \textit{M} & $\rho$ & $\theta$ & \textit{s} & $|U^*_g|$ \\
 & code &  & (M$_\odot$) & (arcsec) & (deg) & ($10^3$\,au) & ($10^{33}$\,J) \\
\noalign{\smallskip}
\hline
\noalign{\smallskip}
\noalign{\smallskip}
\endhead
\noalign{\smallskip}
\hline
\noalign{\smallskip}
\endfoot 
\multicolumn{8}{c}{\it Double systems}\\
\noalign{\smallskip}
 
00016--0102 &  & 2MASS J00013688-0101441 & 0.22$\pm$0.02 &  &  &  & 0.74$\pm$0.10 \\
 & WIS   1 & SIPS J0000-0112 & 0.13$\pm$0.01 & 1135 & 234 & 68.4$\pm$0.2 &  \\
 \noalign{\smallskip} 
01005--1923 &  & HD 5911 & 1.32$\pm$0.13 &  &  &  & 20.6$\pm$2.9 \\
 & SHY 392 & HD 6103 & 1.24$\pm$0.12 & 1725 & 126 & 140.8$\pm$0.3 &  \\
  \noalign{\smallskip}
01066+1353 &  & HD 6566 & 1.32$\pm$0.10 &  &  &  & 3.69$\pm$0.52 \\
 & SHY 396 & HD 5433$^a$ & 1.02$\pm$0.10 & 11089 & 305 & 643.7$\pm$1.2 &  \\
  \noalign{\smallskip}
01134--3932 &  & HD 7382 & 1.32$\pm$0.13 &  &  &  & 7.82$\pm$1.11 \\
\end{longtable}
\normalsize
\tablefoot{
    \tablefoottext{a}{Proper motion anomaly measured by \citet{kervella19}, \citet{brandt21} or both.}
    \tablefoottext{b}{{\tt RUWE} $> 10$;}
    \tablefoottext{c}{Large $\sigma_{Vr}$ for its $G$ magnitude;}
        \tablefoottext{d}{Classified as a white dwarf candidate by \citet{gentilefusillo19}, it has instead absolute magnitudes and colours of an intermediate M dwarf;}
    \tablefoottext{e}{\citet{dasilva15};}
    \tablefoottext{f}{\citet{tokovinin14b};}
    \tablefoottext{g}{Dynamical mass \citep{malkov12};}
    \tablefoottext{h}{\citet{gontcharov10};}
    \tablefoottext{i}{\citet{mitrofanova21};}
    \tablefoottext{j}{\citet{pourbaix16};}
    \tablefoottext{k}{\citet{kervella19};}
    \tablefoottext{l}{\citet{eker18};}
    \tablefoottext{m}{\citet{feuillet16};}
    \tablefoottext{n}{\citet{tokovinin18};}
    \tablefoottext{o}{\citet{montes18a}.}
}

\newpage

\centering
\begin{longtable}{lcclcccc}
\caption{Candidate stars to be part of the $\gamma$ Cas association, in order of the distance to the central star $\gamma$ Cas.}\\
\hline 
\hline
\noalign{\smallskip}

Star & $\alpha$\,(J2000) & $\delta$\,(J2000) & SpT & \textit{M} & \textit{G}$^a$ & \textit{J}$^b$ & $\rho_{\gamma\,\text{Cas}}$  \\
 & (hh:mm:ss.ss) & (dd:mm:ss.s) & & (M$_{\odot}$) & (mag) & (mag) & (deg) \\
\noalign{\smallskip}
\hline
\noalign{\smallskip}
\endfirsthead
\caption{(Continued): Candidates to be part of the moving group, in order of the distance to $\gamma$ Cas.}\\
\hline \hline
\noalign{\smallskip} 
Star & $\alpha$\,(J2000) & $\delta$\,(J2000) & SpT & \textit{M} & \textit{G}$^a$ & \textit{J}$^b$ & $\rho_{\gamma\,\text{Cas}}$  \\
 & (hh:mm:ss.ss) & (dd:mm:ss.s) & & (M$_{\odot}$) & (mag) & (mag) & (deg) \\
\noalign{\smallskip}
\hline
\noalign{\smallskip}
\endhead
\noalign{\smallskip}
\hline
\noalign{\smallskip}
\endfoot 
$\gamma$ Cas & 00:56:42.53 & +60:43:00.3 & B0.5IV & $\sim$ 13$^c$ & 2.3 & 2.0 & 0.000  \\
Gaia DR3 426558563962119808 & 00:56:26.00 & +60:41:55.5 &  & 0.80 $\pm$ 0.08 & 12.3 & 10.8 & 0.038  \\
Gaia DR3 426559693524626816 & 00:55:55.21 & +60:45:44.5 &  & 0.16 $\pm$ 0.02 & 18.7 & 15.0 & 0.107  \\
UCAC4 752-011208 & 00:56:44.80 & +60:22:46.3 & M4Ve & 0.46 $\pm$ 0.05 & 15.4 & 12.4 & 0.337  \\
HD 5408(Aa) & \multirow{3}{*}{\bigg\} 00:56:46.97} & & B7V & 3.40 $\pm$ 0.10$^d$ & \multirow{3}{*}{\bigg\} 6.0} & &  \\
HD 5408(Ab) &  & +60:21:46.2 & B9V & 4.10 $\pm$ 0.10$^d$ & & 5.6 & 0.354  \\
HD 5408(B) & & & A1V & 3.40 $\pm$ 0.10$^d$ & & &  \\
Gaia DR3 426494169508443520 & 00:59:29.59 & +60:28:43.0 &  & 0.14 $\pm$ 0.01 & 19.4 & 15.4 & 0.417  \\
Gaia DR3 426656106950424960 & 00:58:49.75 & +61:07:34.5 &  & 0.19 $\pm$ 0.02 & 18.3 & 14.7 & 0.484  \\
Gaia DR3 426907040426100352 & 00:52:12.81 & +60:28:25.5 &  & 0.45 $\pm$ 0.05 & 15.3 & 12.9 & 0.603  \\
Gaia DR3 426117827290256896 & 00:53:27.10 & +60:01:58.1 &  & 0.75 $\pm$ 0.08 & 12.7 & 11.1 & 0.794  \\
Gaia DR3 426696758828470144 & 00:59:12.34 & +61:38:10.3 &  & 0.28 $\pm$ 0.03 & 17.3 & 14.2 & 0.967  \\
Gaia DR3 426391055933615872 & 01:05:01.74 & +60:30:11.1 &  & 0.11 $\pm$ 0.01 & 20.4 & 16.3 & 1.04  \\
Gaia DR3 427444254930463744 & 00:52:19.64 & +61:37:04.8 &  & 0.24 $\pm$ 0.02 & 17.4 & 13.9 & 1.04  \\
Gaia DR3 426937483155345280 & 00:48:27.00 & +60:20:51.1 &  & 0.43 $\pm$ 0.04 & 15.7 & 12.9 & 1.08  \\
Gaia DR3 427465454889829248 & 00:55:51.79 & +61:52:51.7 &  & 0.35 $\pm$ 0.03 & 16.5 & 13.4 & 1.17  \\
Gaia DR3 426833373147610240 & 00:48:42.59 & +60:03:34.6 &  & 0.39 $\pm$ 0.04 & 15.9 & 12.7 & 1.19  \\
Gaia DR3 426416177204072448 & 01:06:29.72 & +60:53:50.6 &  & 0.40 $\pm$ 0.04 & 16.0 & 12.9 & 1.21  \\
Gaia DR3 426831375980796800 & 00:48:27.33 & +59:58:28.6 &  & 0.11 $\pm$ 0.01 & 20.1 & 16.2 & 1.26  \\
Gaia DR3 522715116413884416 & 01:04:25.80 & +61:38:25.1 &  & 0.38 $\pm$ 0.04 & 16.3 & 13.3 & 1.31  \\
Gaia DR3 426835125494086144 & 00:47:19.08 & +60:00:51.0 &  & 0.14 $\pm$ 0.01 & 19.2 & 15.8 & 1.36  \\
Gaia DR3 426326116027684480 & 01:07:11.44 & +60:15:44.2 &  & 0.25 $\pm$ 0.03 & 17.6 & 14.4 & 1.37  \\
Gaia DR3 425868517332283776 & 01:01:36.89 & +59:29:25.4 &  & 0.41 $\pm$ 0.04 & 15.9 & 12.9 & 1.37  \\
TYC 4021-505-1 & 01:02:27.29 & +62:01:55.3 & G5V & 0.91 $\pm$ 0.09 & 11.4 & 10.0 & 1.48  \\
Gaia DR3 522555034397334912 & 01:06:20.90 & +61:43:55.1 &  & 0.78 $\pm$ 0.08 & 12.4 & 10.9 & 1.54  \\
Gaia DR3 522765865748123648 & 01:02:51.30 & +62:07:44.7 &  & 0.34 $\pm$ 0.03 & 16.6 & 13.4 & 1.59  \\
Gaia DR3 427035030443328000 & 00:44:21.55 & +60:11:20.1 &  & 0.18 $\pm$ 0.02 & 18.7 & 15.6 & 1.61  \\
Gaia DR3 522537644061868544 & 01:08:10.45 & +61:35:06.6 &  & 0.21 $\pm$ 0.02 & 18.2 & 14.9 & 1.63  \\
Gaia DR3 427202843412335104 & 00:43:54.51 & +61:23:23.2 &  & 0.32 $\pm$ 0.03 & 16.9 & 13.8 & 1.69  \\
HD 236617 & 01:05:43.02 & +59:23:26.8 & G5 & 1.28 $\pm$ 0.13 & 9.9 & 8.6 & 1.74  \\
Gaia DR3 427169445745848832 & 00:41:53.36 & +60:57:58.8 &  & 0.23 $\pm$ 0.02 & 17.7 & 14.2 & 1.82  \\
Gaia DR3 523592384950148992 & 00:57:02.17 & +62:39:28.4 &  & 0.22 $\pm$ 0.02 & 17.7 & 14.7 & 1.94  \\
Gaia DR3 426197133869019520 & 01:09:50.64 & +59:30:48.7 &  & 0.27 $\pm$ 0.03 & 17.4 & 14.7 & 2.03  \\
Gaia DR3 523605720834117632 & 00:56:04.78 & +62:46:10.7 &  & 0.34 $\pm$ 0.03 & 16.6 & 13.7 & 2.05  \\
Gaia DR3 523606820345403520 & 00:55:29.74 & +62:49:28.9 &  & 0.48 $\pm$ 0.05 & 15.6 & 13.1 & 2.11  \\
TYC 4021-815-1 & 00:59:20.89 & +62:48:42.6 &  & 0.99 $\pm$ 0.10 & 11.0 & 9.9 & 2.12  \\
Gaia DR3 427108354134806784 & 00:39:17.10 & +60:36:55.3 &  & 0.56 $\pm$ 0.06 & 14.5 & 11.4 & 2.14  \\
V761 Cas & 01:13:09.85 & +61:42:22.3 & B9V & 3.22 $\pm$ 0.32 & 6.5 & 6.2 & 2.21  \\
Gaia DR3 522574580780909440 & 01:13:19.43 & +61:40:00.1 &  & 0.19 $\pm$ 0.02 & 18.4 & 14.8 & 2.22  \\
Gaia DR3 425198223255334400 & 00:48:24.65 & +58:45:19.6 &  & 0.16 $\pm$ 0.02 & 19.1 & 15.8 & 2.22  \\
Gaia DR3 427777647475959680 & 00:41:34.65 & +62:31:42.5 &  & 0.28 $\pm$ 0.03 & 17.3 & 14.2 & 2.55  \\
Gaia DR3 430106477517090560 & 00:35:44.63 & +60:49:57.6 &  & 0.79 $\pm$ 0.08 & 12.5 & 10.9 & 2.56  \\
Gaia DR3 510588362155056384 & 01:16:30.81 & +61:41:01.5 &  & 0.19 $\pm$ 0.02 & 18.4 & 14.9 & 2.57  \\
Gaia DR3 425164005257167232 & 00:43:17.19 & +58:42:44.3 &  & 0.27 $\pm$ 0.03 & 17.5 & 14.4 & 2.62  \\
Gaia DR3 430133793510727040 & 00:35:08.90 & +60:59:03.0 &  & 0.17 $\pm$ 0.02 & 18.8 & 15.5 & 2.64  \\
Gaia DR3 510353788221523584 & 01:18:57.90 & +60:10:51.9 &  & 0.31 $\pm$ 0.03 & 16.8 & 13.7 & 2.80  \\
Gaia DR3 425348547108352768 & 00:41:07.58 & +58:40:29.8 &  & 0.24 $\pm$ 0.02 & 17.6 & 14.6 & 2.83  \\
Gaia DR3 428588949621015296 & 00:33:48.86 & +60:21:42.7 &  & 0.15 $\pm$ 0.01 & 19.4 & 15.8 & 2.84  \\
Gaia DR3 523346541025484160 & 01:03:22.37 & +63:28:35.3 &  & 0.12 $\pm$ 0.01 & 19.9 & 16.2 & 2.87  \\
Gaia DR3 430258798542242176 & 00:33:41.47 & +61:36:28.7 &  & 0.26 $\pm$ 0.03 & 17.6 & 14.5 & 2.91  \\
Gaia DR3 428608599088304512 & 00:32:59.63 & +60:30:54.1 &  & 0.13 $\pm$ 0.01 & 19.6 & 15.0 & 2.92  \\
Cl* NGC 129 SS 525 & 00:32:36.63 & +60:41:19.2 &  & 0.96 $\pm$ 0.10 & 11.2 & 10.1 & 2.95  \\
BD+62 170 & 00:55:40.97 & +63:40:51.8 & F8 & 1.24 $\pm$ 0.12 & 9.7 & 8.8 & 2.97  \\
Gaia DR3 430199734155452160 & 00:32:04.70 & +61:14:49.8 &  & 0.29 $\pm$ 0.03 & 16.9 & 14.1 & 3.03  \\
Gaia DR3 523899664093745408 & 00:43:41.06 & +63:20:19.7 &  & 0.17 $\pm$ 0.02 & 18.8 & 15.6 & 3.03  \\
Gaia DR3 414097283284109952 & 01:16:18.96 & +58:55:15.9 &  & 0.44 $\pm$ 0.04 & 15.7 & 12.7 & 3.05  \\
Gaia DR3 523114788890667776 & 01:11:43.52 & +63:12:14.3 &  & 0.39 $\pm$ 0.04 & 16.1 & 13.2 & 3.05  \\
Gaia DR3 428606507444503552 & 00:31:41.43 & +60:32:29.1 &  & 0.32 $\pm$ 0.03 & 16.8 & 13.7 & 3.07  \\
Gaia DR3 428651789276927872 & 00:31:28.31 & +60:30:36.4 &  & 0.18 $\pm$ 0.02 & 18.8 & 15.1 & 3.10  \\
Gaia DR3 430206674822463744 & 00:31:16.38 & +61:27:47.3 &  & 0.26 $\pm$ 0.03 & 17.3 & 13.9 & 3.16  \\
Gaia DR3 424138538865527168 & 00:57:44.28 & +57:31:32.6 &  & 0.76 $\pm$ 0.08 & 12.8 & 11.1 & 3.19  \\
Gaia DR3 523891933149775616 & 00:46:04.81 & +63:39:46.4 &  & 0.23 $\pm$ 0.02 & 17.8 & 14.8 & 3.20  \\
Gaia DR3 430351707277551616 & 00:33:23.98 & +62:18:00.1 &  & 0.44 $\pm$ 0.04 & 15.7 & 12.9 & 3.20  \\
Gaia DR3 413591954612999296 & 01:06:23.15 & +57:44:02.9 &  & 0.28 $\pm$ 0.03 & 17.2 & 14.0 & 3.23  \\
NGC 129 48 & 00:30:33.45 & +60:17:27.3 & F6V & 1.27 $\pm$ 0.13 & 9.5 & 8.8 & 3.25  \\
Gaia DR3 424242404052127872 & 00:54:06.70 & +57:27:25.2 &  & 0.22 $\pm$ 0.02 & 17.7 & 14.7 & 3.28  \\
Gaia DR3 523918424511630208 & 00:40:28.49 & +63:26:04.7 &  & 0.10 $\pm$ 0.01 & 20.4 & 16.6 & 3.32  \\
Gaia DR3 424738421233795072 & 00:42:39.06 & +57:52:50.3 &  & 0.20 $\pm$ 0.02 & 17.9 & 14.7 & 3.36  \\
BD+61 258 & 01:23:33.25 & +61:46:05.2 & F2 & 1.44 $\pm$ 0.14 & 9.6 & 8.9 & 3.39  \\
Gaia DR3 523413271935393280 & 01:09:59.65 & +63:45:58.1 &  & 0.39 $\pm$ 0.04 & 16.1 & 13.1 & 3.42  \\
Gaia DR3 524005221511877120 & 00:49:59.25 & +64:05:45.1 &  & 0.32 $\pm$ 0.03 & 16.9 & 13.8 & 3.47  \\
Gaia DR3 523228549680118656 & 01:14:29.82 & +63:34:44.6 &  & 0.22 $\pm$ 0.02 & 17.6 & 14.4 & 3.54  \\
Gaia DR3 428677391586484480 & 00:27:30.80 & +60:57:36.1 &  & 0.42 $\pm$ 0.04 & 16.0 & 12.7 & 3.56  \\
Gaia DR3 424897712980359296 & 00:41:06.06 & +57:45:14.0 &  & 0.20 $\pm$ 0.02 & 18.0 & 14.6 & 3.57  \\
Gaia DR3 424073525939801984 & 01:01:54.61 & +57:11:58.5 &  & 0.19 $\pm$ 0.02 & 18.4 & 15.1 & 3.58  \\
Gaia DR3 424031851879846912 & 00:57:35.08 & +57:07:43.0 &  & 0.52 $\pm$ 0.05 & 15.1 & 12.9 & 3.59  \\
Gaia DR3 430861399634552448 & 00:37:36.63 & +63:35:12.6 &  & 0.59 $\pm$ 0.06 & 14.3 & 12.0 & 3.63  \\
Gaia DR3 428484045032883456 & 00:28:38.41 & +59:39:27.0 &  & 0.15 $\pm$ 0.02 & 19.2 & 15.9 & 3.64  \\
Gaia DR3 430518764324231040 & 00:28:32.52 & +62:06:39.1 &  & 0.52 $\pm$ 0.05 & 14.8 & 12.0 & 3.64  \\
HD 4810 & 00:51:11.44 & +64:19:29.7 & A2 & 1.87 $\pm$ 0.19 & 8.4 & 8.1 & 3.66  \\
TYC 4024-250-1 & 00:51:04.61 & +64:20:17.9 &  & 1.07 $\pm$ 0.11 & 10.7 & 9.7 & 3.68  \\
Gaia DR3 413481488050839424 & 01:16:18.59 & +57:58:51.2 &  & 0.80 $\pm$ 0.08 & 12.7 & 10.5 & 3.70  \\
Gaia DR3 510825478709127424 & 01:26:26.82 & +61:49:43.0 &  & 0.33 $\pm$ 0.03 & 17.0 & 14.1 & 3.74  \\
Gaia DR3 423826513780557952 & 00:53:28.82 & +56:54:47.2 &  & 0.11 $\pm$ 0.01 & 20.5 & 16.9 & 3.83  \\
Gaia DR3 424630050620861312 & 00:45:16.82 & +57:06:56.8 &  & 0.23 $\pm$ 0.02 & 17.4 & 14.4 & 3.89  \\
Gaia DR3 524282706455103744 & 01:02:14.82 & +64:37:47.9 &  & 0.23 $\pm$ 0.02 & 17.6 & 14.3 & 3.96  \\
Gaia DR3 423823494422951040 & 00:53:26.81 & +56:44:57.9 &  & 0.57 $\pm$ 0.06 & 14.3 & 12.2 & 3.99  \\
Gaia DR3 524957222478303616 & 01:10:05.51 & +64:25:50.8 &  & 0.19 $\pm$ 0.02 & 18.1 & 14.7 & 4.02  \\
TYC 4015-1647-1 & 00:23:32.15 & +60:38:28.4 &  & 0.97 $\pm$ 0.10 & 11.2 & 10.0 & 4.06  \\
Gaia DR3 427905706211506560 & 00:30:32.34 & +58:20:28.2 &  & 0.14 $\pm$ 0.01 & 19.5 & 16.2 & 4.08  \\
Gaia DR3 512646373042841984 & 01:24:00.84 & +63:21:27.8 &  & 0.79 $\pm$ 0.08 & 12.5 & 11.1 & 4.15  \\
Gaia DR3 524328581007238656 & 00:57:51.06 & +64:52:55.5 &  & 0.40 $\pm$ 0.04 & 16.2 & 13.4 & 4.17  \\
Gaia DR3 423947460059430912 & 01:03:02.76 & +56:37:16.3 &  & 0.10 $\pm$ 0.01 & 20.5 & 16.2 & 4.18  \\
Gaia DR3 524704236026503552 & 01:17:29.40 & +64:09:57.5 &  & 0.26 $\pm$ 0.03 & 17.5 & 14.4 & 4.20  \\
Gaia DR3 526998195239910656 & 00:34:15.41 & +64:02:07.1 &  & 0.39 $\pm$ 0.04 & 16.3 & 13.5 & 4.21  \\
Gaia DR3 526998195239908736 & 00:34:13.63 & +64:02:13.1 &  & 0.79 $\pm$ 0.08 & 12.6 & 11.2 & 4.22  \\
HD 6822 & 01:10:16.31 & +64:38:45.2 & A0 & 1.89 $\pm$ 0.19 & 8.2 & 7.8 & 4.22  \\
Gaia DR3 431081851713650944 & 00:30:03.36 & +63:38:14.5 &  & 0.51 $\pm$ 0.05 & 15.2 & 12.8 & 4.26  \\
TYC 4031-2224-1 & 01:31:39.03 & +60:30:44.6 & F5 & 0.99 $\pm$ 0.10 & 11.0 & 9.9 & 4.29  \\
Gaia DR3 512575969937525376 & 01:27:42.20 & +62:58:26.6 &  & 0.25 $\pm$ 0.03 & 17.3 & 14.2 & 4.29  \\
Gaia DR3 510790844092027008 & 01:31:20.08 & +61:52:29.1 &  & 0.20 $\pm$ 0.02 & 18.4 & 15.7 & 4.31  \\
Gaia DR3 424842569903670656 & 00:34:37.72 & +57:24:57.8 &  & 0.32 $\pm$ 0.03 & 16.8 & 13.9 & 4.35  \\
Gaia DR3 428283418521147392 & 00:23:09.47 & +59:12:54.8 &  & 0.29 $\pm$ 0.03 & 17.1 & 14.0 & 4.45  \\
Gaia DR3 428172299125461120 & 00:24:08.03 & +58:53:49.3 &  & 0.25 $\pm$ 0.03 & 17.4 & 14.1 & 4.48  \\
BD+59 37 & 00:20:47.67 & +60:03:39.4 &  & 1.26 $\pm$ 0.13 & 9.8 & 6.8 & 4.48  \\
Gaia DR3 423769102956471680 & 00:53:28.68 & +56:12:18.6 &  & 0.44 $\pm$ 0.04 & 15.6 & 12.7 & 4.53  \\
Gaia DR3 430603907757603584 & 00:22:05.36 & +62:52:09.2 &  & 0.46 $\pm$ 0.05 & 15.4 & 12.7 & 4.62  \\
Gaia DR3 428342762091655168 & 00:19:49.86 & +59:36:50.2 &  & 0.53 $\pm$ 0.05 & 14.7 & 11.9 & 4.71  \\
Gaia DR3 527493250347583872 & 00:42:12.63 & +65:10:36.1 &  & 0.32 $\pm$ 0.03 & 16.8 & 13.6 & 4.75  \\
Gaia DR3 423535250579327488 & 00:59:56.20 & +55:51:15.4 &  & 0.27 $\pm$ 0.03 & 17.5 & 14.8 & 4.88  \\
Gaia DR3 411948871922993024 & 01:10:06.75 & +56:09:03.8 &  & 0.47 $\pm$ 0.05 & 15.5 & 13.1 & 4.89  \\
Gaia DR3 512501989120374144 & 01:32:08.47 & +63:18:56.2 &  & 0.19 $\pm$ 0.02 & 18.4 & 14.8 & 4.90  \\
Gaia DR3 512789034676716032 & 01:27:08.42 & +64:13:28.1 &  & 0.22 $\pm$ 0.02 & 17.6 & 14.2 & 4.96  \\
Gaia DR3 509912226919850880 & 01:37:26.63 & +60:48:22.2 &  & 0.19 $\pm$ 0.02 & 18.4 & 14.9 & 4.97  \\
Gaia DR3 524616854909023488 & 00:45:39.31 & +65:32:40.3 &  & 0.22 $\pm$ 0.02 & 17.8 & 14.5 & 4.99  \\
Gaia DR3 527148725254317312 & 00:35:20.74 & +65:04:29.5 &  & 0.62 $\pm$ 0.06 & 14.3 & 12.1 & 4.99  \\
Gaia DR3 421818053931122176 & 00:29:34.63 & +57:07:03.5 &  & 0.74 $\pm$ 0.07 & 12.8 & 11.2 & 5.02  \\
Gaia DR3 524934716850804992 & 01:16:26.04 & +65:13:51.0 &  & 0.82 $\pm$ 0.08 & 12.3 & 10.9 & 5.04  \\
Gaia DR3 418461897767667456 & 00:45:39.92 & +55:52:59.2 &  & 0.22 $\pm$ 0.02 & 17.9 & 14.7 & 5.05  \\
Gaia DR3 421811697379570688 & 00:29:48.73 & +57:01:45.2 &  & 0.13 $\pm$ 0.01 & 19.2 & 12.8 & 5.06  \\
Gaia DR3 524938084107230720 & 01:16:01.32 & +65:18:17.2 &  & 0.10 $\pm$ 0.01 & 20.6 & 16.7 & 5.08  \\
Gaia DR3 429704571660226176 & 00:15:13.72 & +61:46:23.3 &  & 0.18 $\pm$ 0.02 & 18.3 & 15.0 & 5.09  \\
Gaia DR3 512451931283034112 & 01:35:47.24 & +63:06:48.3 &  & 0.53 $\pm$ 0.05 & 15.0 & 12.0 & 5.18  \\
HD 4948 & 00:52:36.09 & +65:53:30.3 & A2 & 1.89 $\pm$ 0.19 & 8.3 & 8.1 & 5.20  \\
Gaia DR3 527534653834163072 & 00:41:10.00 & +65:46:21.3 &  & 0.21 $\pm$ 0.02 & 18.0 & 14.8 & 5.35  \\
Gaia DR3 411492437156666368 & 01:05:30.07 & +55:27:14.2 &  & 0.60 $\pm$ 0.06 & 14.3 & 12.3 & 5.39  \\
Gaia DR3 422967112302952192 & 00:15:35.09 & +58:59:21.4 &  & 0.55 $\pm$ 0.06 & 14.9 & 12.7 & 5.44  \\
Gaia DR3 421980919092603392 & 00:22:46.22 & +57:25:23.3 &  & 0.49 $\pm$ 0.05 & 15.2 & 12.6 & 5.46  \\
Gaia DR3 431392360672275584 & 00:15:18.31 & +63:14:19.4 &  & 0.22 $\pm$ 0.02 & 17.9 & 14.9 & 5.47  \\
TYC 4028-969-1 & 00:48:42.53 & +66:07:10.6 &  & 0.86 $\pm$ 0.09 & 11.9 & 10.6 & 5.48  \\
Gaia DR3 421990608538759808 & 00:21:17.48 & +57:27:37.2 &  & 0.30 $\pm$ 0.03 & 17.2 & 14.3 & 5.59  \\
Gaia DR3 512489211600057600 & 01:37:48.00 & +63:36:33.3 &  & 0.53 $\pm$ 0.05 & 15.0 & 12.6 & 5.59  \\
Gaia DR3 422982677264653440 & 00:13:53.86 & +59:04:25.8 &  & 0.36 $\pm$ 0.04 & 16.5 & 13.6 & 5.61  \\
Gaia DR3 508974691392427776 & 01:37:12.24 & +58:23:19.4 &  & 0.39 $\pm$ 0.04 & 16.3 & 13.2 & 5.63  \\
TYC 3673-1289-1 & 01:18:51.10 & +55:49:08.2 &  & 1.28 $\pm$ 0.13 & 9.7 & 8.8 & 5.69  \\
Gaia DR3 429623693140336512 & 00:09:54.06 & +61:10:34.9 &  & 0.21 $\pm$ 0.02 & 18.2 & 12.2 & 5.69  \\
Gaia DR3 421959169377531264 & 00:21:46.47 & +57:08:46.7 &  & 0.15 $\pm$ 0.02 & 18.9 & 15.7 & 5.74  \\
Gaia DR3 421945356762784640 & 00:22:33.71 & +56:59:31.6 &  & 0.34 $\pm$ 0.03 & 16.9 & 13.7 & 5.77  \\
Gaia DR3 526042891426149888 & 01:01:11.94 & +66:32:38.6 &  & 0.11 $\pm$ 0.01 & 20.4 & 16.2 & 5.85  \\
Gaia DR3 421933021616821504 & 00:23:06.48 & +56:47:52.9 &  & 0.64 $\pm$ 0.06 & 14.0 & 11.9 & 5.85  \\
Gaia DR3 526137930469180288 & 00:51:34.57 & +66:35:09.8 &  & 0.71 $\pm$ 0.07 & 13.2 & 11.4 & 5.90  \\
Gaia DR3 509511974622906496 & 01:43:21.47 & +59:34:09.2 &  & 0.21 $\pm$ 0.02 & 18.2 & 14.5 & 5.91  \\
Gaia DR3 509092884599480704 & 01:41:43.57 & +58:49:22.1 &  & 0.14 $\pm$ 0.01 & 19.6 & 16.4 & 5.97  \\
Gaia DR3 431890542519116928 & 00:15:25.61 & +64:20:23.2 &  & 0.53 $\pm$ 0.05 & 14.9 & 12.3 & 5.97  \\
Gaia DR3 422021291786053120 & 00:18:16.50 & +57:20:46.3 &  & 0.47 $\pm$ 0.05 & 15.4 & 12.6 & 5.97  \\
\label{tab:gamCas_candidates}
\normalsize
\end{longtable}
\tablefoot{
    \tablefoottext{a}{The maximum error in $G$ provided by {\it Gaia} DR3 is less than 0.005\,mag;}
    \tablefoottext{b}{The maximum error in $J$ provided by 2MASS is less than 0.2\,mag;}
    \tablefoottext{c}{\citet{nemravova12};}
    \tablefoottext{d}{\citet{tokovinin21}.}
    }
    
\end{document}